\documentclass[a4paper,twocolumn,english,aps,reprint,superscriptaddress,showpacs,showkeys]{revtex4-1}
\usepackage[latin9]{inputenc}
\usepackage{babel}
\usepackage{mathtools}
\usepackage{amsmath}
\usepackage{amssymb}
\usepackage{graphicx}
\usepackage{float}
\usepackage[colorlinks=true, linkcolor=red, citecolor=blue, urlcolor=blue]{hyperref}
\usepackage[usenames,dvipsnames]{color}
\usepackage{bm}
\usepackage{bbold}
\usepackage{braket}
\usepackage{comment}
\usepackage{dcolumn,extarrows}
\usepackage{soul}
\usepackage{footmisc}
\usepackage{pdfpages}
\usepackage{pgf,tikz}
\usepackage{times}
\usepackage{newtxtext}
\usepackage[libertine]{newtxmath}

\definecolor{mygreen}{rgb}{0,0.5,0}
\definecolor{myblue}{rgb}{0,0,0.75}
\definecolor{mymagenta}{cmyk}{0,1,0,0.12}

\newcommand{\minus}{
  \setbox0=\hbox{-}
  \vcenter{
    \hrule width\wd0 height \the\fontdimen8\textfont3
  }%
}

\def\inner(#1,#2,#3,#4,#5,#6){\ensuremath\left(\begin{array}{ccc} #1 & #2 & #3 \\ #4 & #5 & #6 \end{array}\right)}

\def\innerv(#1,#2,#3,#4,#5,#6){\ensuremath\left\{\begin{array}{ccc} #1 & #2 & #3 \\ #4 & #5 & #6 \end{array}\right\}}

\definecolor{mygreen}{rgb}{0,0.5,0}\definecolor{myblue}{rgb}{0,0,0.75}\definecolor{mymagenta}{cmyk}{0,1,0,0.12}

\newcommand{\mbp}[1]{{\color{green}#1}}

\newcommand{\ketbra}[2]{\ket{ #1}\bra{ #2} }
\newcommand{\bla}[1]{\left( #1 \right)}

\makeatletter
\renewcommand{\fnum@figure}{{\bf Fig.\thefigure}}
\makeatother

\makeatletter
\AtBeginDocument{\let\LS@rot\@undefined}
\makeatother

\begin{document}

\title{Nanoscale magnetic resonance spectroscopy using a carbon nanotube
double quantum dot}
\author{Wanlu Song}
\thanks{These authors contributed equally.}
\affiliation{School of Physics, Huazhong University of Science and Technology, Wuhan
430074, China}
\affiliation{International Joint Laboratory on Quantum Sensing and Quantum
Metrology, Huazhong University of Science and Technology, Wuhan, 430074,
China}
\author{Tianyi Du}
\thanks{These authors contributed equally.}
\affiliation{School of Physics, Huazhong University of Science and Technology, Wuhan
430074, China}
\author{Haibin Liu}
\email{liuhb@hust.edu.cn}
\affiliation{School of Physics, Huazhong University of Science and Technology, Wuhan
430074, China}
\affiliation{International Joint Laboratory on Quantum Sensing and Quantum
Metrology, Huazhong University of Science and Technology, Wuhan, 430074,
China}
\author{Martin B. Plenio}
\affiliation{Institut f\"{u}r Theoretische Physik \& IQST, Albert-Einstein Allee 11,
Universit\"{a}t Ulm, D-89081 Ulm, Germany}
\affiliation{International Joint Laboratory on Quantum Sensing and Quantum
Metrology, Huazhong University of Science and Technology, Wuhan, 430074,
China}
\author{Jianming Cai}
\email{jianmingcai@hust.edu.cn}
\affiliation{School of Physics, Huazhong University of Science and Technology, Wuhan
430074, China}
\affiliation{International Joint Laboratory on Quantum Sensing and Quantum
Metrology, Huazhong University of Science and Technology, Wuhan, 430074,
China}

\begin{abstract}
Quantum sensing exploits fundamental features of quantum mechanics and quantum control to realise sensing devices with potential applications in a broad range of scientific fields ranging from basic science to applied technology. The ultimate goal are devices that combine unprecedented sensitivity with excellent spatial resolution. Here, we propose a new platform for all-electric nanoscale quantum sensing based on a carbon nanotube double quantum dot. Our analysis demonstrates that the platform can achieve sensitivities that allow for the implementation of single-molecule magnetic resonance spectroscopy and therefore opens a promising route towards integrated on-chip quantum sensing devices.
\end{abstract}

\date{\today }

\maketitle

\section{Introduction}

Quantum systems embodying fundamental quantum features offer an appealing perspective in sensing and metrology \cite{Degen_RMP2017,Vahala2007,Kukura2016,Apkarian2018}. Ultra-small quantum sensors provide the possibility for locating them in very close proximity to the target to realise strong sensor-target interaction. This facilitates sensing with both ultra-high measurement sensitivity combined with a nanoscale resolution thus allowing for the identification of nanoscale objects or the detection of signals carrying, for example, magnetic information of nano-structures. The thereof emerging technology of nanoscale magnetic resonance spectroscopy provides a versatile experimental tool to investigate a wide range of physical, chemical and biophysical phenomena in minute sample volumes \cite{glenn2018high,schmitt2017submillihertz,Lukin2016,Shi15,Muller14,Sus14,Stau13,Mamin13,Stau13,Cai2014,Zhang2018_NC}.

There are two key challenges for the implementation of nanoscale magnetic resonance spectroscopy. First, the smallest possible probe-target distance is generally limited by the size of the quantum sensor. Remarkably, nanoscale quantum sensors based on nitrogen-vacancy (NV) center in diamond \cite{Doherty13,Schir14,Wu2016} can achieve sizes of a few nanometers. However, perturbations from the surface then start to significantly affect its sensing capabilities thus limiting further miniaturisation \cite{Tisler09,Rosskopf2014,Myers2014,Kim2015}. Secondly, a scalable architecture of an integrated on-chip quantum sensing device would represent fundamental progress in the development of nanoscale magnetic resonance spectroscopy with appealing practical applications.
In this work, we address both challenges and propose a new type of quantum sensor based on a
valley-spin qubit of a carbon nanotube double quantum dot \cite{Laird2015,Burkard2012,Burkard2009,Briggs2011,Marcus2009,Laird2012,Ishibashi2005,Fujisawa2012} aiming for on-chip nanoscale magnetic resonance spectroscopy. By applying continuous electrical driving on a double quantum dot, the system can efficiently identify the frequency of weak external signals. Due to the nanometer diameter of single walled carbon nanotubes, the valley-spin quantum sensor can be brought extremely close to the target which promises ultra-high sensitivity. Our detailed analysis based on realistic experimental parameters demonstrates that such a carbon nanotube quantum sensor is able to identify the species of individual external nuclei, thus going well beyond both the detection of external ensembles of nuclei \cite{Vander2006} and the detection of a single strongly coupled intrinsic nucleus \cite{Wern2014}, and thereby provides a new platform for nanoscale magnetic resonance spectroscopy. The system can be controlled coherently \cite{Marcus2010,Laird2013} and efficiently readout \cite{Koppens2005,Nazarov2006} electrically. Such all-electric manipulation without requiring optical elements facilitates the integration of on-chip carbon nanotube quantum sensor arrays \cite{Fujisawa2008}. The present result is expected to extend the scope of quantum technologies based on a carbon nanotube double quantum dot system from quantum information processing to nanoscale magnetic resonance spectroscopy.
\section{Model of a nanotube quantum sensor} 
\begin{figure}[t]
\centering
\includegraphics[width=1\linewidth]{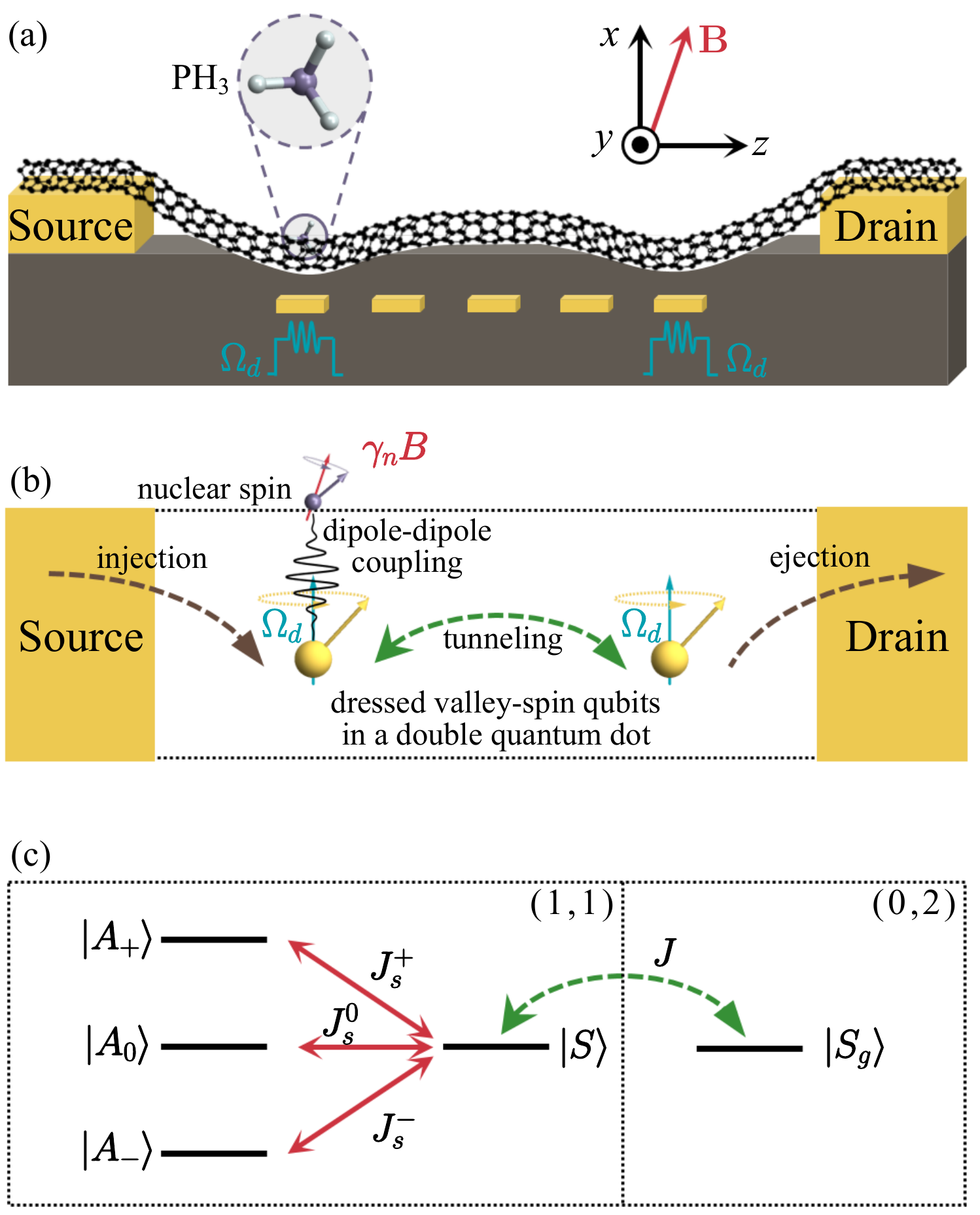}
\caption{{\bf A nanotube quantum sensor for nanoscale magnetic resonance spectroscopy}. (\textbf{a}) Model of a (bent) nanotube quantum sensor for nanoscale magnetic resonance spectroscopy (a PH$_{3}$ molecule shown as an illustrative example). Five local gate electrodes create an electrically driven double quantum dot and control the electron tunneling rates \cite{Laird2013,Marcus2010}. (\textbf{b}) shows the mechanism of a nanotube quantum sensor. Both electrons are electrically driven with a Rabi frequency $\Omega_{d}$, the magnetic dipole-dipole coupling between one electron spin and the external target spin (with a Larmor frequency $\gamma_{n} B$) will lift the Pauli blockade, see (\textbf{c}), and lead to a resonant leakage current when $\Omega_{d}=\gamma_{n} B$. (\textbf{c}) Due to a change in the local environment of the left or right quantum dot, Pauli blockade is lifted by three additional tunnelling channels as denoted by $\left\vert A_{0,\pm}\right\rangle \leftrightarrow\left\vert S\right\rangle \leftrightarrow \left\vert S_{g}\right\rangle $ with the corresponding tunneling rates $J_{S}^{0,\pm}$ and $J$.}
\label{model}
\end{figure}

Our quantum sensor is based on a carbon nanotube double quantum dot, as shown in Fig.\ref{model}(a). In a single-wall carbon nanotube, an electron has two angular momentum quantum numbers, arising from spin and orbital motions. The orbital motion has two flavours known as the $K$ and $K^{\prime }$ valleys, which correspond to the clockwise and counterclockwise motions around the nanotube. Due to the anisotropy of orbital magnetic moment \cite{Lu1995}, the energy levels of electron in carbon nanotube become sensitive to the direction of a magnetic field, which has been applied into the detection of static magnetic fields \cite{Palyi2015,Palyi2017} and electrically driven electron spin resonance \cite{Marcus2010,Laird2013}. Although it has been demonstrated that nuclear magnetic fields may influence electron transport \cite{Nazarov2006} and electron spin resonance \cite{Vander2006} in a double quantum dot confined in the GaAs heterostructure, it is not clear how the mechanism can be engineered for nanoscale magnetic resonance spectroscopy.

The goal of the present work is to design a quantum sensor based on a carbon nanotube double quantum dot system that can achieve a sensitivity on the order of 10 nT$/ \sqrt{\mbox{Hz}}$ for weak oscillating magnetic fields, which is sufficient for achieving nuclear magnetic resonance spectroscopy at the single-molecule level. The key idea which enables us to achieve such a goal is continuous electrical driving on a carbon nanotube double quantum dot which leads to resonant leakage current when the driving Rabi frequency matches the fingerprint frequency of a weak signal (e.g. arising from nuclei), see Fig.\ref{model}(b-c). This in turn allows to obtain the relevant information on the weak signal from the electron transport spectroscopy in Pauli blockade regime \cite{Laird2012}.
In a static magnetic field $\mathbf{B}$, the Hamiltonian of an electron in nanotube is
given by (for simplicity we set $\hbar =1$) \cite{Marcus2010,Laird2014}
\begin{gather}
\hat{H}\left( z\right) =-\frac{1}{2}\Delta _{SO}\hat{\tau}_{3}\mathbf{n}%
\left( z\right) \cdot \boldsymbol{\hat{\sigma}}-\frac{1}{2}\Delta _{KK^{\prime
}}\left( \hat{\tau}_{1}\cos \varphi +\hat{\tau}_{2}\sin \varphi \right)
\notag \\
+\frac{1}{2}g_{s}\mu _{B}\mathbf{B}\cdot \boldsymbol{\hat{\sigma}}+g_{orb}\mu
_{B}\mathbf{B}\cdot \mathbf{n}\left( z\right) \hat{\tau}_{3},
\label{eq-Hamiltonian}
\end{gather}%
where $\boldsymbol{\hat{\tau}}=%
\begin{pmatrix}
\hat{\tau}_{1}, & \hat{\tau}_{2}, & \hat{\tau}_{3}%
\end{pmatrix}%
$ and $\boldsymbol{\hat{\sigma}}=%
\begin{pmatrix}
\hat{\sigma}_{x}, & \hat{\sigma}_{y}, & \hat{\sigma}_{z}%
\end{pmatrix}%
$ are the Pauli operators of valley and spin, $\mathbf{n}(z)=\cos
\theta \left( z\right) \hat{z}+\sin \theta \left( z\right) \hat{x}$ is the
local tangent unit vector with $\theta \left( z\right) $ the angle between $%
\mathbf{n}(z)$ and $\hat{z}$, $\Delta _{SO}$ is the spin-orbit
coupling strength \cite{Kuemmeth2008}, $\Delta _{KK^{\prime }}$ and $\varphi
$ are the magnitude and phase of valley mixing \cite{Burkard2010}, $g_{s}$
and $g_{orb}$ are the $g$ factors of spin and valley respectively. At \textbf{\ }$\theta \left( z_{0}\right) =0$ and $\mathbf{B}=0$, four eigenstates form two Kramers doublets $\{\left\vert
\Uparrow ^{\ast }\right\rangle ,\left\vert \Downarrow ^{\ast }\right\rangle
\}$ and $\{\left\vert \Uparrow \right\rangle ,\left\vert \Downarrow
\right\rangle \}$ which are separated by an energy gap $\Delta
E_{0}=\left( \Delta _{SO}^{2}+\Delta _{KK^{\prime }}^{2}\right) ^{1/2}$ . Each doublet can serve as a valley-spin qubit which shows different
energy splittings in the parallel ($\mathbf{B}=B_{z}\hat{z}$) and
perpendicular ($\mathbf{B}=B_{x}\hat{x}$) magnetic field due to the
anisotropic magnetic moment. As mediated by a bent nanotube \cite{Hels2016}, the qubit can be
electrically driven while the quantum dot is driven back and forth with
frequency $\omega $ and amplitude $\Delta z_{m}$ by applying a microwave frequency
gate voltage.
The effective Hamiltonian of a driven valley-spin qubit in
the magnetic field $\mathbf{B}=\left\{ B_{x},0,B_{z}\right\} $ is \cite%
{Laird2014}
$\hat{H}_{e}=\frac{1}{2}\bla{\omega _{x}\hat{s}_{x}+\omega _{z}\hat{s}_{z}}%
+\Omega _{x}\cos \left( \omega t\right) \hat{s}_{x}+\Omega _{z}\cos \left(
\omega t\right) \hat{s}_{z}$,
where $\hat{s}_{x,y,z}$ are Pauli operators of the valley-spin qubit and $%
\omega _{x}=g_{\perp }\mu _{B}B_{x}$, $\omega _{z}=g_{\parallel }\mu
_{B}B_{z}$, with $g_{\perp }=g_{s}\sin \zeta $, $g_{\parallel
}=g_{s}-2ag_{orb}\cos \zeta $. The characteristic parameter $\zeta $ is
defined as $\tan \zeta =\Delta _{KK^{\prime }}/\Delta _{SO}$, and $a$ takes the value $\pm 1$ for the upper
and lower Kramers doublets respectively. We choose $\omega =\omega _{0}\equiv (\omega _{x}^{2}+\omega _{z}^{2})^{1/2}$ and  obtain a dressed valley-spin qubit under the conditions $\Omega _{x},\Omega _{z}\ll
\omega _{0}$ as described by (see more details in Appendix \ref{AA})
\begin{equation}
\hat{H}_{d}=\frac{1}{2}\Omega _{d}\hat{S}_{x}
\end{equation}
where $\hat{S}_{x}$ is Pauli operator in the eigenbasis of $\hat{H}_{0}=\bla{1/2}\bla{\omega _{x}\hat{s}_{x}+\omega _{z}\hat{s}_{z}}$ and the driving Rabi frequency is $\Omega _{d}= \Omega _{x}\cos \gamma - \Omega _{z}\sin \gamma $ with $\tan \gamma={\omega_x/\omega_z}$. Note that the effect
of fluctuation in the driving fields can be mitigated by concatenated driving schemes \cite{Cai_NJP_2012}.

We consider a double quantum dot in the \emph{n-p} region and encode a valley-spin qubit in the lower Kramers doublet for both quantum dots. In the Pauli blockade regime, electron tunneling is forbidden when two electrons in the $\left(1,1\right)$ configuration are in a triplet state \cite{Hanson2007}. The leakage current can be obtained from the quantum transport master equation (see more details in Appendix \ref{AB}). When the Rabi frequency of an applied continuous driving field on the valley-spin qubits matches the frequency of local signal fields, e.g. from the hyperfine coupling between left quantum dot and a single molecule, additional electron tunnelling channels open up, see Fig.\ref{model}(c). In the following, we show that the change in the leakage current through such a nanotube quantum dot system can serve as a highly sensitive probe for selective detection of localised external signals.

\begin{figure}[b]
\centering
 \includegraphics[width=1\linewidth]{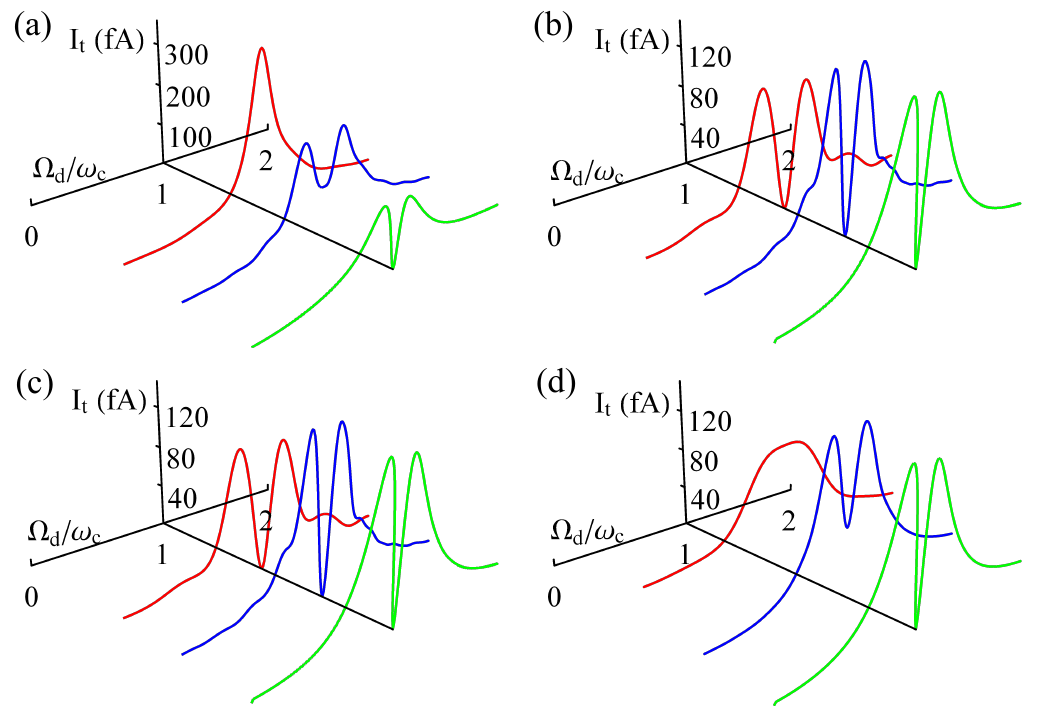} \centering
\caption{{\bf Illustration of quantum sensing mechanism}. (\textbf{a})-(\textbf{d}) Leakage current $I_{t}$ as a function of the driving Rabi frequency $\Omega _{d}/\protect\omega _{c}$ with the initial state $\left\vert A_{0}\right\rangle $ (\textbf{a}), $\left\vert A_{+}\right\rangle $ (\textbf{b}), $\left\vert A_{-}\right\rangle$ (\textbf{c}) and the unpolarized electron spin state (\textbf{d}) in $\left(0,1\right)$ subspace at time $t=0.5 $ $\protect\mu $s (red), $1$ $\protect\mu $s (blue), $10$ $\protect\mu $s (green) respectively. The parameters are $g_{orb}=12$, $\Delta_{SO}=0.8$ meV, $\Delta _{KK^{\prime }}=0.2$ meV, $\protect\varphi =0$ for both electrons \cite{Laird2014,Laird2013}, and $\protect\omega _{c} =\left(2\protect\pi\right)5$ MHz, $b=6$ $\protect\mu $T for the oscillating signal field, the electron injection and ejection rate are $\Gamma _{L} =\Gamma _{R}=\left(2\protect\pi\right) 8$ MHz and the tunneling rate is $J=\left(2\protect\pi\right) 2$ MHz.}
\label{A}
\end{figure}
\section{Sensing of a weak oscillating field} To illustrate the working principle of nanoscale magnetic resonance spectroscopy using a nanotube quantum sensor, we first consider the measurement of an oscillating magnetic field (e.g. arising from a local magnetic moment) $\mathbf{b}\left( t\right) =b\cos \left( \omega _{c}t\right) \hat{z}$ acting on left quantum dot, where the right quantum dot is out of the nanoscale field due to the much larger distance from the left quantum dot. The effective Hamiltonian in the $\left( 1,1\right) $ subspace is
\begin{equation}
\hat{H}_{sb}=\frac{1}{2}\Omega _{d}\hat{S}_{x}^{\left( 1\right) }+\Omega
_{c}\cos \left( \omega _{c}t\right) \hat{S}_{z}^{\left( 1\right) }+\frac{1}{2%
}\Omega _{d}\hat{S}_{x}^{\left( 2\right) },
\end{equation}%
where $\hat{S}_{x,z}^{\left( j\right) }$ are the Pauli operators of left ($j=1$) and right ($j=2$) dressed valley-spin qubit, and $\Omega _{c}=g_{\parallel }\mu _{B}b/2$
represents the coupling strength of left dressed valley-spin qubit to the weak oscillating magnetic field. The Hamiltonian in the $\left( 0,2\right) $ subspace is $H_{\Delta}=\Delta \ketbra{S_g}{S_g}$ with the energy detuning $\Delta$, and the tunnelling Hamiltonian is $H_t=J(\ketbra{S}{S_g}+\ketbra{S_g}{S})$. In the new picture after making a transformation $\hat{S}_{x}\leftrightarrow \hat{S}_{z}$ and using rotating wave approximation, we introduce the basis states including
\begin{gather}
\left\vert A_{0}\right\rangle =\frac{1}{\sqrt{2}}%
\begin{pmatrix}
\cos\vartheta\\
\sin\vartheta \\
\sin\vartheta \\
-\cos\vartheta%
\end{pmatrix}%
, \quad \left\vert A_{\pm}\right\rangle =\frac{1}{2}
\begin{pmatrix}
{\pm}1+\sin\vartheta\\
-\cos\vartheta\\
-\cos\vartheta\\
{\pm}1-\sin\vartheta
\end{pmatrix}%
\end{gather}%
and the singlet state $\left\vert S\right\rangle =(1/{\sqrt{2}})%
\begin{pmatrix}
0 & -1 & 1 & 0%
\end{pmatrix}%
^{T}$
with $\cos\vartheta=\Omega_c/\lambda$, $\sin\vartheta=2\delta/\lambda$,
$\delta =\Omega _{d}-\omega _{c}$ and $\lambda =\bla{4\delta
^{2}+\Omega _{c}^{2}}^{1/2}$, to rewrite the Hamiltonian $\hat{H}_{sb}$
as
\begin{equation}
\hat{H}_{sb}^{\prime \prime }=
\begin{pmatrix}
0 & 0 & 0 & J_{S}^{0} \\
0 & \lambda /2 & 0 & J_{S}^{+} \\
0 & 0 & -\lambda /2 & J_{S}^{-}\\
J_{S}^{0} & J_{S}^{+}& J_{S}^{-}& 0%
\end{pmatrix}%
,
\end{equation}%
where the local field induced tunneling rates are $J_{S}^{0}=-{\Omega _{c}^{2}}/{\left(2\lambda\right) }$ and $J_{S}^{+}=J_{S}^{-}=\delta \Omega _{c}/{\left(\sqrt{2}\lambda\right) }$ (see more details in Appendix \ref{AB}).

%
\begin{figure}[t]
\centering
 \includegraphics[width=1\linewidth]{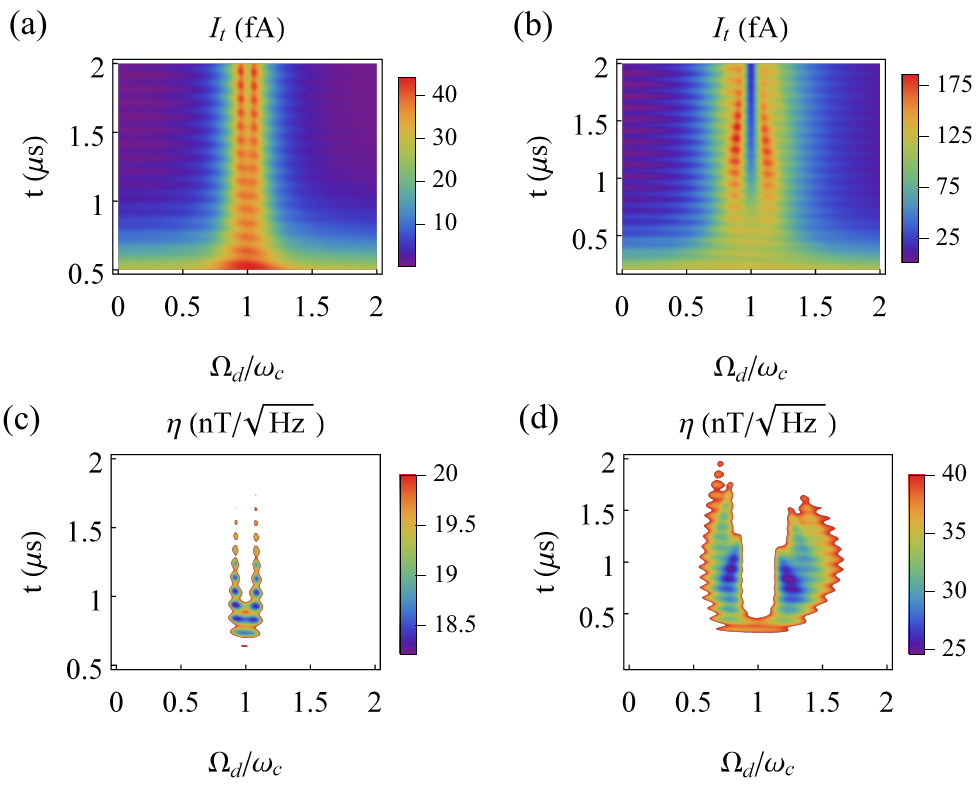}
\caption{{\bf Performance of measurement sensitivity}. (\textbf{a})-(\textbf{b}) Leakage current $I_{t}$ and (\textbf{c})-(\textbf{d}) estimated sensitivity $\eta$ (at the resonant peaks and dip of leakage current) as a function of the driving Rabi frequency $\Omega _{d}/\protect\omega _{c}$ and the evolution time $t$ for an oscillating magnetic field with different amplitudes: (\textbf{a}, \textbf{c}) $b=2$ $\protect\mu $T and (\textbf{b}, \textbf{d}) $b=6$ $\protect\mu $T. The system starts from one unpolarised electron located in the right quantum dot as the initial state. The other parameters are the same as Fig.\protect\ref{A}.}\label{IOmega}
\end{figure}
%

%
The above Hamiltonian reveals two essential ingredients of the present nanotube quantum sensor. Firstly, in the absence of an oscillating magnetic field, all of the channels to the state $\left\vert S\right\rangle $ are closed, and the leakage current is only contributed by the state $\left\vert S\right\rangle $. The external oscillating field opens up three additional channels $\left\vert
A_{0,\pm}\right\rangle \leftrightarrow \left\vert S\right\rangle \leftrightarrow
\left\vert S_{g}\right\rangle $ for electron tunnelling, see Fig.\ref{model}(c), and thus can significantly influence the leakage current. Secondly, the transition $\left\vert A_{0}\right\rangle\leftrightarrow\left\vert S\right\rangle $ is most efficient when $\Omega _{d}=\omega _{c}$, as shown in Fig.\ref{A}(a). In contrast, the
transitions $\left\vert A_{\pm}\right\rangle\leftrightarrow \left\vert S\right\rangle $ play a most significant role with a slight detuning between $\Omega _{d}$ and $\omega _{c}$, which is verified by the resonant dip of leakage current in Fig.\ref{A}(b)-(c). As the electron injected from the source is unpolarised, the total leakage current reflects an overall contribution of all tunnelling channels. As the system evolves, the transitions $\left\vert A_{\pm}\right\rangle \leftrightarrow \left\vert S\right\rangle $ becomes dominant, which leads to a pronounced resonant dip as evident in Fig.\ref{A}(d). These features demonstrate the feasibility of using such a nanotube quantum sensor to selectively detect a weak oscillating magnetic field from driving field induced variations in the leakage current.

By sweeping the Rabi frequency $\Omega _{d}$ of the driving field, a resonance appears in the leakage current when it matches the frequency of the weak oscillating magnetic field emanating from the target (i.e. $\Omega _{d}=\omega _{c}$), as shown in Fig.\ref{IOmega}(a-b). Such a resonance measurement offers an efficient way to identify the frequency of the external signal, which provides a basis for  single-molecule nuclear magnetic resonance spectroscopy. We further analyse the shot-noise limited sensitivity for the measurement of the amplitude of a weak oscillating field from the instantaneous leakage current $I$ at time $t$, which is defined by $\eta =\Delta I \sqrt{t}\left(\partial_{b}I\right) ^{-1}$. We estimate the achievable sensitivity from the measurement of the resonant leakage current in the weak field regime as shown in Fig.\ref{IOmega}(c-d), which implies that the sensitivity can reach the order of 10 nT$/\sqrt{\mbox{Hz}}$ by measuring the instantaneous leakage current after an evolution time of a few microseconds using the feasible experimental parameters given in Fig.\protect\ref{A}.
\section{Nanoscale magnetic resonance spectroscopy} Based on the idea presented and analysed above in the scenario of measuring a weak oscillating signal field, we proceed to demonstrate the applicability of the present scheme for nanoscale magnetic resonance spectroscopy at a single-molecule level. Without loss of generality, we assume that a target molecule is attached on the surface of the nanotube close to the left quantum dot. The interaction strength of magnetic dipole-dipole coupling between the nuclear spins of the target molecule and the valley-spin qubit is $h_{n}=\mu _{0}\mu _{B}\mu _{N}g_{n}g_{\parallel }/\left(4\pi r^{3}\right) $ where $r$ represents the distance from the valley-spin qubit and individual nuclear spins. Two unique features of the present proposal are responsible for its excellent performance, namely a large value of $g_{\parallel }$ (due to a much more prominent orbital $g$-factor $g_{orb}$) and the achievable small sensor-target distance $r$ (which benefits from the compact dimension of nanotube).
%

%
\begin{figure}[t]
\hspace{-0.2cm}
\includegraphics[width=1\linewidth]{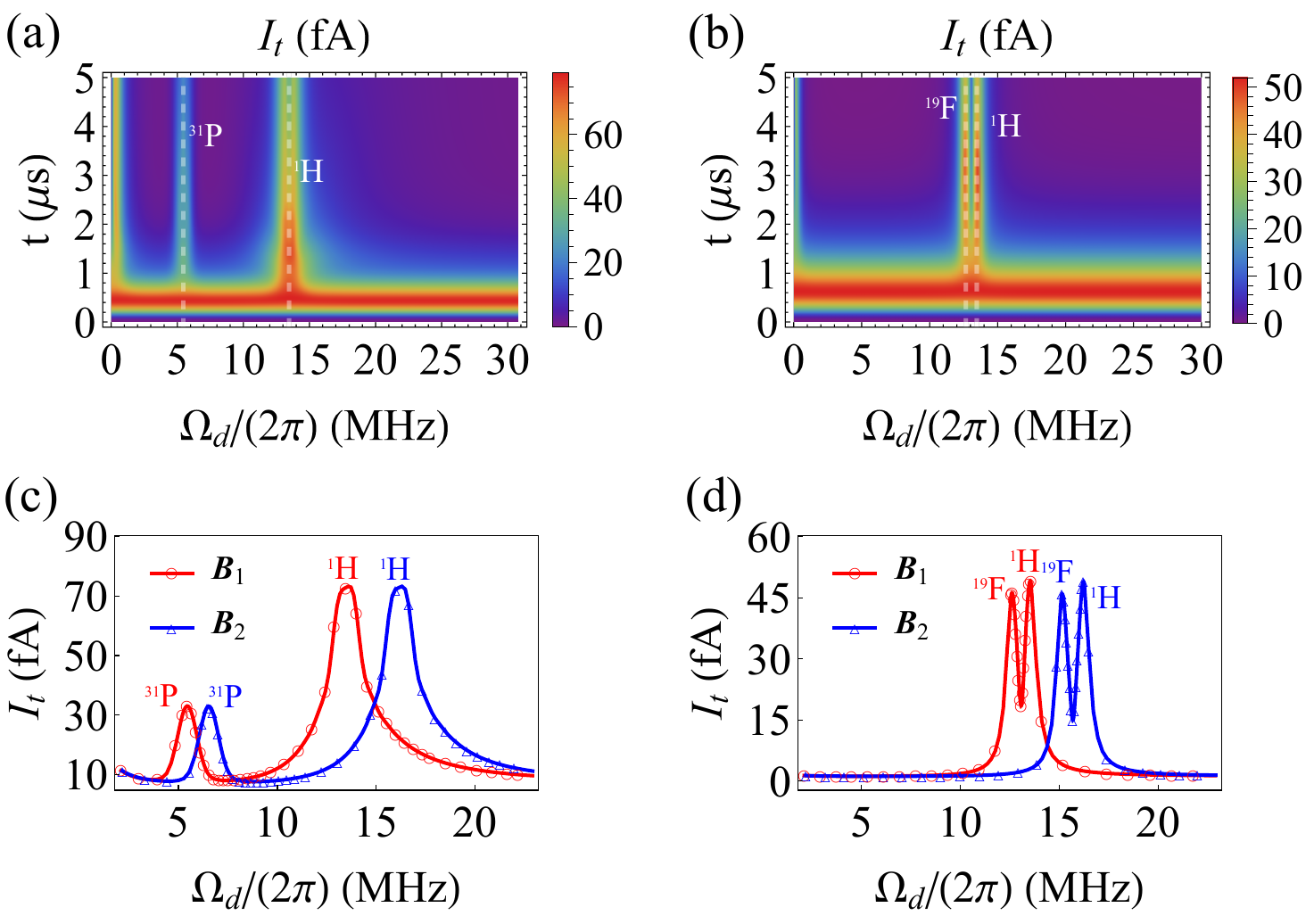}
\caption{{\bf Single-molecule magnetic resonance spectroscopy}. (\textbf{a})-(\textbf{b}) Leakage current $I_{t}$ as a function of the driving Rabi frequency $\Omega _{d}$ and the evolution time $t$ for the detection of a single PH$_{3}$ molecule with $\Gamma _{L}=\left(2\protect\pi\right) 1$ MHz, $\Gamma _{R}=\left(2\protect\pi\right)  0.6$ MHz, $J=\left(2\protect\pi\right) 0.55$ MHz (\textbf{a}) and a single HF molecule with $\Gamma _{L}=\left(2\protect\pi\right) 1$ MHz, $\Gamma _{R}=\left(2\protect\pi\right) 0.4$ MHz, $J=\left(2\protect\pi\right) 0.35$ MHz (\textbf{b}) in the magnetic field $\mathbf{B}=\left\{300,0,100\right\}$ mT. The dash lines (white) in (\textbf{a})-(\textbf{b}) represents the corresponding Larmor frequencies. (\textbf{c})-(\textbf{d}) Leakage current $I_{t}$ as a function of the driving Rabi frequency $\Omega_{d}$ at time $t=1.5$ $\mu$s for the detection of a single PH$_{3}$ molecule (\textbf{c}) and at time $t=3$ $\mu$s for the detection of a single HF molecule (\textbf{d}) in different magnetic fields with $\mathbf{B}_{1}=\left\{300,0,100\right\}$ mT and $\mathbf{B}_{2}=\left\{360,0,120\right\}$ mT. The other parameters are the same as Fig.\ref{A}.
}
\label{molecule}
\end{figure}
%

%
A single molecule is characterized by different species of nuclear spins with multiple Larmor frequencies. For each nuclear spin, an effective magnetic field introduced by its Larmor precession influences the energy levels of left quantum dot through the magnetic dipole-dipole coupling. It leads to individual resonance signals of leakage current that are detected by the present driven nanotube quantum sensor. The identification of these characteristic Larmor frequencies, as implemented by sweeping the Rabi frequency of continuous driving, provides a fingerprint for the detection of single molecules. As an example, we consider phosphine (PH$_{3}$)  and hydrogen fluoride (HF) molecule, both of which are toxic gases. As shown in Fig.\ref{molecule}, owing to the half-integer nuclear spins $^{1}$H, $^{19}$F and $^{31}$P, the leakage current clearly exhibits resonances when the Rabi frequency of continuous driving matches the condition $\Omega _{d}=\gamma _{n}B$, where $\gamma _{n}$ is the gyromagnetic ratio corresponding to individual nuclear species. We remark that electron injection (ejection) rates and the tunneling rate can be tuned by local gate voltages in order to optimize the performance of the protocol (see more details in Appendix \ref{AD}).

\section{Feasibility of experimental realisation} Current experimental advances in fabricating nanotube quantum dot and electrically driven spin resonance quantum control facilitates the implementation of our proposed scheme. The key ingredient for experimental realization is the tuneable Rabi frequency of electric continuous driving  $\Omega _{d}$. The driving Rabi frequency depends on the bending parameter $ \left( \partial _{z}\theta \right) _{z=z_{0}}$ and the oscillation amplitude $\Delta z_{m}$ of the of quantum dot, the required value of which is feasible with the state-of-the-art experiment capability \cite{Laird2013,Laird2014}. The scheme prefers two valley-spin qubits that have the same valley mixing $\Delta_{KK^{\prime }}^{\left( j\right) }$ thereby the same characteristic parameter $\zeta _{j}$ for two valley-spin qubits. In order to compensate for non-uniformity in quantum dots and achieve the best sensing performance, we adopt a bent arc shape nanotube with an appropriate tilted angle. By applying a magnetic field in $\hat{x}$-$\hat{z}$ plane with the proper components of $B_{x}$, $B_{z}$, we find that two valley-spin qubits can have identical parameters $\zeta _{j}$ (see more details in Appendix \ref{AB}).

The main decoherence that may affect the performance of the present carbon nanotube quantum sensor arises from the thermal phonons, the environmental nuclear spins and the charge fluctuation. At low temperature $T=100$ K, the dominated bending-mode phonon-mediated spin relaxation time is about $1$ $\mu$s \cite{Churchill2009}, which is much longer than the electron tunnelling time in quantum dot. The influence of nuclei may be mitigated by synthesizing carbon nanotube with isotopically purified $^{12}$CH$_{4}$, allowing for the fabrication of almost nuclei-spin-free devices \cite{Laird2015,Bulaev2008,Rudner2010}. In addition, the driving via continuous control fields serves to suppress noise effects from nuclear impurities in the device, which underlines their importance in our scheme. Our numerical simulation shows that the performance is robust given feasible isotopic engineering (see more details in Appendix \ref{AC}). As the carbon nanotube quantum dot is gate-defined, the charge noise would modify the energy levels of the quantum dot, i.e., inducing fluctuations of the energy detuning $\Delta$ between the singlet states $\left\vert S\right\rangle$ and $\left\vert S_{g}\right\rangle$ \cite{Palyi2017}, the role of which in our scheme is mainly the suppression of effective tunneling. The charge noise is slow \cite{Jiang2016} and its influence can be compensated by optimizing the parameters $\Gamma_{L}$, $\Gamma_{R}$ and $J$ to sustain the leakage current (see more details in Appendix \ref{AC}). We stress that this is quite different from the dephasing effect on the coherence time of qubit involving the singlet state $\ket{S_g}$ in the (0,2) subspace \cite{Laird2013}, where the energy splitting of the qubit relies on the energy detuning $\Delta$.

We note that the higher-order tunneling (cotunneling) processes \cite{Coish_2011_PRB,Lai_2011_SR1} may also have influence on our scheme. As the cotunneling corrections $\varpropto \Gamma_{L}\Gamma_{R}$ \cite{Coish_2011_PRB,Hanson2004},it is helpful to make the tunnel rates small for the suppression of cotunneling current. In our scheme, the tunnel rates $\Gamma_{L},\Gamma_{R}$ are $1 \sim 2$ orders of magnitude smaller than the values in Ref.\cite{Lai_2011_SR1}, thus the cotunneling current here is estimated to be much less than $3$ fA, which would not significantly influence the performance of our scheme. Overall, we remark that the implementation of our proposal may require new experimental efforts, the feasibility of which appears promising.
\section{Conclusions \& Outlook} To summarise, we propose a new platform for nanoscale magnetic resonance spectroscopy using a continuously driven carbon nanotube double quantum dot as a quantum sensor. The system allows to achieve a high sensitivity due to its unique features of a large valley $g$-factor and ultra-small dimension. In particular, our simulation demonstrates that such a quantum sensor may identify individual nuclear spin and detect a single molecule. The all-electric control and readout techniques make it appealing for on-chip quantum sensing device integration. Assisted by the functionalized carbon nanotube \cite{FCN1,FCN2}, such a quantum sensor can serve as a nanoscale probe to capture the target molecule selectively and provide a new route to implement nanoscale magnetic resonance spectroscopy at a single-molecule level with a wide range of potential applications both in basic science and applied technology.

\section*{Acknowledgments} We thank Ying Li, Guido Burkard and Andras Palyi for valuable discussions and suggestions. The work is supported by National Natural Science Foundation of China (11874024, 11574103, 11690030, 11690032). W.S. is supported by the Postdoctoral Innovation Talent Program, H.L. is also supported by the China Postdoctoral Science Foundation grant (2016M602274). M.B.P. is supported by the EU projects ASTERIQS and HYPERDIAMOND, the ERC Synergy grant BioQ and the BMBF via DiaPol and NanoSpin.

\begin{appendix}

\section{Derivation of effective Hamiltonian}\label{AA}

Electrons in a nanotube have two angular momentum quantum numbers, arising from
the spin and the valley degree of freedom. These two degrees of freedom are coupled
via spin-orbit interaction \cite{Kuemmeth2008}. In addition, two valley states
are coupled to each other by electrical disorder and contact electrodes \cite%
{Burkard2010}. We introduce the identity and Pauli matrices in the spin
space $\hat{\sigma}_{i}$ with $i=\left\{ 0,x,y,z\right\} $ and the
three-dimensional spin vector $\boldsymbol{\hat{\sigma}\equiv }\left\{ \hat{%
\sigma}_{x},\hat{\sigma}_{y},\hat{\sigma}_{z}\right\} $. The positive and
negative projections of $\hat{\sigma}_{z}$ (component of $\mathbf{\hat{\sigma%
}}$ along $\hat{z}$-axis) are denoted by $\left\{ \left\vert \uparrow
\right\rangle ,\left\vert \downarrow \right\rangle \right\} $. Similarly,
the identity and Pauli matrices in the valley space are denoted as $\hat{\tau%
}_{j}$ with $j=\left\{ 0,1,2,3\right\} $ and the three-dimensional valley
vector $\boldsymbol{\hat{\tau}}\equiv \left\{ \hat{\tau}_{1},\hat{\tau}_{2},\hat{%
\tau}_{3}\right\} $, where we choose $\left\{ \left\vert K^{\prime
}\right\rangle ,\left\vert K\right\rangle \right\} $ as the positive and
negative projections of $\hat{\tau}_{3}$ (component of $\mathbf{\hat{\tau}}$
along $\mathbf{n}\left( z\right) =\cos \theta \left( z\right) \mathbf{z}%
+\sin \theta \left( z\right) \mathbf{x}$ which is a local tangent unit
vector of the nanotube with $\theta \left( z\right) $ the angle between $%
\mathbf{n}\left( z\right) $ and $\mathbf{z}$). In a static magnetic field $%
\mathbf{B}$, the Hamiltonian of an electron can be written as \cite%
{Marcus2010,Laird2014}
\begin{eqnarray}
  \hat{H} &=& -\frac{1}{2}\Delta _{SO}\hat{\tau}_{3}\mathbf{n}\left( z\right)
\cdot \boldsymbol{\hat{\sigma}}-\frac{1}{2}\Delta _{KK^{\prime }}\left( \hat{\tau%
}_{1}\cos \varphi +\hat{\tau}_{2}\sin \varphi \right) \nonumber\\
    & & +\frac{1}{2}g_{s}\mu
_{B}\mathbf{B}\cdot \boldsymbol{\hat{\sigma}}+g_{orb}\mu _{B}\mathbf{B}\cdot
\mathbf{n}\left( z\right) \hat{\tau}_{3},
\end{eqnarray}\label{eq-si:Ham}
where $\Delta _{SO}$ is the spin-orbit coupling strength, $\Delta
_{KK^{\prime }}$ and $\varphi $ are the magnitude and phase of valley
mixing, $g_{s}$ and $g_{orb}$ are the spin and orbital $g$ factors
respectively. We remark that $g_{orb}$ is much larger than $g_{s}$, and
would provide an advantage for magnetic field sensing \cite{Palyi2017}. At%
\textbf{\ }$\theta \left( z_{0}\right) =0$ and $\mathbf{B}=0$, four
eigenstates are separated by an energy gap $\Delta E_{0}=\left( \Delta
_{SO}^{2}+\Delta _{KK^{\prime }}^{2}\right) ^{1/2}$ and form two Kramers
doublets $\{\left\vert \Uparrow ^{\ast }\right\rangle ,\left\vert \Downarrow
^{\ast }\right\rangle \}$ and $\{\left\vert \Uparrow \right\rangle
,\left\vert \Downarrow \right\rangle \}$ with%
\begin{eqnarray}
\left\vert \Uparrow ^{\ast }\right\rangle &=&-\cos \left( \zeta /2\right)
\left\vert K^{\prime }\right\rangle \left\vert \downarrow \right\rangle
+\sin \left( \zeta /2\right) \left\vert K\right\rangle \left\vert \downarrow
\right\rangle ,  \label{s1} \\
\left\vert \Downarrow ^{\ast }\right\rangle &=&-\sin \left( \zeta /2\right)
\left\vert K^{\prime }\right\rangle \left\vert \uparrow \right\rangle +\cos
\left( \zeta /2\right) \left\vert K\right\rangle \left\vert \uparrow
\right\rangle , \\
\left\vert \Uparrow \right\rangle &=&\cos \left( \zeta /2\right) \left\vert
K^{\prime }\right\rangle \left\vert \uparrow \right\rangle +\sin \left(
\zeta /2\right) \left\vert K\right\rangle \left\vert \uparrow \right\rangle ,
\\
\left\vert \Downarrow \right\rangle &=&\sin \left( \zeta /2\right)
\left\vert K^{\prime }\right\rangle \left\vert \downarrow \right\rangle
+\cos \left( \zeta /2\right) \left\vert K\right\rangle \left\vert \downarrow
\right\rangle ,  \label{s4}
\end{eqnarray}%
with $\tan \zeta =\Delta _{KK^{\prime }}/\Delta_{SO}$ (without loss of generality we consider $\varphi =0$), either of which can serve as a valley-spin qubit.

\vspace{0.05in}

Electrons in nanotube can be longitudinally confined to form a quantum dot by introducing tunnel barriers which can be created by modifying the electrostatic potential with gate voltages. For a double quantum dot, even the tunnelling of a single electron is permitted by Coulomb blockade, the
transition from a ground $\left( 1,1\right) $-triplet state with one
electron in each dot to a ground $\left( 0,2\right) $-singlet state with
both electrons in the right dot is blocked by Pauli exclusion principle,
hence the leakage current is zero. In carbon nanotube, the energy difference between
an excited $\left( 0,2\right) $-triplet state and a ground $\left(
0,2\right) $-singlet state can be one or two orders of magnitude smaller
than in III-V materials, which gives rise to the transition from a ground $%
\left( 1,1\right) $-triplet state to an excited $\left( 0,2\right) $-triplet
state, hence the Pauli blockade does not work perfectly. A robust Pauli blockade in
carbon nanotube is most evident with a double quantum dot tuned into the
n-p region, where the first shells of electrons and holes are separated by a
large gap \cite{Laird2012}.

\vspace{0.05in}

Based on Pauli blockade in a double quantum dot, we consider two valley-spin
qubits both of which are encoded in the lower Kramers doublet $\{\left\vert
\Uparrow \right\rangle ,\left\vert \Downarrow \right\rangle \}$, then the
leakage current can be regarded as a {\it meter} of the right valley-spin qubit and the left valley-spin qubit servers as a quantum {\it probe} interacting with a target. In our scheme, two dressed
valley-spin qubits in a double quantum dot can be used as a nanotube quantum
sensor to detect e.g. a local magnetic field or a locally interacting spin.

\subsection{Dressed valley-spin qubit} \label{sec-si:dressed_qubit}

A valley-spin qubit in a static magnetic field $\mathbf{B}$ can be electrically
driven when the quantum dot in a bent nanotube is driven by an microwave gate voltage \cite{Marcus2010,Laird2013}. We denote the
frequency and the amplitude of the driven motion of the quantum dot as $%
\omega $ and $\Delta z_{m}$. The effective Hamiltonian of such a driven
valley-spin qubit can be written as follows \cite{Laird2014}
\begin{equation}
\hat{H}_{e}=\frac{1}{2}\mathbf{g}^{\ast }\mu _{B}\cdot \mathbf{B}\cdot
\mathbf{\hat{s}}+\Omega _{x}\cos \left( \omega t\right) \hat{s}_{x}+\Omega
_{z}\cos \left( \omega t\right) \hat{s}_{z},  \label{Heff1}
\end{equation}%
where $\mathbf{\hat{s}=}\left\{ \hat{s}_{x},\hat{s}_{y},\hat{s}_{z}\right\} $
is the Pauli operator of a valley-spin qubit in the basis of the lower
Kramers doublet $\left\{ \left\vert \Uparrow \right\rangle ,\left\vert
\Downarrow \right\rangle \right\} $ with%
\begin{eqnarray}
\left\vert \Uparrow \right\rangle &=&\cos \left( \zeta /2\right) \left\vert
K^{\prime }\right\rangle \left\vert \uparrow \right\rangle +\sin \left(
\zeta /2\right) \left\vert K\right\rangle \left\vert \uparrow \right\rangle ,
\\
\left\vert \Downarrow \right\rangle &=&\sin \left( \zeta /2\right)
\left\vert K^{\prime }\right\rangle \left\vert \downarrow \right\rangle
+\cos \left( \zeta /2\right) \left\vert K\right\rangle \left\vert \downarrow
\right\rangle .
\end{eqnarray}%
The effective $g$ tensor is
\begin{equation}
\mathbf{g}^{\ast }=%
\begin{pmatrix}
g_{\perp } & 0 & 0 \\
0 & g_{\perp } & 0 \\
0 & 0 & g_{\parallel }%
\end{pmatrix}%
\end{equation}%
with $g_{\perp }=g_{s}\sin \zeta $, $g_{\parallel }=g_{s}+2g_{orb}\cos \zeta
$. The effective driving Rabi frequencies are
\begin{eqnarray}
\Omega _{x} &=&\sin\left( 2\zeta\right) g_{orb}\mu _{B}B_{z}\delta _{\theta }/2, \\
\Omega _{z} &=&\left( 2 g_{orb}\cos \zeta+g_{s}\cos ^{2}\zeta\right) \mu
_{B}B_{x}\delta _{\theta }/2,
\end{eqnarray}%
with $\delta _{\theta }=\left( \partial _{z}\theta \right) _{z=z_{0}}\Delta
z_{m}$. Considering a magnetic field $\mathbf{B}=\left\{
B_{x},0,B_{z}\right\} $ in the $x$-$z$ plane, the effective Hamiltonian can
be written as%
\begin{equation}
\hat{H}_{e}=\frac{1}{2}\left(\omega _{x}\hat{s}_{x}+\omega _{z}\hat{s}%
_{z}\right)+\Omega _{x}\cos \left( \omega t\right) \hat{s}_{x}+\Omega _{z}\cos
\left( \omega t\right) \hat{s}_{z}  \label{Heff2}
\end{equation}%
where
\begin{eqnarray}
\omega _{x} &=&g_{\perp }\mu _{B}B_{x}, \\
\omega _{z} &=&g_{\parallel }\mu _{B}B_{z}.
\end{eqnarray}%
The eigenvalues of $\hat{H}_{0}=\frac{1}{2}\left(\omega _{x}\hat{s}_{x}+%
\omega _{z}\hat{s}_{z}\right)$ are
\begin{equation}
\epsilon _{1,2}=\pm \sqrt{\omega _{x}^{2}+\omega _{z}^{2}}/2,
\end{equation}%
and the corresponding eigenstates are
\begin{eqnarray}
\left\vert \psi _{1}\right\rangle &=&\cos \left( \gamma /2\right) \left\vert
\Uparrow \right\rangle +\sin \left( \gamma /2\right) \left\vert \Downarrow
\right\rangle ,  \label{psi1} \\
\left\vert \psi _{2}\right\rangle &=&-\sin \left( \gamma /2\right)
\left\vert \Uparrow \right\rangle +\cos \left( \gamma /2\right) \left\vert
\Downarrow \right\rangle  \label{psi2}
\end{eqnarray}%
with
\begin{eqnarray}
\cos \gamma &=&\omega _{z}/\sqrt{\omega _{x}^{2}+\omega _{z}^{2}}, \\
\sin \gamma &=&\omega _{x}/\sqrt{\omega _{x}^{2}+\omega _{z}^{2}}.
\end{eqnarray}%
We can rewrite the Hamiltonian $\hat{H}_{e}$ in the basis of $\{\ket{\psi_1},\ket{\psi_2}\}$ as follows
\begin{equation}
\hat{H}_{e}^{\prime }=\frac{1}{2}\omega _{0}\hat{S}_{z}+g_{z}\cos \left(
\omega t\right) \hat{S}_{z}+g_{x}\cos \left( \omega t\right) \hat{S}_{x}
\end{equation}%
with%
\begin{eqnarray}
\omega _{0} &=&\sqrt{\omega _{x}^{2}+\omega _{z}^{2}}, \\
g_{z} &=&\Omega _{z}\cos \gamma +\Omega _{x}\sin \gamma , \\
g_{x} &=&\Omega _{x}\cos \gamma -\Omega _{z}\sin \gamma ,
\end{eqnarray}%
where $\hat{S}_{z}$ and $\hat{S}_{x}$ are Pauli matrices in the basis of $%
\left\vert \psi _{1}\right\rangle $ and $\left\vert \psi _{2}\right\rangle $. We choose $\omega =\omega _{0}$ and use rotating-wave approximation under
the conditions $g_{x},g_{z}\ll \omega _{0}$, which leads to a dressed valley-spin qubit
system with the following effective Hamiltonian as
\begin{equation}
\hat{H}_{d}=\frac{1}{2}g_{x}\hat{S}_{x}.
\end{equation}

%
\begin{figure*}
\centering
\includegraphics[width=12cm]{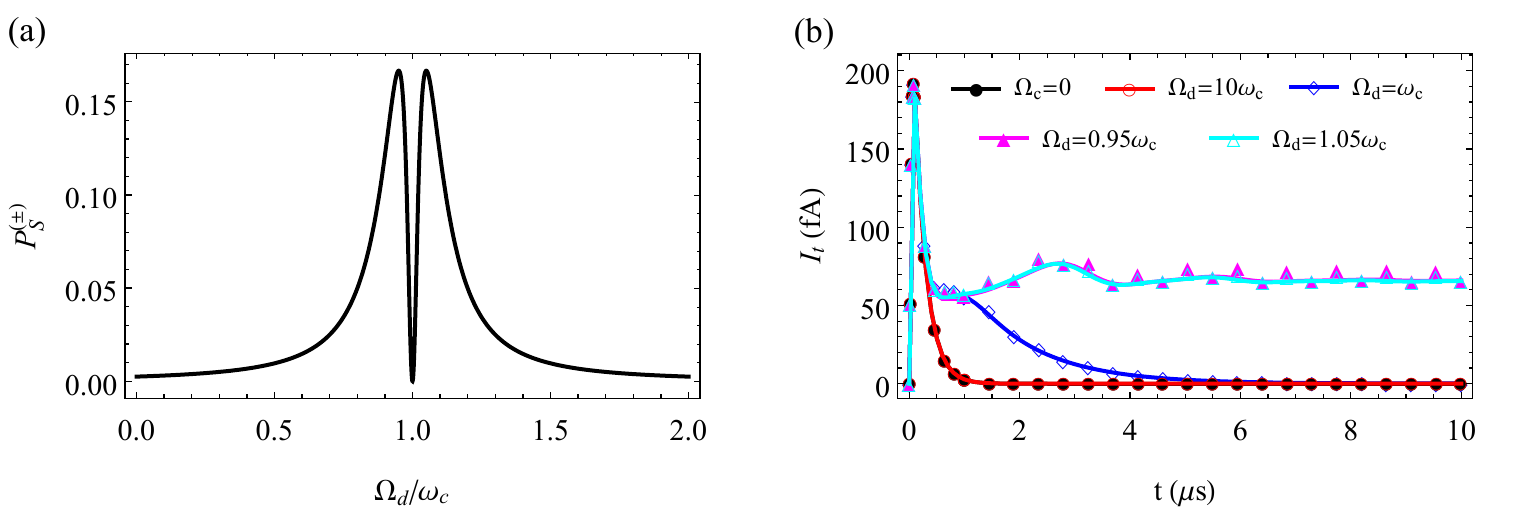}
\caption{(Color online) ({a}) The average population of the singlet state $
\ket{S}$ in the steady state as a function of the driving Rabi
frequency $\Omega _{d}/\omega_{c}$. $P_{S}^{\left(\pm\right) }$
correspond to the initial state $\ket{A_\pm}$ respectively.
({b}) Leakage current $I_{t}$ as a function of the evolution time $t$ in the following situations: $\Omega _{c}=0$ (no oscillating magnetic field), $\Omega _{d}=10\protect\omega _{c}$ (far detuning), $\Omega _{d}=%
\protect\omega _{c}$ (on resonance), $\Omega _{d}=0.95\protect\omega _{c}$
and $\Omega _{d}=1.05\protect\omega _{c}$ (near resonance), with one unpolarised electron located in the right quantum dot as the initial state. The other
parameters are $\protect\omega _{c}=\left(2\protect\pi\right) 5$ MHz, $\Omega _{c}=\left(2
\protect\pi\right) 0.5$ MHz, $\Gamma _{L}= \Gamma _{R}=\left(2 \protect\pi\right) %
8$ MHz and $J=\left(2\protect\pi\right) 2$ MHz.}
\label{It}
\end{figure*}
%

\subsection{Coupling between a dressed valley-spin qubit and a local oscillating signal field}

We first consider the situation in which a driven valley-spin qubit is
coupled to a weak oscillating signal field $\mathbf{b}=b\cos \left(
\omega _{c}t\right) \mathbf{z}$ in additional to the static magnetic field $%
\mathbf{B}=\left\{ B_{x},0,B_{z}\right\} $ with $\left\vert \mathbf{b}%
\right\vert \ll \left\vert \mathbf{B}\right\vert $. According to Eqs.\ref%
{Heff1}-\ref{Heff2}, the total Hamiltonian can be written as
\begin{eqnarray}
  \hat{H}_{eb} &=& \frac{1}{2}\left(\omega _{x}\hat{s}_{x}+\omega _{z}\hat{s}%
_{z}\right)+\Omega _{x}\cos \left( \omega t\right) \hat{s}_{x} \nonumber\\
    & & +\Omega _{z}\cos
\left( \omega t\right) \hat{s}_{z}+\Omega _{c}\cos \left( \omega
_{c}t\right) \hat{s}_{z},
\end{eqnarray}
with $\Omega _{c}={g_{\parallel }\mu _{B}b}/{2}$, where $\mathbf{b}$
only contributes to the first-order perturbative approximation of $\hat{H}$ (Eq.\ref{eq-si:Ham}).
The above Hamiltonian $\hat{H}_{eb}$ can be written in the eigenbases of $\hat{H}_{0}=\frac{1}{2}\left(\omega _{x}\hat{s}_{x}+\omega _{z}\hat{s}%
_{z}\right)$ as follows
\begin{eqnarray}
  \hat{H}_{eb}^{\prime } &=& \frac{1}{2}\omega _{0}\hat{S}_{z}+g_{z}\cos \left(
\omega t\right) \hat{S}_{z}+g_{x}\cos \left( \omega t\right) \hat{S}%
_{x}+\Omega _{c}\cos \gamma  \nonumber\\
    & & \cos \left( \omega _{c}t\right) \hat{S}%
_{z}-\Omega _{c}\sin \gamma \cos \left( \omega _{c}t\right) \hat{S}_{x}.
\end{eqnarray}
By choosing $\omega =\omega _{0}$ and using rotating-wave approximation under
the conditions $\Omega _{c}\ll \omega _{c},g_{x},g_{z}\ll \omega _{0}$, we
can obtain the following effective Hamiltonian for a dressed valley-spin
qubit coupled to an oscillating magnetic field as
\begin{equation}
\hat{H}_{db}=\frac{1}{2}g_{x}\hat{S}_{x}+\Omega _{c}\cos \left( \omega
_{c}t\right) \hat{S}_{z},
\end{equation}%
where we assume $\cos \gamma \approx 1$ that is valid when $\omega _{x}\ll \omega _{z}$.

\subsection{Coupling between a dressed valley-spin qubit and a nuclear spin}

We proceed to consider the situation in which a driven valley-spin qubit is
coupled to a nuclear spin via magnetic dipole-dipole interaction. The
interaction strength is usually much weaker than the static magnetic field $%
\mathbf{B}=\left\{ B_{x},0,B_{z}\right\} $. The Hamiltonian of the total
system is
\begin{eqnarray}
\hat{H}_{en} &=&\frac{1}{2}\left(\omega _{x}\hat{s}_{x}+\omega _{z}\hat{%
s}_{z}\right)+\Omega _{x}\cos \left( \omega t\right) \hat{s}_{x}+\Omega _{z}\cos
\left( \omega t\right) \hat{s}_{z}  \notag \\
&+& g_{n}\mu _{N}\mathbf{B}\cdot \mathbf{\hat{%
I}}+\left( \frac{\mu _{0}\mu _{B}\mu _{N}}{4\pi r^{3}}\right) \left\{ \left(
\mathbf{g}^{\ast }\cdot \mathbf{\hat{s}}/2\right) \cdot \left( g_{n}\mathbf{%
\hat{I}}\right)\right. \notag\\
&-&\left.3\left[ \left( \mathbf{g}^{\ast }\cdot \mathbf{\hat{s}}%
/2\right) \cdot \mathbf{n}_{\mathbf{r}}\right] \left[ \left( g_{n}\mathbf{%
\hat{I}}\right) \mathbf{\cdot n}_{\mathbf{r}}\right] \right\} ,
\end{eqnarray}%
where $g_{n}$ is the $g$ factor of the nuclear spin, $\mathbf{\hat{I}=}%
\left\{ \hat{I}_{x},\hat{I}_{y},\hat{I}_{z}\right\} $ is the spin operator
of the nuclear spin and $\mathbf{r}=r\mathbf{n}_{\mathbf{r}}$ is the vector
connecting the valley-spin qubit and the nuclear spin with a distance $r$
and a unit vector $\mathbf{n}_{\mathbf{r}}=\left\{ n_{x},n_{y},n_{z}\right\}
$. Written in the eigenbases of $\hat{H}_{0}$, one can obtain
\begin{eqnarray}
\hat{H}_{en}^{\prime } &=&\frac{1}{2}\omega _{0}\hat{S}_{z}+g_{z}\cos \left(
\omega t\right) \hat{S}_{z}+g_{x}\cos \left( \omega t\right) \hat{S}%
_{x}+g_{n}\mu _{N}\mathbf{B}\cdot \mathbf{\hat{I}}  \notag \\
&+&\left( \frac{\mu _{0}\mu _{B}\mu _{N}g_{n}}{8\pi r^{3}}\right) \left\{ %
\left[ g_{\perp }\left( \sin \gamma \hat{S}_{z}+\cos \gamma \hat{S}%
_{x}\right) \hat{I}_{x}+g_{\perp }\hat{S}_{y}\hat{I}_{y}\right.\right. \notag\\
&+& \left.\left.g_{\parallel}\left( \cos \gamma \hat{S}_{z}-\sin \gamma \hat{S}_{x}\right) \hat{I}_{z}%
\right]-3\left[ g_{\perp }n_{x}\left( \sin \gamma \hat{S}_{z}+\cos \gamma
\hat{S}_{x}\right)\right.\right. \notag\\
&+& \left.\left.g_{\perp }n_{y}\hat{S}_{y}+g_{\parallel }n_{z}\left(
\cos \gamma \hat{S}_{z}-\sin \gamma \hat{S}_{x}\right) \right]\right.\notag\\
& &\left.\left( n_{x}%
\hat{I}_{x}+n_{y}\hat{I}_{y}+n_{z}\hat{I}_{z}\right) \right\}.
\end{eqnarray}%
Similarly, we choose $\omega =\omega _{0}$ and use rotating wave
approximation under the conditions $g_{x},g_{z}\ll \omega _{0}$ and $\left( {%
\mu _{0}\mu _{B}\mu _{N}g_{n}}\right) /\left( 8\pi r^{3}\right) \ll \omega
_{0}-\left( g_{n}\mu _{N}\left\vert \mathbf{B} \right\vert \right)$, thereby
obtain the following effective Hamiltonian for a dressed valley-spin qubit
coupled with a nuclear spin as described by
\begin{eqnarray}
  \hat{H}_{dn} &=& \frac{1}{2}g_{x}\hat{S}_{x}+g_{n}\mu _{N}\mathbf{B}\cdot
\mathbf{\hat{I}}+\frac{h_{n}}{2}\left[ \hat{S}_{z}\hat{I}_{z}\right. \notag\\
    &-& \left.3n_{z}\hat{S}_{z}\left( n_{x}\hat{I}%
_{x}+n_{y}\hat{I}_{y}+n_{z}\hat{I}_{z}\right) \right]
\end{eqnarray}
with $h_{n}={\mu _{0}\mu _{B}\mu _{N}g_{n}g_{\parallel }}/\left({4\pi
r^{3}}\right)$,
where we assume $\omega _{x}\ll \omega _{z}$ and thus $\cos \gamma \approx 1$. We remark that the above Hamiltonian can be straightforwardly generalised
to the scenario of multiple nuclear spins.

\section{Detailed mechanism of a nanotube quantum sensor}\label{AB}

To illustrate the basic idea, here we present further details on the sensing
mechanism for the detection of a weak oscillating magnetic field using a
nanotube quantum sensor. The system dynamics is governed by the following quantum transport master equation as \cite{Gurvitz1996,Li2005}
\begin{equation}
\dot{\rho}_{t}=-i\left[ \mathcal{\hat{H}},\rho
_{t}\right] +\mathcal{L}\rho_{t},
\end{equation}
with
\begin{equation}
\mathcal{\hat{H}}=
\begin{pmatrix}
\mathcal{\hat{H}}_{I} & 0\\
0 & \mathcal{\hat{H}}_{II}
\end{pmatrix},
\end{equation}
where $\mathcal{\hat{H}}_{I}$ and $\mathcal{\hat{H}}_{II}$ correspond to the $\left( 0,1\right) $ and $%
\left( 1,1\right) \oplus \left( 0,2\right) $ subspaces respectively, namely
\begin{equation}
\mathcal{\hat{H}}_{I}=\frac{1}{2}\Omega _{d}\hat{S}_{x}^{\left( 2\right) },
\end{equation}
and
\begin{eqnarray}
  \mathcal{\hat{H}}_{II} &=& \frac{1}{2}\Omega _{d}\hat{S}_{x}^{\left( 1\right) }+\Omega
_{c}\cos \left( \omega _{c}t\right) \hat{S}_{z}^{\left( 1\right) }+\frac{1}{2}\Omega _{d}\hat{S}_{x}^{\left( 2\right) } \notag\\
    &+& \Delta \left\vert S_{g}\right\rangle\left\langle S_{g}\right\vert+J\left(\left\vert S\right\rangle\left\langle S_{g}\right\vert+\left\vert S_{g}\right\rangle\left\langle S\right\vert\right),
\end{eqnarray}
where $\left\vert S\right\rangle$ and $\left\vert S_{g}\right\rangle$ are the singlet states in $\left(1,1\right)$ and $\left(0,2\right)$ subspaces respectively. We note that $H_{t}=J\left(\left\vert S\right\rangle\left\langle S_{g}\right\vert+\left\vert S_{g}\right\rangle\left\langle S\right\vert\right)$ represents the tunneling between two quantum dots, and $H_{\Delta}=\Delta \left\vert S_{g}\right\rangle\left\langle S_{g}\right\vert$ is the Hamiltonian in the $\left( 0,2\right)$ subspace. The superoperator $\mathcal{L}$ is generated by Lindblad operators $L_1=\sqrt{%
\Gamma _{L}}\hat{a}_{1\psi }^{\dag }$ and $L_2=\sqrt{\Gamma _{R}}\hat{a}%
_{2\psi }$ describing the processes, by which an unpolarised electron is injected
from the source at a rate $\Gamma _{L}$ and is ejected to the drain at a
rate $\Gamma _{R}$, where $\psi $ denotes a set of complete and orthogonal basis states of a valley-spin qubit. The leakage current at time $t$ can be calculated as follows
\begin{equation}
I\left( t\right) =\bla{e\Gamma_{R}}\sum_{\psi}\mbox{Tr}\bla{\hat{a}_{2\psi }^{\dag }\hat{a}_{2\psi }\rho _{t}}.
\label{eq:current}
\end{equation}
On the other hand, we derive the shot noise of the leakage current (see Eq.\ref{eq:current}) as follows
\begin{equation}
\Delta I^{2}=(e\Gamma _{R})^{2}\sum_{\psi}\left\{\mbox{Tr}\left[ \left( \hat{a}_{2\psi }^{\dag }\hat{a}_{2\psi}\right) ^{2}\rho _{t}\right] -\left[ \mbox{Tr}\left( \hat{a}
_{2\psi}^{\dag }\hat{a}_{2\psi}\rho _{t}\right) \right] ^{2}\right\}.
\end{equation}
Therefore, the shot-noise limited measurement sensitivity for an evolution time $t$ is given by
\begin{equation}
\eta =\Delta I \sqrt{t}/\left( \frac{\partial I}{\partial b}%
\right).
\end{equation}%

\subsection{Tunnelling channels for leakage current}

\begin{figure}
\centering\includegraphics[width=8cm]{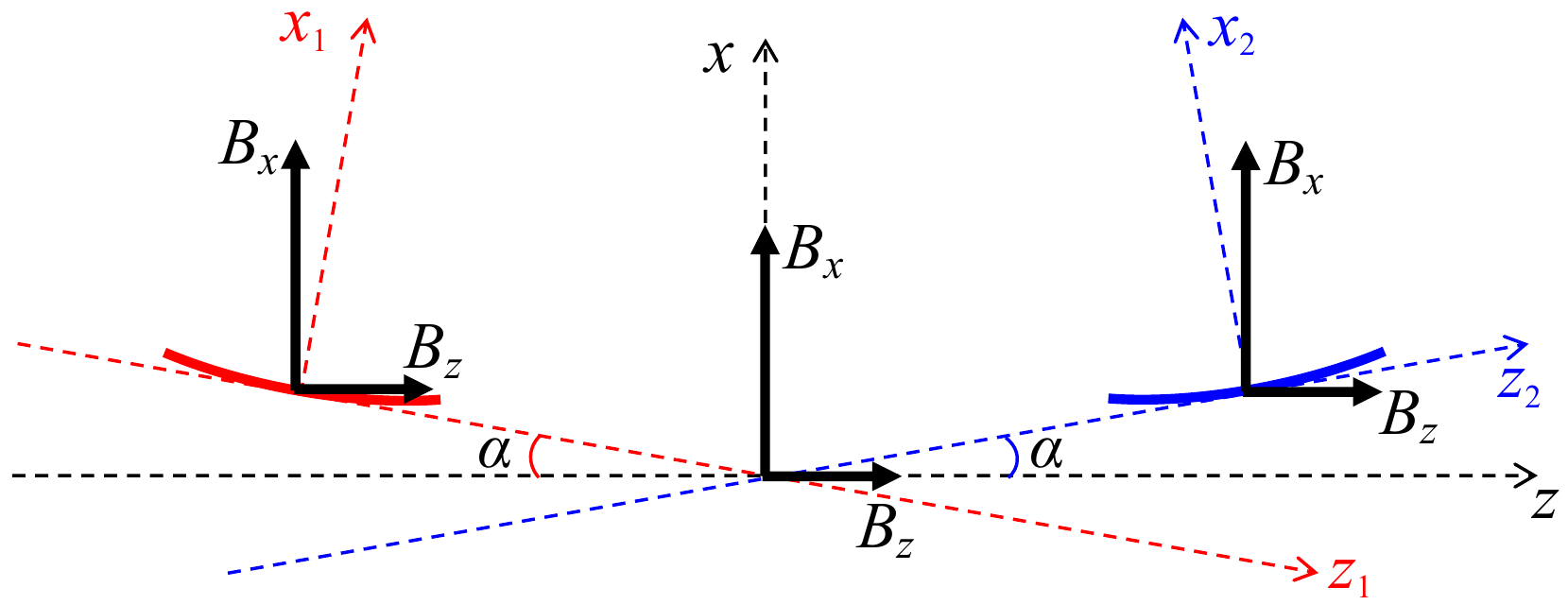}
\caption{(Color online) Schematic diagram of coordinates for two quantum dots in a bent arc shape nanotube with a tilted angle $\protect\alpha $. Local coordinates $x_{1}$-$z_{1}$ and coordinates $x_{2}$-$z_{2}$
are used to describe the Hamiltonian of electron in the left and right
quantum dot respectively. The magnetic field $\mathbf{B}=B_x\mathbf{x}+B_z\mathbf{z} $ is applied in the $x$-$z$
plane.}
\label{tilted}
\end{figure}
\begin{figure*}
\centering\includegraphics[width=12cm]{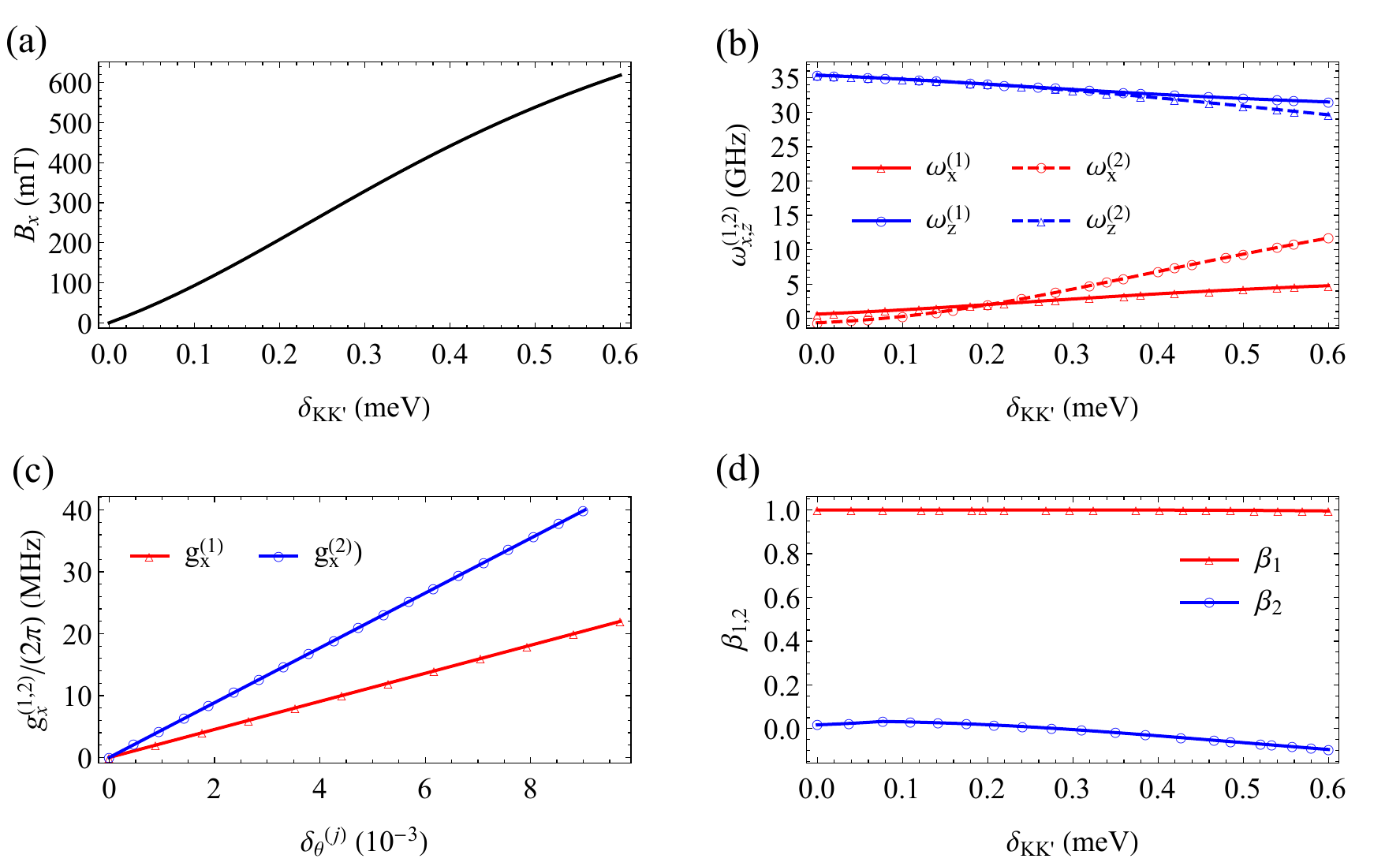}
\caption{(Color online) The parameters to achieve the
condition $\protect\omega _{0}^{\left( 1\right) }=\protect\omega %
_{0}^{\left( 2\right) }$ and $g_{x}^{\left( 1\right) }=g_{x}^{\left( 2\right) }$.
({a}) $B_{x}$ as a function of the intervalley
scattering difference $\protect\delta _{KK^{\prime }}=\Delta _{KK^{\prime
}}^{\left( 2\right) }-\Delta _{KK^{\prime }}^{\left( 1\right) }$. ({b}) $%
\protect\omega _{x,z}^{\left( 1,2\right) }$ as a function of the intervalley
scattering difference $\protect\delta _{KK^{\prime }}$. ({c}) For a certain
value of $\protect\delta _{KK^{\prime }}=0.2$ meV, $g_{x}^{\left( 1,2\right)
}$ as a function of the driving parameter $\protect\delta _{\protect%
\theta }^{\left( j\right) }$. (d) $\protect\beta _{1,2}$ as a function of the
intervalley scattering difference $\protect\delta _{KK^{\prime }}$. The other
parameters are: $\Delta _{KK^{\prime }}^{\left( 1\right) }
=0.2$ meV, $\protect\alpha =1^{\circ }$, $B_{z}=100$ mT.}
\label{values}
\end{figure*}

The system of a carbon nanotube double quantum dot in the $\left( 1,1\right) $ charge configuration can be described by the following Hamiltonian as
\begin{equation}
\hat{H}_{sb}=\frac{1}{2}\Omega _{d}\hat{S}_{x}^{\left( 1\right) }+\Omega
_{c}\cos \left( \omega _{c}t\right) \hat{S}_{z}^{\left( 1\right) }+\frac{1}{2%
}\Omega _{d}\hat{S}_{x}^{\left( 2\right) },  \label{Hc}
\end{equation}%
where $g_{x}^{(1)}=g_{x}^{(2)}\equiv \Omega _{d}$ and $\hat{S}_{x,z}^{(j)}$
represent the Pauli operators of the left ($j=1$) and right ($j=2$) dressed
valley-spin qubits. Here, we assume that both dressed valley-spin qubits are
identical. We remark that two dressed valley-spin qubits may not be
completely identical due to e.g. local disorder. We will address this issue
in detail in the next section. Using rotating-wave approximation under the
conditions $\Omega _{c}\ll \omega _{c},\Omega _{d}$, in the interaction
picture with respect to $\hat{H}_{sb}^{(0)}=(\omega _{c}/2)\left[ \hat{S}%
_{x}^{\left( 1\right) }+\hat{S}_{x}^{\left( 2\right) }\right] $, the
Hamiltonian $\hat{H}_{sb}$ can be simplified into%
\begin{equation}
\hat{H}_{sb}^{\prime }=\frac{1}{2}\delta \hat{S}_{z}^{\left( 1\right) }-%
\frac{1}{2}\Omega _{c}\hat{S}_{x}^{\left( 1\right) }+\frac{1}{2}\delta \hat{S%
}_{z}^{\left( 2\right) }
\end{equation}%
with $\delta =\Omega _{d}-\omega _{c}$, where we adopt a transformation $\hat{S%
}_{x}\leftrightarrow \hat{S}_{z}$  for simplicity and use rotating wave approximation. In the basis of $\left\{ \left\vert
A_{0}\right\rangle ,\left\vert A_{+}\right\rangle ,\left\vert
A_{-}\right\rangle ,\left\vert S\right\rangle \right\} $, one can partially
diagonalize the Hamiltonian $\hat{H}_{sb}^{\prime }$ into the following form
\begin{equation}
\hat{H}_{sb}^{^{\prime \prime }}=%
\begin{pmatrix}
0 & 0 & 0 & J_{S}^{0} \\
0 & \lambda /2 & 0 & J_{S}^{+} \\
0 & 0 & -\lambda /2 & J_{S}^{-} \\
J_{S}^{0} & J_{S}^{+} & J_{S}^{-} & 0%
\end{pmatrix}%
\end{equation}%
with $\lambda =\sqrt{4\delta ^{2}+\Omega _{c}^{2}}$, $J_{S}^{0}=-\Omega_{c}^{2}/\left(2\lambda\right)$, $J_{S}^{\pm}=\delta\Omega_{c}/\left(\sqrt{2}\lambda\right)$, where%
\begin{equation}
\left\vert A_{0}\right\rangle =\frac{1}{\sqrt{2}}%
\begin{pmatrix}
\cos{\vartheta} \\
\sin{\vartheta} \\
\sin{\vartheta} \\
-\cos{\vartheta}%
\end{pmatrix}%
, \quad \left\vert A_{\pm}\right\rangle =\frac{1}{2}%
\begin{pmatrix}
{\pm}1+\sin{\vartheta} \\
-\cos{\vartheta} \\
-\cos{\vartheta} \\
{\pm}1-\sin{\vartheta}%
\end{pmatrix}%
\end{equation}%
with $\cos{\vartheta}=\Omega_{c}/\lambda$, $\sin{\vartheta}=2\delta/\lambda$ and $\left\vert S\right\rangle =\left({1}/{\sqrt{2}}\right)%
\begin{pmatrix}
0 & -1 & 1 & 0%
\end{pmatrix}%
^{T}$ is the singlet state, which is the only unblocked state allowing
electron tunnelling to the $\left( 0,2\right) $-singlet state\ $%
\left\vert S_{g}\right\rangle $. The other three states $\ket{A_0,\pm}$ are
blocked, nevertheless they couple with the singlet state $\ket{S}$ which may
open three tunnelling channels for leakage current. The transitions $%
\left\vert A_{0,\pm}\right\rangle \leftrightarrow \left\vert S\right\rangle $ are characterised by the following reduced effective Hamiltonian
as
\begin{eqnarray}
  \hat{H}_{A_{0}} &=& \begin{pmatrix}
0 & -\Omega _{c}^{2}/\left( 2\lambda \right) \\
-\Omega _{c}^{2}/\left( 2\lambda \right) & 0%
\end{pmatrix}, \\
  \quad \hat{H}_{A_{\pm}} &=& \begin{pmatrix}
\pm{\lambda}/{2} & \delta \Omega _{c}/\left( \sqrt{2}\lambda \right) \\
\delta \Omega _{c}/\left( \sqrt{2}\lambda \right) & 0%
\end{pmatrix},
\end{eqnarray}
respectively. In the absence of a weak oscillating magnetic field (namely $%
\Omega _{c}=0$), all of the channels to the state $\left\vert S\right\rangle
$ are closed. In this case, the $\ket{S}$ state fraction leads to electron
tunnelling and a prominent leakage current, see Fig.\ref{It}(b). The other
three states are blocked which results in exponentially decay of leakage current.
The presence of the weak oscillating field would open three tunnelling
channels via the transitions $\left\vert A_{0,\pm}\right\rangle \leftrightarrow
\left\vert S\right\rangle $. To qualitatively understand the role of
frequency detuning $\delta $ in these tunnelling channels, we assume that
two electrons are initialised in the state $\left\vert A_{0,\pm}\right\rangle $
respectively. One can obtain that the average population of the singlet
state $\left\vert S\right\rangle $ is
\begin{eqnarray}
P_{S}^{\left( 0\right) } &=&\frac{1}{2}, \\
P_{S}^{\left(\pm\right) } &=&\frac{4\delta ^{2}\Omega _{c}^{2}}{\left(
4\delta ^{2}+\Omega _{c}^{2}\right) ^{2}+8\delta ^{2}\Omega _{c}^{2}}.
\end{eqnarray}%
For the transition $\left\vert A_{0}\right\rangle \leftrightarrow \left\vert
S\right\rangle $, these two states $\ket{S}$ and $\ket{A_0}$ are on resonance, therefore the transition rate $\Omega _{c}^{2}/\left( 2\lambda \right) $ is maximized
when $\delta =0$. In contrast, the transitions $\left\vert
A_{\pm}\right\rangle \leftrightarrow \left\vert S\right\rangle $ rely on a
non-zero frequency detuning, otherwise the transition rate $\delta \Omega
_{c}/\left( \sqrt{2}\lambda \right) $ would instead be zero. Thus, these two
tunnelling channels would make most significant contribution to leakage
current with an appropriate non-zero frequency detuning. This is evident by two
symmetric peaks in the average singlet state population when the initial
states are $\ket{A_\pm}$, as shown in Fig.\ref{It}(a). It can
also be seen from Fig.\ref{It}(b) that a small frequency detuning can
sustain a relatively large leakage current in the steady state.

\vspace{0.02in}

\subsection{Compensation of non-uniformity between two nanotube quantum dots}

As the intervalley scattering is induced by electric disorder, it is
usually hard to fabricate two valley-spin qubits that have uniform
parameters. To be more specific, two nanotube quantum dots may have
different valley mixing parameter $\Delta _{KK^{\prime }}^{\left( j\right) }$ (see Eq.\ref{eq-si:Ham} and Eq.\ref{Heff1}) which results in different values of the characteristic parameter $%
\zeta _{j}$ for two valley-spin qubits (see Eqs.\ref{s1}-\ref{s4}). In
order to compensate such a non-uniformity, we consider a bent arc shape
nanotube with an tilted angle $\alpha $ as shown in Fig.\ref{tilted}. When
applying a magnetic field in $x$-$z$ plane, the magnetic fields for both electrons in the left ($
j=1$) and right ($j=2$) nanotube quantum dot written in
their local coordinates $x_{j}$-$z_{j}$ are
\begin{eqnarray}
B_{x_{j}}^{\left( j\right) } &=&B_{x}\cos \alpha +\left( -1\right)
^{j+1}B_{z}\sin \alpha , \\
B_{z_{j}}^{\left( j\right) } &=&B_{z}\cos \alpha -\left( -1\right)
^{j+1}B_{x}\sin \alpha .
\end{eqnarray}%
The effective Hamiltonian of the driven valley-spin qubit in the left
nanotube quantum dot, which interacts with an oscillating magnetic
field, as written in its local coordinates $x_{1}$-$z_{1}$ is
\begin{eqnarray}
  \hat{H}_{eb}^{\left( 1\right) } &=& \frac{1}{2}\left(\omega_{x_{1}}^{\left( 1\right) }\hat{s}_{x_{1}}^{\left( 1\right) }+\omega _{z_{1}}^{\left(1\right) }\hat{s}_{z_{1}}^{\left( 1\right) }\right)+\Omega _{x_{1}}^{\left(1\right) }\cos \left( \omega t\right) \hat{s}_{x_{1}}^{\left( 1\right)} \notag\\
  &+& \Omega _{z_{1}}^{\left( 1\right) }\cos \left( \omega t\right) \hat{s}_{z_{1}}^{\left( 1\right) }+\Omega _{c}\cos \left( \omega _{c}t\right) \hat{s}_{z_{1}}^{\left( 1\right)},
\end{eqnarray}
where $\delta _{\theta }^{\left( 1\right) }=\left( \partial _{z_{1}}\theta
_{1}\right) _{z_{1}=z_{0}}\Delta z_{m}^{\left( 1\right) }$, $\omega
_{x_{1}}^{\left( 1\right) }$, $\omega _{z_{1}}^{\left( 1\right) }$, $\Omega
_{x_{1}}^{\left( 1\right) }$, $\Omega _{z_{1}}^{\left( 1\right) }$ are
defined as in the Section {\textcolor{blue}{S1-1}}, in which the parameter
$\zeta _{1}$ is given by $\tan \zeta _{1}=\Delta _{KK^{\prime }}^{\left(
1\right) }/\Delta_{SO}^{\left( 1\right) }$.
The Pauli operators of the left valley-spin qubit $\hat{s}%
_{x_{1}}^{\left( 1\right) },\hat{s}_{y_{1}}^{\left( 1\right) },\hat{s}%
_{z_{1}}^{\left( 1\right) }$ are defined in the following basis as
\begin{eqnarray}
\left\vert \Uparrow _{1}^{\left( 1\right) }\right\rangle &=&\cos \left(
\zeta _{1}/2\right) \left\vert K^{\prime }\right\rangle \left\vert \uparrow
\right\rangle +\sin \left( \zeta _{1}/2\right) \left\vert K\right\rangle
\left\vert \uparrow \right\rangle , \\
\left\vert \Downarrow _{1}^{\left( 1\right) }\right\rangle &=&\sin \left(
\zeta _{1}/2\right) \left\vert K^{\prime }\right\rangle \left\vert
\downarrow \right\rangle +\cos \left( \zeta _{1}/2\right) \left\vert
K\right\rangle \left\vert \downarrow \right\rangle .
\end{eqnarray}

Similarly, the effective Hamiltonian for the driven valley-spin qubit in the
right nanotube quantum dot as written in its local coordinates $x_{2}$-$%
z_{2} $ is
\begin{eqnarray}
  \hat{H}_{e}^{\left( 2\right) } &=& \frac{1}{2}\left(\omega _{x_{2}}^{\left( 2\right) }%
\hat{s}_{x_{2}}^{\left( 2\right) }+\omega _{z_{2}}^{\left(
2\right) }\hat{s}_{z_{2}}^{\left( 2\right) }\right)+\Omega _{x_{2}}^{\left(
2\right) }\cos \left( \omega t\right) \hat{s}_{x_{2}}^{\left( 2\right)
} \notag\\
    &+& \Omega _{z_{2}}^{\left( 2\right) }\cos \left( \omega t\right) \hat{s}%
_{z_{2}}^{\left( 2\right) }
\end{eqnarray}
The corresponding Pauli operators $\hat{s}_{x_{2}}^{\left( 2\right) },\hat{s}%
_{y_{2}}^{\left( 2\right) },\hat{s}_{z_{2}}^{\left( 2\right) }$ are written
in the following basis as
\begin{eqnarray}
\left\vert \Uparrow _{2}^{\left( 2\right) }\right\rangle &=&\cos \left(
\zeta _{2}/2\right) \left\vert K^{\prime }\right\rangle \left\vert \uparrow
\right\rangle +\sin \left( \zeta _{2}/2\right) \left\vert K\right\rangle
\left\vert \uparrow \right\rangle , \\
\left\vert \Downarrow _{2}^{\left( 2\right) }\right\rangle &=&\sin \left(
\zeta _{2}/2\right) \left\vert K^{\prime }\right\rangle \left\vert
\downarrow \right\rangle +\cos \left( \zeta _{2}/2\right) \left\vert
K\right\rangle \left\vert \downarrow \right\rangle ,
\end{eqnarray}%
with the characteristic parameter $\zeta _{2}$ as defined by $\tan \zeta
_{2}=\Delta _{KK^{\prime }}^{\left( 2\right) }/\Delta _{SO}^{\left( 2\right) }$.
\begin{figure}
\centering
\includegraphics[width=8cm]{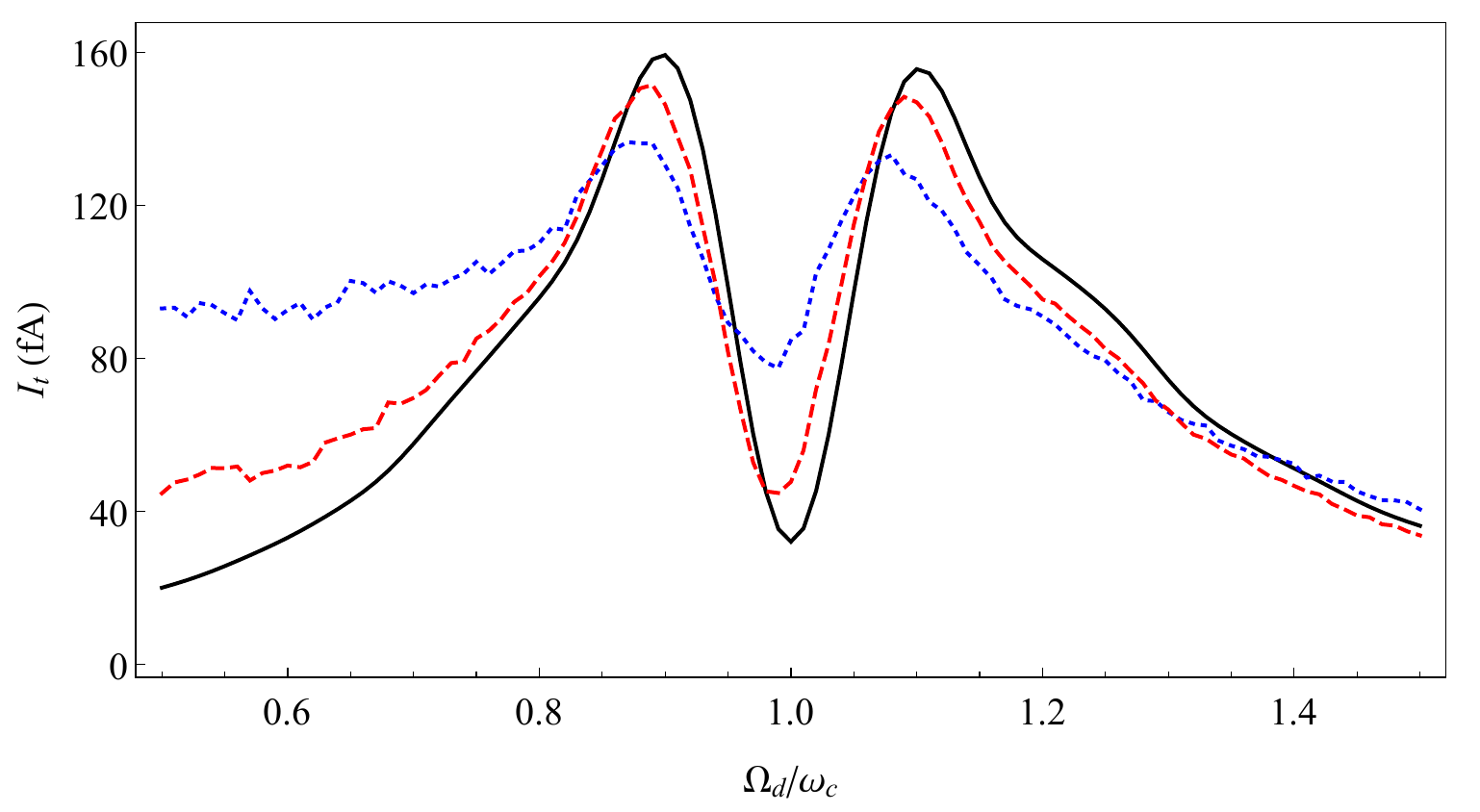}
\caption{(Color online)(a)-(c) Leakage current $I_{t}$ as a function of the driving Rabi frequency $\Omega_{d}/\omega_{c}$ for an oscillating magnetic field with amplitude $b=5.65$ $\mu$T [$\Omega_{c}=\left(2\pi\right)1 $ MHz] at time $t=1.5$ $\mu$s, where solid, dashed, dotted lines represent the results simulated with $\Delta f=0,5,10$ MHz and $\tau=100$ ms respectively. The other parameters are the same as Figure.\textcolor{red}{2} in the main text.}
\label{effect1}
\end{figure}
Therefore, the total effective Hamiltonian in the common coordinates $x$-$z$ is
\begin{eqnarray}
\hat{H}_{tb} &=&\exp \left[ -i\alpha \hat{s}_{y_{1}}^{\left( 1\right) }/2%
\right] \hat{H}_{eb}^{\left( 1\right) }\exp \left[ i\alpha \hat{s}%
_{y_{1}}^{\left( 1\right) }/2\right] \notag\\
&&+ \exp \left[ i\alpha \hat{s}%
_{y_{2}}^{\left( 2\right) }/2\right] \hat{H}_{e}^{\left( 2\right) }\exp %
\left[ -i\alpha \hat{s}_{y_{2}}^{\left( 2\right) }/2\right]  \notag \\
&=&\left[ \cos \left( \frac{\alpha }{2}\right) -i\hat{s}_{y_{1}}^{\left(
1\right) }\sin \left( \frac{\alpha }{2}\right) \right] \hat{H}_{eb}^{\left(
1\right) }\left[ \cos \left( \frac{\alpha }{2}\right) + i\hat{s}
_{y_{1}}^{\left( 1\right) }\sin \left( \frac{\alpha }{2}\right) \right] \notag\\
&& +
\left[ \cos \left( \frac{\alpha }{2}\right) +i\hat{s}_{y_{2}}^{\left(
2\right) }\sin \left( \frac{\alpha }{2}\right) \right] \hat{H}_{e}^{\left(
2\right) }\left[ \cos \left( \frac{\alpha }{2}\right) -i\hat{s}%
_{y_{2}}^{\left( 2\right) }\sin \left( \frac{\alpha }{2}\right) \right]
\notag \\
&=&\cos ^{2}\left( \frac{\alpha }{2}\right) \hat{H}_{eb}^{\left( 1\right)
}-i\sin \left( \frac{\alpha }{2}\right) \cos \left( \frac{\alpha }{2}\right)
\hat{s}_{y_{1}}^{\left( 1\right) }\hat{H}_{eb}^{\left( 1\right) }\notag\\
&&+i\sin
\left( \frac{\alpha }{2}\right) \cos \left( \frac{\alpha }{2}\right) \hat{H}%
_{eb}^{\left( 1\right) }\hat{s}_{y_{1}}^{\left( 1\right) }+\sin ^{2}\left(
\frac{\alpha }{2}\right) \hat{s}_{y_{1}}^{\left( 1\right) }\hat{H}%
_{eb}^{\left( 1\right) }\hat{s}_{y_{1}}^{\left( 1\right) }  \notag \\
&&+\cos ^{2}\left( \frac{\alpha }{2}\right) \hat{H}_{e}^{\left( 2\right)
}+i\sin \left( \frac{\alpha }{2}\right) \cos \left( \frac{\alpha }{2}\right)
\hat{s}_{y_{2}}^{\left( 2\right) }\hat{H}_{e}^{\left( 2\right) }\notag\\
&&-i\sin
\left( \frac{\alpha }{2}\right) \cos \left( \frac{\alpha }{2}\right) \hat{H}%
_{e}^{\left( 2\right) }\hat{s}_{y_{2}}^{\left( 2\right) }+\sin ^{2}\left(
\frac{\alpha }{2}\right) \hat{s}_{y_{2}}^{\left( 2\right) }\hat{H}%
_{e}^{\left( 2\right) }\hat{s}_{y_{2}}^{\left( 2\right) }  \notag \\
&=&\frac{1}{2}\left(\omega _{x}^{\left( 1\right) }\hat{s}_{x}^{\left( 1\right) }+%
\omega _{z}^{\left( 1\right) }\hat{s}_{z}^{\left( 1\right)
}\right) \notag \\
&&+\Omega _{x}^{\left( 1\right) }\cos \left( \omega t\right) \hat{s}%
_{x}^{\left( 1\right) }+\Omega _{z}^{\left( 1\right) }\cos \left( \omega
t\right) \hat{s}_{z}^{\left( 1\right) }\notag\\
&&+\Omega _{c}\sin \alpha \cos \left(
\omega _{c}t\right) \hat{s}_{x}^{\left( 1\right) }+\Omega _{c}\cos \alpha
\cos \left( \omega _{c}t\right) \hat{s}_{z}^{\left( 1\right) }  \notag \\
&&+\frac{1}{2}\left(\omega _{x}^{\left( 2\right) }\hat{s}_{x}^{\left( 2\right) }+%
\omega _{z}^{\left( 2\right) }\hat{s}_{z}^{\left( 2\right)
}\right)+\Omega _{x}^{\left( 2\right) }\cos \left( \omega t\right) \hat{s}%
_{x}^{\left( 2\right) }\notag\\
&&+\Omega _{z}^{\left( 2\right) }\cos \left( \omega
t\right) \hat{s}_{z}^{\left( 2\right) }
\end{eqnarray}%
with%
\begin{eqnarray}
\omega _{x}^{\left( 1\right) } &=&\omega _{x_{1}}^{\left( 1\right) }\cos
\alpha +\omega _{z_{1}}^{\left( 1\right) }\sin \alpha , \\
\omega _{z}^{\left( 1\right) } &=&\omega _{z_{1}}^{\left( 1\right) }\cos
\alpha -\omega _{x_{1}}^{\left( 1\right) }\sin \alpha , \\
\Omega _{x}^{\left( 1\right) } &=&\Omega _{x_{1}}^{\left( 1\right) }\cos
\alpha +\Omega _{z_{1}}^{\left( 1\right) }\sin \alpha , \\
\Omega _{z}^{\left( 1\right) } &=&\Omega _{z_{1}}^{\left( 1\right) }\cos
\alpha -\Omega _{x_{1}}^{\left( 1\right) }\sin \alpha ,
\end{eqnarray}%
and
\begin{eqnarray}
\omega _{x}^{\left( 2\right) } &=&\omega _{x_{2}}^{\left( 2\right) }\cos
\alpha -\omega _{z_{2}}^{\left( 2\right) }\sin \alpha , \\
\omega _{z}^{\left( 2\right) } &=&\omega _{z_{2}}^{\left( 2\right) }\cos
\alpha +\omega _{x_{2}}^{\left( 2\right) }\sin \alpha , \\
\Omega _{x}^{\left( 2\right) } &=&\Omega _{x_{2}}^{\left( 2\right) }\cos
\alpha -\Omega _{z_{2}}^{\left( 2\right) }\sin \alpha , \\
\Omega _{z}^{\left( 2\right) } &=&\Omega _{z_{2}}^{\left( 2\right) }\cos
\alpha +\Omega _{x_{2}}^{\left( 2\right) }\sin \alpha .
\end{eqnarray}%

\begin{figure*}
\centering
\includegraphics[width=12cm]{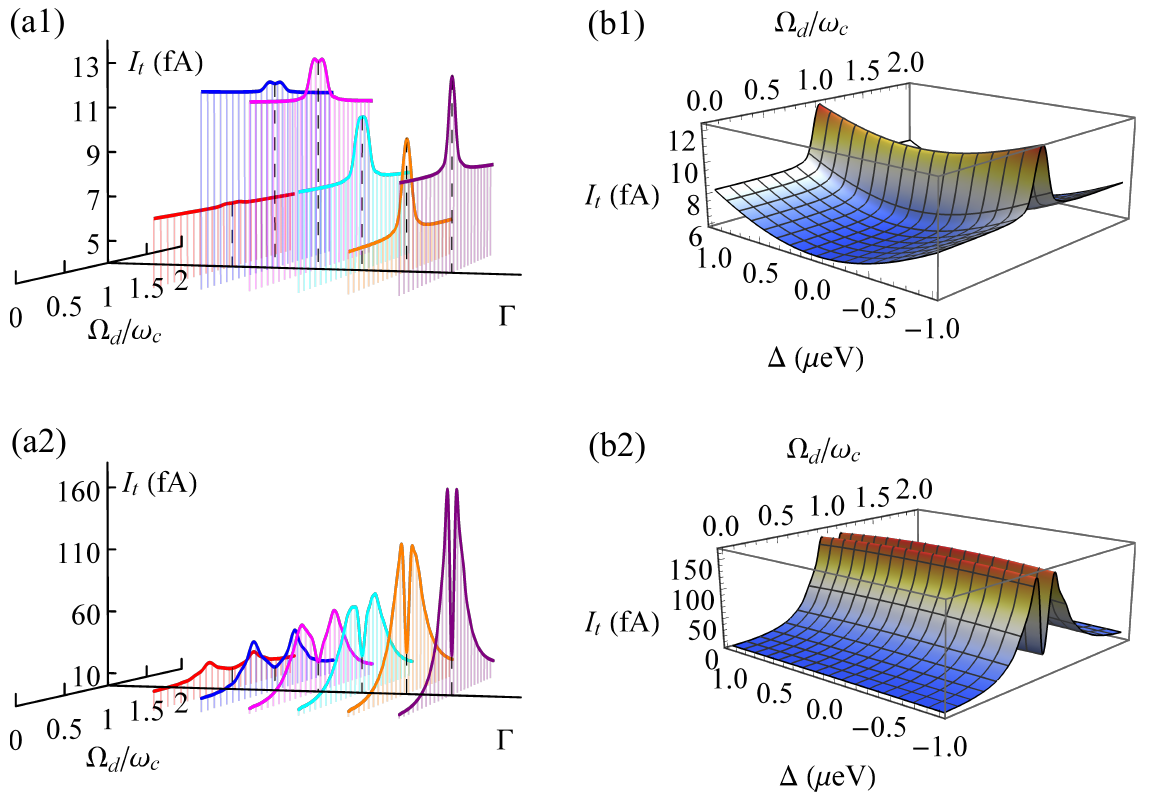}
\caption{(Color online)(a1)-(a2) Leakage current $I_{t}$ as a function of the driving Rabi frequency $\Omega_{d}/\omega_{c}$ at time $t=1.5$ $\mu$s for an oscillating magnetic field with frequency $\omega_{c}=\left(2\pi\right)5$ MHz and different amplitudes: (a1) $b=0.565$ $\mu$T [$\Omega_{c}=\left(2\pi\right)0.1 $ MHz] simulated with $J=\left(2\pi\right)10$ MHz; (a2) $b=5.65$ $\mu$T [$\Omega_{c}=\left(2\pi\right)1 $ MHz] simulated with $J=\left(2\pi\right)20$ MHz, in the presence of energy detuning $\Delta=1$ $\mu$eV, where the red, blue, magenta, cyan, orange and purple lines represent the results simulated with the transition rate $\Gamma/\left(2\pi\right)=10,25,75,125,250,500$ MHz respectively. (b1)-(b2) Leakage current $I_{t}$ as a function of the driving Rabi frequency $\Omega_{d}/\omega_{c}$ and the energy detuning $\Delta$ at time $t=1.5$ $\mu$s for an oscillating magnetic field with frequency $\omega_{c}=\left(2\pi\right)5$ MHz and different amplitudes: (b1) $b=0.565$ $\mu$T [$\Omega_{c}=\left(2\pi\right)0.1 $ MHz] simulated with $J=\left(2\pi\right)10$ MHz; (b2) $b=5.65$ $\mu$T [$\Omega_{c}=\left(2\pi\right)1 $ MHz] simulated with $J=\left(2\pi\right)20$ MHz. The injection and ejection rates are $\Gamma_L=\Gamma_R=\Gamma =(2\pi) 500$ MHz.}
\label{effect2}
\end{figure*}

The Pauli matrices $\hat{s}_{x,z}^{\left( j\right) }$ are written in the
following basis as
\begin{eqnarray}
\left\vert \Uparrow ^{\left( j\right) }\right\rangle &=&\exp \left[ -\left(
-1\right) ^{j+1}i\alpha \hat{s}_{y_{j}}^{\left( j\right) }/2\right]
\left\vert \Uparrow _{j}^{\left( j\right) }\right\rangle \notag\\&=&\cos \left( \alpha
/2\right) \left\vert \Uparrow _{j}^{\left( j\right) }\right\rangle -\left(
-1\right) ^{j+1}i\sin \left( \alpha /2\right) \hat{s}_{y_{j}}^{\left(
j\right) }\left\vert \Uparrow _{j}^{\left( j\right) }\right\rangle  \notag \\
&=&\cos \left( \alpha /2\right) \left\vert \Uparrow _{j}^{\left( j\right)
}\right\rangle +\left( -1\right) ^{j+1}\sin \left( \alpha /2\right)
\left\vert \Downarrow _{j}^{\left( j\right) }\right\rangle , \\
\left\vert \Downarrow ^{\left( j\right) }\right\rangle &=&\exp \left[
-\left( -1\right) ^{j+1}i\alpha \hat{s}_{y_{j}}^{\left( j\right) }/2\right]
\left\vert \Downarrow _{j}^{\left( j\right) }\right\rangle \notag\\&=&\cos \left(
\alpha /2\right) \left\vert \Downarrow _{j}^{\left( j\right) }\right\rangle
-\left( -1\right) ^{j+1}i\sin \left( \alpha /2\right) \hat{s}%
_{y_{j}}^{\left( j\right) }\left\vert \Downarrow _{j}^{\left( j\right)
}\right\rangle  \notag \\
&=&\cos \left( \alpha /2\right) \left\vert \Downarrow _{j}^{\left( j\right)
}\right\rangle -\left( -1\right) ^{j+1}\sin \left( \alpha /2\right)
\left\vert \Uparrow _{j}^{\left( j\right) }\right\rangle .
\end{eqnarray}%
The eigenstates of $\hat{H}_{0}=\left({1}/{2}\right)\left[\omega _{x}^{\left( 1\right) }%
\hat{s}_{x}^{\left( 1\right) }+\omega _{z}^{\left( 1\right) }\hat{%
s}_{z}^{\left( 1\right) }\right]+\left({1}/{2}\right)\left[\omega _{x}^{\left( 2\right) }\hat{s}%
_{x}^{\left( 2\right) }+\frac{1}{2}\omega _{z}^{\left( 2\right) }\hat{s}%
_{z}^{\left( 2\right) }\right]$ are
\begin{eqnarray}
\left\vert L_{0}\right\rangle &=&\cos \left( \gamma _{1}/2\right) \left\vert
\Uparrow ^{\left( 1\right) }\right\rangle +\sin \left( \gamma _{1}/2\right)
\left\vert \Downarrow ^{\left( 1\right) }\right\rangle , \\
\left\vert L_{1}\right\rangle &=&-\sin \left( \gamma _{1}/2\right)
\left\vert \Uparrow ^{\left( 1\right) }\right\rangle +\cos \left( \gamma
_{1}/2\right) \left\vert \Downarrow ^{\left( 1\right) }\right\rangle , \\
\left\vert R_{0}\right\rangle &=&\cos \left( \gamma _{2}/2\right) \left\vert
\Uparrow ^{\left( 2\right) }\right\rangle +\sin \left( \gamma _{2}/2\right)
\left\vert \Downarrow ^{\left( 2\right) }\right\rangle , \\
\left\vert R_{1}\right\rangle &=&-\sin \left( \gamma _{2}/2\right)
\left\vert \Uparrow ^{\left( 2\right) }\right\rangle +\cos \left( \gamma
_{2}/2\right) \left\vert \Downarrow ^{\left( 2\right) }\right\rangle ,
\end{eqnarray}%
with $\cos \gamma _{j}=\omega _{z}^{\left( j\right) }/\omega _{0}^{\left( j\right) }$ and $\sin \gamma _{j}=\omega _{x}^{\left( j\right) }/%
\omega _{0}^{\left( j\right) }$. In this set of bases, $\hat{H}_{tb}$ can be rewritten by%
\begin{eqnarray}
\hat{H}_{tb}^{\prime } &=&\frac{1}{2}\omega _{0}^{\left( 1\right) }\hat{S}%
_{z}^{\left( 1\right) }+g_{z}^{\left( 1\right) }\cos \left( \omega t\right)
\hat{S}_{z}^{\left( 1\right) }+g_{x}^{\left( 1\right) }\cos \left( \omega
t\right) \hat{S}_{x}^{\left( 1\right) }\notag\\
&+&\Omega _{cz}\cos \left( \omega
_{c}t\right) \hat{S}_{z}^{\left( 1\right) }+\Omega _{cx}\cos \left( \omega
_{c}t\right) \hat{S}_{x}^{\left( 1\right) }+\frac{1}{2}\omega _{0}^{\left( 2\right) }\hat{S}_{z}^{\left( 2\right)
}  \notag \\
&+&g_{z}^{\left( 2\right) }\cos \left( \omega t\right) \hat{S}_{z}^{\left(
2\right) }+g_{x}^{\left( 2\right) }\cos \left( \omega t\right) \hat{S}%
_{x}^{\left( 2\right) }
\end{eqnarray}%
with%
\begin{eqnarray}
\omega _{0}^{\left( j\right) } &=&\sqrt{\left( \omega _{x}^{\left( j\right)
}\right) ^{2}+\left( \omega _{z}^{\left( j\right) }\right) ^{2}}, \\
g_{z}^{\left( j\right) } &=&\Omega _{z}^{\left( j\right) }\cos \gamma
_{j}+\Omega _{x}^{\left( j\right) }\sin \gamma _{j}, \\
g_{x}^{\left( j\right) } &=&\Omega _{x}^{\left( j\right) }\cos \gamma
_{j}-\Omega _{z}^{\left( j\right) }\sin \gamma _{j}, \\
\Omega _{cz} &=&\Omega _{c}\cos \left( \alpha -\gamma _{1}\right) , \\
\Omega _{cx} &=&\Omega _{c}\sin \left( \alpha -\gamma _{1}\right) .
\end{eqnarray}%
In Fig.\ref{values}(a), it can be seen that for a certain value of $\delta
_{KK^{\prime }}$, it is possible to satisfy the condition $\omega
_{0}^{\left( 1\right) }=\omega _{0}^{\left( 2\right) }$ by choosing an
appropriate magnetic field $B_{x}$. For example, one shall apply $B_{x}=207$ mT for $%
\delta _{KK^{\prime }}=0.2$ meV. And we plot the corresponding values of $%
\omega _{x,z}^{\left( j\right) }$ in Fig.\ref{values}(b), which depend on
$\delta _{KK^{\prime }}$ and $B_{x}$. The driving Rabi
frequencies $g_{x}^{\left( j\right) }$ depend on the driven motion parameter
$\delta _{\theta }^{\left( j\right) }$, see Fig.\ref{values}(c), which shows that $
g_{x}^{\left( 1\right) }=g_{x}^{\left( 2\right) }=\Omega _{d}$ can be
achieved by choosing appropriate $\delta _{\theta }^{\left( j\right) }$.
Therefore, it is reasonable to consider the case of $\omega =\omega _{0}^{\left( 1\right) }=\omega
_{0}^{\left( 2\right) }$. Using rotating-wave approximation under the
conditions $\Omega _{cx},\Omega _{cz}\ll \omega _{c},g_{x}^{\left( j\right)
},g_{z}^{\left( j\right) }\ll \omega _{0}^{\left( j\right) }$, the total
effective Hamiltonian can be simplified as follows
\begin{equation}
\hat{H}_{sb}=\frac{1}{2}\Omega _{d}\hat{S}_{x}^{\left( 1\right) }+\Omega
_{c}\cos \left( \omega _{c}t\right) \hat{S}_{z}^{\left( 1\right) }+\frac{1}{2%
}\Omega _{d}\hat{S}_{x}^{\left( 2\right) },
\end{equation}%
where $\cos \left( \alpha -\gamma _{1}\right) \approx 1$ in the limit of $%
\omega _{x}^{\left( 1\right) }\ll \omega _{z}^{\left( 1\right) }$ and $%
\alpha \ll 1$.

\vspace{0.05in}

Two electrons both in the right quantum dot can be considered as
identical particles, thus the antisymmetric state can be written as%
\begin{equation}
\left\vert S_{g}\right\rangle =\frac{1}{\sqrt{2}}\left( \left\vert
R_{1}\right\rangle _{2}\left\vert R_{0}\right\rangle _{2}-\left\vert
R_{0}\right\rangle _{2}\left\vert R_{1}\right\rangle _{2}\right),
\end{equation}%
where the subscript $1$ and $2$ denotes the left and right quantum dot respectively.
As the tunnelling between the left and right quantum dot would not change
the electron state, the antisymmetric state of two electrons separated in
the left and right quantum dot can be written as follows
\begin{equation}
\left\vert S\right\rangle =\frac{1}{\sqrt{2}}\left( \left\vert
R_{1}\right\rangle _{1}\left\vert R_{0}\right\rangle _{2}-\left\vert
R_{0}\right\rangle _{1}\left\vert R_{1}\right\rangle _{2}\right).
\label{SS}
\end{equation}%
Projecting Eq.\ref{SS} into the basis of $\left\{ \left\vert
L_{0}\right\rangle \otimes \left\vert R_{0}\right\rangle ,\left\vert
L_{0}\right\rangle \otimes \left\vert R_{1}\right\rangle ,\left\vert
L_{1}\right\rangle \otimes \left\vert R_{0}\right\rangle ,\left\vert
L_{1}\right\rangle \otimes \left\vert R_{1}\right\rangle \right\} $, one can
obtain%
\begin{align}
\left\vert S\right\rangle & =\frac{1}{\sqrt{2}}\beta _{1}\cos \left[ \left(
\zeta _{1}-\zeta _{2}\right) /2\right] \left( \left\vert L_{1}\right\rangle
\left\vert R_{0}\right\rangle -\left\vert L_{0}\right\rangle \left\vert
R_{1}\right\rangle \right)  \notag \\
& +\frac{1}{\sqrt{2}}\beta _{2}\cos \left[ \left( \zeta _{1}-\zeta
_{2}\right) /2\right] \left( \left\vert L_{0}\right\rangle \left\vert
R_{0}\right\rangle +\left\vert L_{1}\right\rangle \left\vert
R_{1}\right\rangle \right)
\end{align}
with%
\begin{eqnarray}
\left\langle L_{0}\right. \left\vert R_{0}\right\rangle &=&\beta _{1}\cos
\left[ \left( \zeta _{1}-\zeta _{2}\right) /2\right] , \\
\left\langle L_{1}\right. \left\vert R_{1}\right\rangle &=&\beta _{1}\cos
\left[ \left( \zeta _{1}-\zeta _{2}\right) /2\right] , \\
\left\langle L_{0}\right. \left\vert R_{1}\right\rangle &=&\beta _{2}\cos
\left[ \left( \zeta _{1}-\zeta _{2}\right) /2\right] , \\
\left\langle L_{1}\right. \left\vert R_{0}\right\rangle &=&-\beta _{2}\cos
\left[ \left( \zeta _{1}-\zeta _{2}\right) /2\right] ,
\end{eqnarray}%
and
\begin{eqnarray}
\beta _{1} &=&\cos \left[ \left( \gamma _{1}-\gamma _{2}\right) /2+\alpha %
\right] , \\
\beta _{2} &=&\sin \left[ \left( \gamma _{1}-\gamma _{2}\right) /2+\alpha %
\right] .
\end{eqnarray}%
As shown in Fig.\ref{values}(d), $\beta _{2}\ll \beta _{1}$ is valid for a wide range of $\delta _{KK^{\prime }}$ under the condition $\omega_{0}^{\left( 1\right) }=\omega _{0}^{\left( 2\right) }$. Therefore, after
normalization, the antisymmetric state turns out to be%
\begin{equation}
\left\vert S\right\rangle =\frac{1}{\sqrt{2}}\left( \left\vert
L_{1}\right\rangle \left\vert R_{0}\right\rangle -\left\vert
L_{0}\right\rangle \left\vert R_{1}\right\rangle \right) .
\end{equation}

\section{Influence of decoherence}\label{AC}

Gate-defined quantum dots in a carbon nanotube, suffers from decoherence mainly due to the hyperfine coupling with the environmental nuclear spins and the electric noise. The influence of nuclei may be mitigated by synthesizing carbon nanotube with isotopically purified $^{12}$CH$_{4}$, allowing for the fabrication of devices without nuclear spins \cite{Laird2015}. Under ideal conditions, the coherence time of quantum dot in such a isotopically purified device is predicted to be of the order of seconds \cite{Bulaev2008,Rudner2010}. However, the charge noise due to the electric potential fluctuations would modify the parameters of the quantum dot, namely inducing fluctuations of the energy difference $\Delta$ between the singlet states $\left\vert S\right\rangle$ and $\left\vert S_{g}\right\rangle$ \cite{Palyi2017}, as $H_{\Delta}=\Delta \ketbra{S_g}{S_g}$. In this section, we provide detailed analysis of the influence of decoherence on our proposed carbon nanotube quantum sensor.

\begin{figure*}
\centering
\includegraphics[width=12cm]{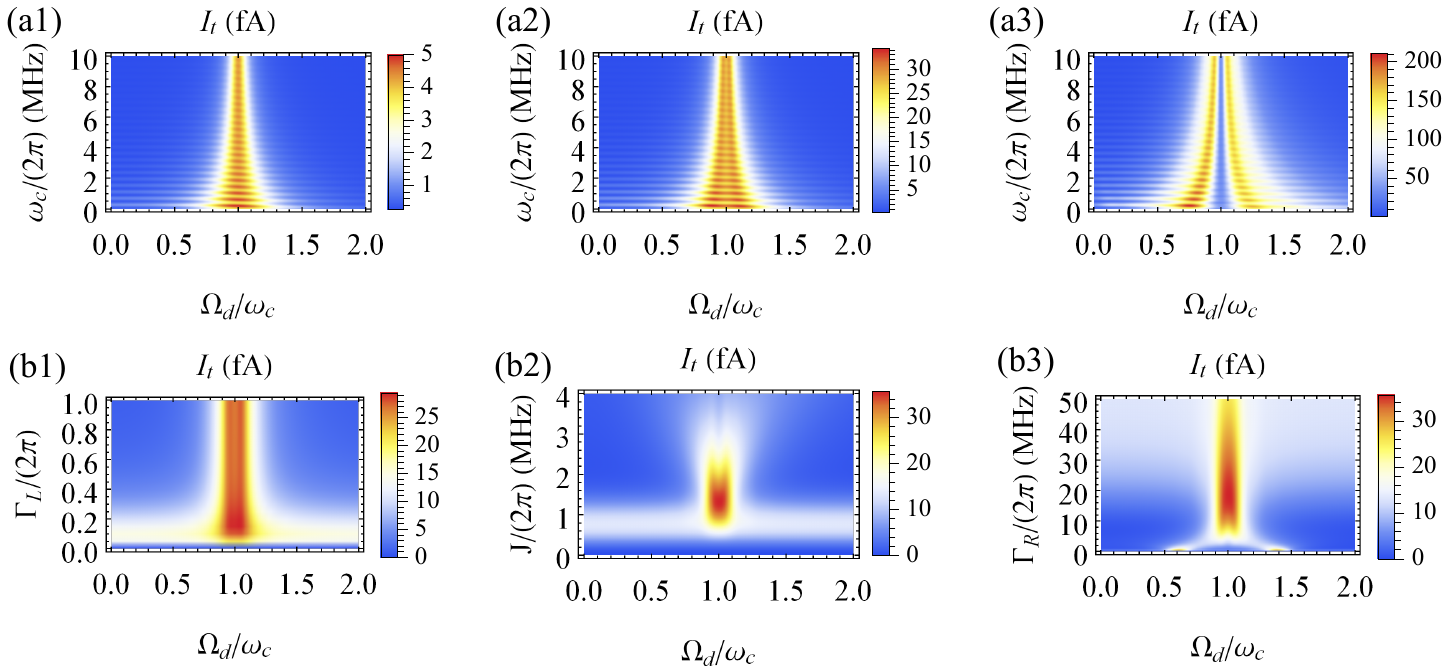}
\caption{(Color online) (a1)-(a3) Leakage current $I_{t}$ at time $t$ as a function of
the driving Rabi frequency $\Omega _{d}$ and the frequency $\protect\omega _{c}$ of the oscillating magnetic field
with different amplitudes: (a1) $b_{z}=0.565$ $\protect\mu $T [$\Omega _{c}=\left(2%
\protect\pi\right) 0.1$ MHz], (a2) $b_{z}=1.70$ $\protect\mu $T [$\Omega _{c}=\left(2%
\protect\pi\right) 0.3$ MHz], (a3) $b_{z}=5.65$ $\protect\mu $T [$\Omega _{c}=\left(2%
\protect\pi\right) 1$ MHz]. The other parameters are $J=\left(2\protect\pi\right) 2$ MHz and $%
\Gamma _{L} =\Gamma _{R}=\left(2\protect\pi\right) 8$ MHz. (b1) Leakage
current $I_{t}$ at time $t$ as a function of the driving Rabi frequency $\Omega _{d}%
$ and the electron injection rate $\Gamma _{L}$ with $%
\protect\omega _{c}=\left(2\protect\pi\right) 5$ MHz, $\Omega _{c}=\left(2\protect\pi\right) 0.3$
MHz, $J=\left(2\protect\pi\right) 2$ MHz and $\Gamma _{R}=\left(2\protect\pi\right) 8$ MHz. (b2)
Leakage current $I_{t}$ at time $t$ as a function of the driving Rabi frequency $\Omega
_{d}$ and the tunnelling rate $J$ with $\protect\omega %
_{c}=\left(2\protect\pi \right)5$ MHz, $\Omega _{c}=\left(2\protect\pi\right) 0.3$ MHz and $\Gamma
_{L}= \Gamma _{R}=\left(2\protect\pi\right) 8$ MHz. (b3) Leakage current at time $t$ $%
I_{t}$ as a function of the driving Rabi frequency $\Omega _{d}%
$ and the electron ejection rate $\Gamma _{R}$ with $\protect%
\omega _{c}=\left(2\protect\pi\right) 5$ MHz, $\Omega _{c}=\left(2\protect\pi\right) 0.3$ MHz, $J=\left(2%
\protect\pi\right) 2$ MHz and $\Gamma _{L}=\left(2\protect\pi\right) 8$ MHz. Taking one unpolarised electron located in the right quantum dot as the initial state. The time is set
as $t=1.5\protect\mu s$.}
\label{b}
\end{figure*}
\begin{figure*}
\centering
\includegraphics[width=12cm]{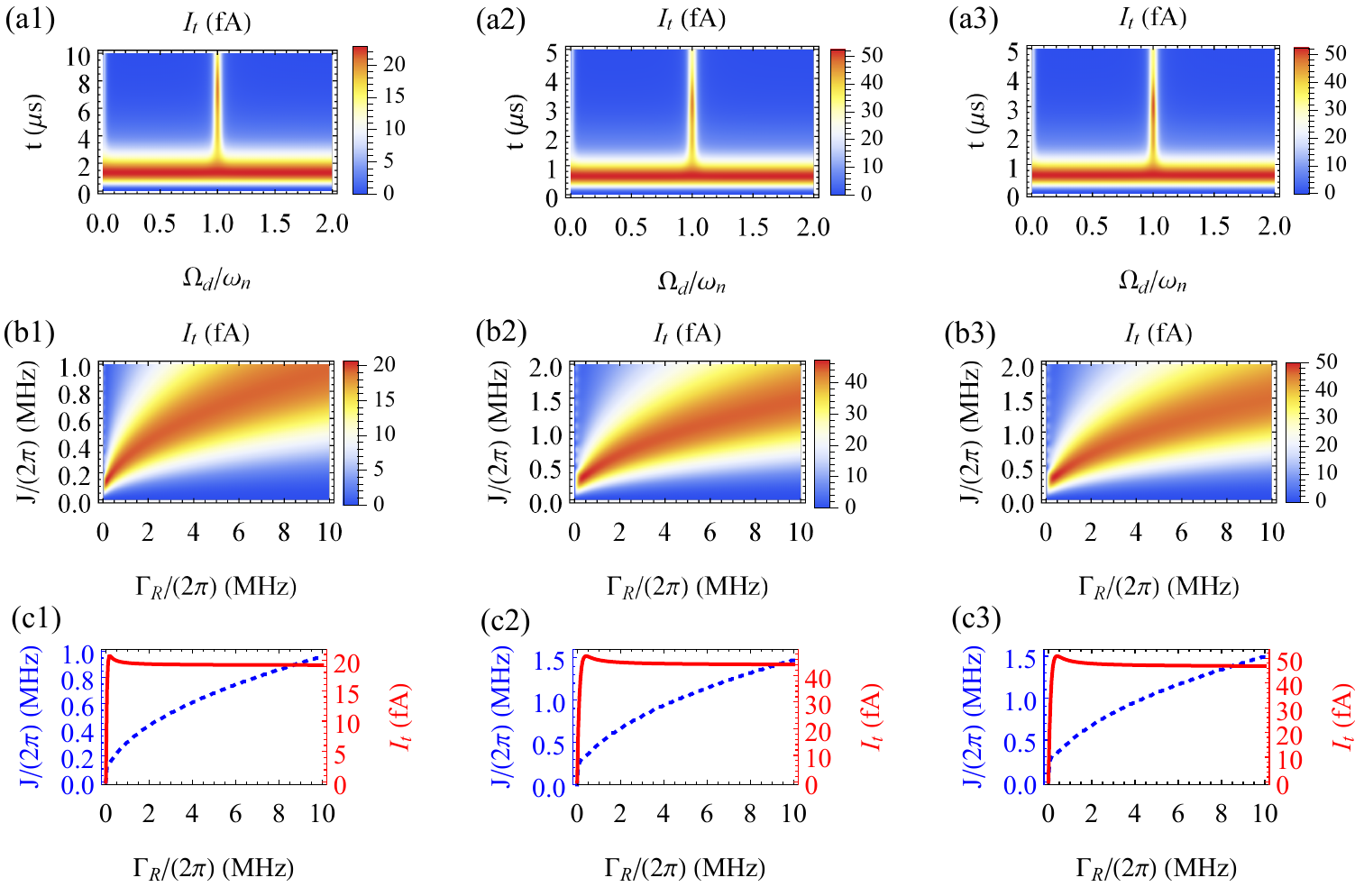}
\caption{(Color online) (a1)-(a3) Leakage current $I_{t}$ as a function of
the driving Rabi frequency $\Omega _{d}/\omega_{n}$ and the evolution time $t$ for the
nuclear spin $^{31}$P (a1) with $\Gamma _{R}=\left(2\protect\pi\right) 0.2$ MHz, $J=\left(2%
\protect\pi\right) 0.15$ MHz; the nuclear spin $^{19}$F (a2) with $\Gamma _{R}=\left(2%
\protect\pi\right) 0.4$ MHz, $J=\left(2\protect\pi\right) 0.35$ MHz; and the nuclear spin $%
^{1}$H (a3) with $\Gamma _{R}=\left(2\protect\pi\right) 0.4$ MHz, $J=\left(2\protect\pi\right) 0.35$
MHz respectively, where $\omega_{n}=\gamma_{n}B$ is the Larmor frequency of nuclear spin. (b1)-(b3) Leakage current $I_{t}$ as a function of the
electron ejection rate $\Gamma _{R}$ and the tunnelling rate $J$ on resonance of
the nuclear spin $^{31}$P (b1) at time $t=7$ $\protect\mu $s, the nuclear
spin $^{19}$F (b2) at time $t=5$ $\protect\mu $s and the nuclear spin $^{1}$%
H (b3) at time $t=3$ $\protect\mu $s. (c1)-(c3) The optimised leakage current (red solid lines) and the optimal parameters (blue dashed lines) of the electron ejection rate $\Gamma _{R}$ and the tunnelling rate $J$ as shown in (b1)-(b3) respectively. In (a1)-(a3), (b1)-(b3),(c1)-(c3), the electron
injection rate is $\Gamma _{L}=\left(2\protect\pi\right) 1$ MHz, the magnetic field is $%
B_{x}=300$ mT and $B_{z}=100$ mT. The other parameters are the same as Fig.4
in the main text.}
\label{n}
\end{figure*}
\begin{figure*}
\centering
 \includegraphics[width=12cm]{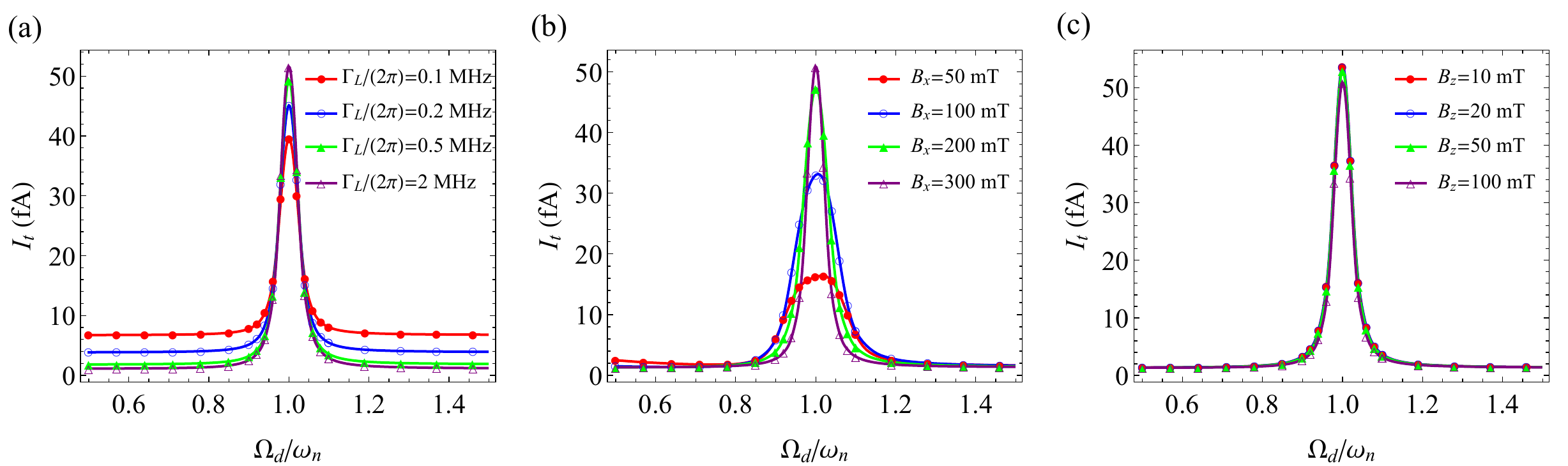}
\caption{(Color online) Leakage current $I_{t}$ as a function of the driving
Rabi frequency $\Omega _{d}/\omega_{n}$ for the nuclear spin $^{1}$H, where $\omega_{n}=\gamma_{n}B$ is the Larmor frequency of nuclear spin. (a) under the
situations with different injection rate $\Gamma _{L}$ where $B_{x}=300$ mT
and $B_{z}=100$ mT. (b) under the situations with different magnetic field $%
B_{x}$ where $\Gamma _{L}=\left(2\protect\pi\right) 1$ MHz and $B_{z}=100$ mT. (c) under
the situations with different magnetic field $B_{z}$ where $\Gamma _{L}=\left(2%
\protect\pi\right) 1$ MHz and $B_{x}=300$ mT. In (a)-(c) $\Gamma _{R}=\left(2%
\protect\pi\right) 0.4$ MHz, $J=\left(2\protect\pi\right) 0.35$ MHz, and $t=3$ $\mu s$. The other parameters are the same as Fig.4 in the main text.}
\label{Bxz}
\end{figure*}

\subsection{Hyperfine coupling with nuclei}

Nanotubes synthesized from natural hydrocarbons consist of $99\%$ $^{12}$C (nuclear spin $I=0$) and $1\%$ $^{13}$C (nuclear spin $I=\frac{1}{2}$). The abundance of $^{13}$C nuclei may be reduced to $0.01\%$ or even lower by using isotopically purified CH$_{4}$ during the growth of nanotubes. The hyperfine coupling to the $^{13}$C nuclear spins induces magnetic field noise on the confined quantum dot, leading to the slow fluctuation of frequency detuning of the encoded valley-spin qubit. The effective Hamiltonian for detecting a weakly oscillating magnetic field in the $\left(1,1\right)$ charge configuration, incorporating the noise in  frequency detunings $\delta\omega_{i}\left(t\right) (i=1,2)$ ,  can be written as
\begin{eqnarray}
  \hat{H}_{sb}^{*} &=& \frac{1}{2}\left(\Omega_{d}\hat{S}_{x}^{\left(1\right)}
+\delta\omega_{1}\left(t\right)\hat{S}_{z}^{\left(1\right)}\right)
+\frac{1}{2}\left(\Omega_{d}\hat{S}_{x}^{\left(2\right)}
+\delta\omega_{2}\left(t\right)\hat{S}_{z}^{\left(2\right)}\right)\notag \\
    &+& \Omega_{c}\cos\left(\omega_{c}t\right)\hat{S}_{z}^{\left(1\right)}
\end{eqnarray}
For numerical simulation, we assume that $\delta\omega_{i}\left(t\right)$ ($i=1,2$) fulfills the Gaussian distribution
\begin{equation}
    D\left(\delta\omega_{i}\right)=\frac{1}{\sqrt{2\pi}\Delta_f}e^{-{\delta\omega_{i}^{2}/2\Delta_f^{2}}},
\end{equation}
and evolves following the Ornstein-Uhlenbeck process with
\begin{equation}
    \delta\omega_{i}\left(t+d t\right)=\delta\omega_{i}\left(t\right)e^{-d t/\tau}+\left[\Delta_f^{2}\left(1-e^{-2d t/\tau}\right)\right]^{1/2}n,
\end{equation}
where $\tau$ is the correlation time which depends on the nuclear spin dynamics and relaxation, $n$
represents a sample value of the unit normal random variable and $\Delta_f=\sqrt{2}/T_{2}^*$ is the standard
deviation of $\delta\omega_{i}$. As $T_{2}^*$ scales as $\sqrt{N/p}$ where $N$ is the number of nuclei in the
quantum dot and $p$ is the isotopic fraction of $^{13}$C we can estimate  $\Delta f$ from the upper limit
of $T_{2}^*$ that was measured in an $99\%$ isotopically purified $^{13}$C nanotube quantum dot with $N\approx7\times10^4$ nuclei in the interaction range \cite{Churchill2009}. For an isotopic fraction
of $^{13}$C of $0.0127\%$ and $0.05\%$ percents we predict $\Delta f=5$ MHz and $\Delta f=10$ MHz respectively with the same value of $N$.  As shown in Figs.\ref{effect1}, the fluctuation of frequency detuning induced by the hyperfine coupling would reduce the contrast of the resonance signal in a certain extent. Nevertheless, the influence can be well mitigated by isotopically engineering the nanotubes. The resonance signal of the oscillating magnetic field with larger amplitude demonstrates more robust feature. Similar results can be obtained in the case of nanoscale magnetic resonance spectroscopy.

\subsection{Charge noise}

As the carbon nanotube quantum dot is gate-defined, the electric noise would modify the energy levels of the quantum dot, i.e., inducing fluctuations of the energy detuning $\Delta$ between the singlet states $\left\vert S\right\rangle$ and $\left\vert S_{g}\right\rangle$ \cite{Palyi2017}, as $H_{\Delta}=\Delta \ketbra{S_g}{S_g}$, the role of which in our scheme is mainly the suppression of effective tunneling. Thus, the influence may be compensated by choosing proper values of $\Gamma_{L}$, $\Gamma_{R}$ and $J$. This is in contrast to the dephasing effect on the coherence time of qubit involving the singlet state $\ket{S_g}$ in the (0,2) subspace \cite{Laird2013}, where the energy splitting of the qubit relies on the energy detuning $\Delta$.

\vspace{0.05in}

As shown in Fig.\ref{effect2}(a1)-(a2), when the energy detuning $\Delta$ up to 1$\mu$eV is considered,
the resonant signal (dip or peak) of leakage current is degraded in the region with a small injection (ejection) rate of $\Gamma$ ($\Gamma_{L}=\Gamma_{R}=\Gamma$). The influence of the energy detuning on the resonant signal becomes negligible for larger but still reasonable rates $\Gamma$ (with $\Gamma_{L}=\Gamma_{R}=\Gamma$). The results hold without relying on a large amplitude of the oscillating magnetic field. In addition, the resonance signal of the leakage current is tolerant to the energy detuning varying from $-1$ $\mu$eV to $1$ $\mu$eV when the transition rate is chosen as $\Gamma/\left(2\pi\right)=500$ MHz \cite{Palyi2017}, see Fig.\ref{effect2}(b1)-(b2). Therefore, improving the transition rate of
the electron can efficiently compensate for the energy detuning between singlet states $\left\vert S\right\rangle$
and $\left\vert S_{g}\right\rangle$.

\section{More details on the response of leakage current}\label{AD}

\subsection{Coupling to a weak oscillating signal field}

As what we have presented in the main text, the leakage current always shows
resonance features characterized by either a peak or a dip when the nanotube
double quantum dot sensor couples with an oscillating magnetic field. The
width of the resonance signal is critical for the spectra resolution. As
shown in Fig.\ref{b}(a1)-(a3), it can be seen that the frequency of an
oscillating magnetic field can also influence the width of the resonance
signal. In particular, a smaller ratio $\Omega_{c}/\omega _{c}$ will lead to
a sharper resonance signal.

\vspace{0.05in}

In addition, there are another three important parameters $\Gamma _{L}$, $J$
and $\Gamma _{R}$ that are involved in the process of electron transport
directly. $\Gamma _{L}$ ($\Gamma _{R}$) determines the rate at which
electrons are continuously pumped into (out of) the left (right) quantum
dot. $J$ represents the electron tunnelling rate between two quantum dots.
Due to Coulomb blockade, these three parameters play an important role in
the dynamic behaviour of the leakage current when Pauli blockade is lift by
an oscillating magnetic field. As shown in Fig.\ref{b}(b1), the leakage
current is saturated when the injection rate $\Gamma _{L}$ becomes large,
because the residual electrons in the source lead are (Coulomb) blocked by
the electron in the left quantum dot. As shown in Fig.\ref{b}(b2)-(b3), a
large $\Gamma _{R}$ and a suitable value of $J$ would lead to a more
prominent leakage current. The dependence of leakage current on these
parameters may help to optimise the performance of the proposed nanotube quantum
sensor.

\subsection{Identifying nucleus species}
Based on the essential idea for the detection of an oscillating field, we propose to realize single molecule detection in the main text. A single molecule is usually characterized by different species of nuclear
spins with multiple Larmor frequencies. For each individual nuclear spin,
the magnetic moment determines not only the Larmor frequency in an external
magnetic field which indicates the driving Rabi frequency required to achieve
a resonance signal but also the magnetic dipole-dipole coupling strength
between the nuclear spin and the valley-spin qubit. As shown in Fig.\ref{n}(a1)-(a3), once the driving Rabi frequency matches the Larmor frequency of
each nuclear spin ($^{31}$P, $^{19} $F, $^{1}$H), the leakage current would
demonstrate a resonance feature. Given the same distance between the
valley-spin qubit and the nuclear spin, the magnitude of leakage current
induced by $^{31}$P nuclear spin is much smaller than that of $^{31}$F and $%
^{1}$H nuclear spins due to its smaller magnetic moment. We remark that the
values of parameters $J$ and $\Gamma _{R}$ have been optimised according the
results as shown in Fig.\ref{n}(b1)-(b3).

\vspace{0.05in}

We also investigate the effect of the parameters $\Gamma_{L}$, $B_{x}$, $B_{z}$ on the observed resonance signal. In Fig.\ref{Bxz}(a), the leakage current on resonance increases
with the electron injection rate $\Gamma _{L}$ up to a saturation value when
$\Gamma _{L}/2\pi$ is about 1 MHz. In Fig.\ref{Bxz}(b)-(c), we plot the
leakage current for different values of transverse ($B_{x}$) and
longitudinal ($B_{z}$) magnetic field components. It can be seen that the
peak value of leakage current is almost independent on $B_{z}$, while it
increases as $B_{x}$ becomes larger.

\end{appendix}

\begin{thebibliography}{56}%
\makeatletter
\providecommand \@ifxundefined [1]{%
 \@ifx{#1\undefined}
}%
\providecommand \@ifnum [1]{%
 \ifnum #1\expandafter \@firstoftwo
 \else \expandafter \@secondoftwo
 \fi
}%
\providecommand \@ifx [1]{%
 \ifx #1\expandafter \@firstoftwo
 \else \expandafter \@secondoftwo
 \fi
}%
\providecommand \natexlab [1]{#1}%
\providecommand \enquote  [1]{``#1''}%
\providecommand \bibnamefont  [1]{#1}%
\providecommand \bibfnamefont [1]{#1}%
\providecommand \citenamefont [1]{#1}%
\providecommand \href@noop [0]{\@secondoftwo}%
\providecommand \href [0]{\begingroup \@sanitize@url \@href}%
\providecommand \@href[1]{\@@startlink{#1}\@@href}%
\providecommand \@@href[1]{\endgroup#1\@@endlink}%
\providecommand \@sanitize@url [0]{\catcode `\\12\catcode `\$12\catcode
  `\&12\catcode `\#12\catcode `\^12\catcode `\_12\catcode `\%12\relax}%
\providecommand \@@startlink[1]{}%
\providecommand \@@endlink[0]{}%
\providecommand \url  [0]{\begingroup\@sanitize@url \@url }%
\providecommand \@url [1]{\endgroup\@href {#1}{\urlprefix }}%
\providecommand \urlprefix  [0]{URL }%
\providecommand \Eprint [0]{\href }%
\providecommand \doibase [0]{http://dx.doi.org/}%
\providecommand \selectlanguage [0]{\@gobble}%
\providecommand \bibinfo  [0]{\@secondoftwo}%
\providecommand \bibfield  [0]{\@secondoftwo}%
\providecommand \translation [1]{[#1]}%
\providecommand \BibitemOpen [0]{}%
\providecommand \bibitemStop [0]{}%
\providecommand \bibitemNoStop [0]{.\EOS\space}%
\providecommand \EOS [0]{\spacefactor3000\relax}%
\providecommand \BibitemShut  [1]{\csname bibitem#1\endcsname}%
\let\auto@bib@innerbib\@empty
\bibitem [{\citenamefont {Degen}\ \emph {et~al.}(2017)\citenamefont {Degen},
  \citenamefont {Reinhard},\ and\ \citenamefont {Cappellaro}}]{Degen_RMP2017}%
  \BibitemOpen
  \bibfield  {author} {\bibinfo {author} {\bibfnamefont {C.~L.}\ \bibnamefont
  {Degen}}, \bibinfo {author} {\bibfnamefont {F.}~\bibnamefont {Reinhard}}, \
  and\ \bibinfo {author} {\bibfnamefont {P.}~\bibnamefont {Cappellaro}},\
  }\href {\doibase 10.1103/RevModPhys.89.035002} {\bibfield  {journal}
  {\bibinfo  {journal} {Rev. Mod. Phys.}\ }\textbf {\bibinfo {volume} {89}},\
  \bibinfo {pages} {035002} (\bibinfo {year} {2017})}\BibitemShut {NoStop}%
\bibitem [{\citenamefont {Armani}\ \emph {et~al.}(2007)\citenamefont {Armani},
  \citenamefont {Kulkarni}, \citenamefont {Fraser}, \citenamefont {Flagan},\
  and\ \citenamefont {Vahala}}]{Vahala2007}%
  \BibitemOpen
  \bibfield  {author} {\bibinfo {author} {\bibfnamefont {A.~M.}\ \bibnamefont
  {Armani}}, \bibinfo {author} {\bibfnamefont {R.~P.}\ \bibnamefont
  {Kulkarni}}, \bibinfo {author} {\bibfnamefont {S.~E.}\ \bibnamefont
  {Fraser}}, \bibinfo {author} {\bibfnamefont {R.~C.}\ \bibnamefont {Flagan}},
  \ and\ \bibinfo {author} {\bibfnamefont {K.~J.}\ \bibnamefont {Vahala}},\
  }\href {\doibase 10.1126/science.1145002} {\bibfield  {journal} {\bibinfo
  {journal} {Science}\ }\textbf {\bibinfo {volume} {317}},\ \bibinfo {pages}
  {783} (\bibinfo {year} {2007})}\BibitemShut {NoStop}%
\bibitem [{\citenamefont {Arroyo}\ and\ \citenamefont
  {Kukura}(2015)}]{Kukura2016}%
  \BibitemOpen
  \bibfield  {author} {\bibinfo {author} {\bibfnamefont {J.~O.}\ \bibnamefont
  {Arroyo}}\ and\ \bibinfo {author} {\bibfnamefont {P.}~\bibnamefont
  {Kukura}},\ }\href {\doibase 10.1038/nphoton.2015.251} {\bibfield  {journal}
  {\bibinfo  {journal} {Nature Photonics}\ }\textbf {\bibinfo {volume} {10}},\
  \bibinfo {pages} {11} (\bibinfo {year} {2015})}\BibitemShut {NoStop}%
\bibitem [{\citenamefont {Lee}\ \emph {et~al.}(2018)\citenamefont {Lee},
  \citenamefont {Tallarida}, \citenamefont {Chen}, \citenamefont {Jensen},\
  and\ \citenamefont {Apkarian}}]{Apkarian2018}%
  \BibitemOpen
  \bibfield  {author} {\bibinfo {author} {\bibfnamefont {J.}~\bibnamefont
  {Lee}}, \bibinfo {author} {\bibfnamefont {N.}~\bibnamefont {Tallarida}},
  \bibinfo {author} {\bibfnamefont {X.}~\bibnamefont {Chen}}, \bibinfo {author}
  {\bibfnamefont {L.}~\bibnamefont {Jensen}}, \ and\ \bibinfo {author}
  {\bibfnamefont {V.~A.}\ \bibnamefont {Apkarian}},\ }\href
  {http://advances.sciencemag.org/content/4/6/eaat5472} {\bibfield  {journal}
  {\bibinfo  {journal} {Science Advances}\ }\textbf {\bibinfo {volume} {4}}
  (\bibinfo {year} {2018})}\BibitemShut {NoStop}%
\bibitem [{\citenamefont {Glenn}\ \emph {et~al.}(2018)\citenamefont {Glenn},
  \citenamefont {Bucher}, \citenamefont {Lee}, \citenamefont {Lukin},
  \citenamefont {Park},\ and\ \citenamefont {Walsworth}}]{glenn2018high}%
  \BibitemOpen
  \bibfield  {author} {\bibinfo {author} {\bibfnamefont {D.}~\bibnamefont
  {Glenn}}, \bibinfo {author} {\bibfnamefont {D.}~\bibnamefont {Bucher}},
  \bibinfo {author} {\bibfnamefont {J.}~\bibnamefont {Lee}}, \bibinfo {author}
  {\bibfnamefont {M.}~\bibnamefont {Lukin}}, \bibinfo {author} {\bibfnamefont
  {H.}~\bibnamefont {Park}}, \ and\ \bibinfo {author} {\bibfnamefont
  {R.}~\bibnamefont {Walsworth}},\ }\href
  {https://www.nature.com/articles/nature25781} {\bibfield  {journal} {\bibinfo
   {journal} {Nature}\ }\textbf {\bibinfo {volume} {555}},\ \bibinfo {pages}
  {351} (\bibinfo {year} {2018})}\BibitemShut {NoStop}%
\bibitem [{\citenamefont {Schmitt}\ \emph {et~al.}(2017)\citenamefont
  {Schmitt}, \citenamefont {Gefen}, \citenamefont {St{\"u}rner}, \citenamefont
  {Unden}, \citenamefont {Wolff}, \citenamefont {M{\"u}ller}, \citenamefont
  {Scheuer}, \citenamefont {Naydenov}, \citenamefont {Markham}, \citenamefont
  {Pezzagna} \emph {et~al.}}]{schmitt2017submillihertz}%
  \BibitemOpen
  \bibfield  {author} {\bibinfo {author} {\bibfnamefont {S.}~\bibnamefont
  {Schmitt}}, \bibinfo {author} {\bibfnamefont {T.}~\bibnamefont {Gefen}},
  \bibinfo {author} {\bibfnamefont {F.~M.}\ \bibnamefont {St{\"u}rner}},
  \bibinfo {author} {\bibfnamefont {T.}~\bibnamefont {Unden}}, \bibinfo
  {author} {\bibfnamefont {G.}~\bibnamefont {Wolff}}, \bibinfo {author}
  {\bibfnamefont {C.}~\bibnamefont {M{\"u}ller}}, \bibinfo {author}
  {\bibfnamefont {J.}~\bibnamefont {Scheuer}}, \bibinfo {author} {\bibfnamefont
  {B.}~\bibnamefont {Naydenov}}, \bibinfo {author} {\bibfnamefont
  {M.}~\bibnamefont {Markham}}, \bibinfo {author} {\bibfnamefont
  {S.}~\bibnamefont {Pezzagna}},  \emph {et~al.},\ }\href
  {http://science.sciencemag.org/content/356/6340/832} {\bibfield  {journal}
  {\bibinfo  {journal} {Science}\ }\textbf {\bibinfo {volume} {356}},\ \bibinfo
  {pages} {832} (\bibinfo {year} {2017})}\BibitemShut {NoStop}%
\bibitem [{\citenamefont {Lovchinsky}\ \emph {et~al.}(2016)\citenamefont
  {Lovchinsky}, \citenamefont {Sushkov}, \citenamefont {Urbach}, \citenamefont
  {de~Leon}, \citenamefont {Choi}, \citenamefont {De~Greve}, \citenamefont
  {Evans}, \citenamefont {Gertner}, \citenamefont {Bersin}, \citenamefont
  {M{\"u}ller}, \citenamefont {McGuinness}, \citenamefont {Jelezko},
  \citenamefont {Walsworth}, \citenamefont {Park},\ and\ \citenamefont
  {Lukin}}]{Lukin2016}%
  \BibitemOpen
  \bibfield  {author} {\bibinfo {author} {\bibfnamefont {I.}~\bibnamefont
  {Lovchinsky}}, \bibinfo {author} {\bibfnamefont {A.~O.}\ \bibnamefont
  {Sushkov}}, \bibinfo {author} {\bibfnamefont {E.}~\bibnamefont {Urbach}},
  \bibinfo {author} {\bibfnamefont {N.~P.}\ \bibnamefont {de~Leon}}, \bibinfo
  {author} {\bibfnamefont {S.}~\bibnamefont {Choi}}, \bibinfo {author}
  {\bibfnamefont {K.}~\bibnamefont {De~Greve}}, \bibinfo {author}
  {\bibfnamefont {R.}~\bibnamefont {Evans}}, \bibinfo {author} {\bibfnamefont
  {R.}~\bibnamefont {Gertner}}, \bibinfo {author} {\bibfnamefont
  {E.}~\bibnamefont {Bersin}}, \bibinfo {author} {\bibfnamefont
  {C.}~\bibnamefont {M{\"u}ller}}, \bibinfo {author} {\bibfnamefont
  {L.}~\bibnamefont {McGuinness}}, \bibinfo {author} {\bibfnamefont
  {F.}~\bibnamefont {Jelezko}}, \bibinfo {author} {\bibfnamefont {R.~L.}\
  \bibnamefont {Walsworth}}, \bibinfo {author} {\bibfnamefont {H.}~\bibnamefont
  {Park}}, \ and\ \bibinfo {author} {\bibfnamefont {M.~D.}\ \bibnamefont
  {Lukin}},\ }\href {\doibase 10.1126/science.aad8022} {\bibfield  {journal}
  {\bibinfo  {journal} {Science}\ }\textbf {\bibinfo {volume} {351}},\ \bibinfo
  {pages} {836} (\bibinfo {year} {2016})}\BibitemShut {NoStop}%
\bibitem [{\citenamefont {Shi}\ \emph {et~al.}(2015)\citenamefont {Shi},
  \citenamefont {Zhang}, \citenamefont {Wang}, \citenamefont {Sun},
  \citenamefont {Wang}, \citenamefont {Rong}, \citenamefont {Chen},
  \citenamefont {Ju}, \citenamefont {Reinhard}, \citenamefont {Chen},
  \citenamefont {Wrachtrup}, \citenamefont {Wang},\ and\ \citenamefont
  {Du}}]{Shi15}%
  \BibitemOpen
  \bibfield  {author} {\bibinfo {author} {\bibfnamefont {F.}~\bibnamefont
  {Shi}}, \bibinfo {author} {\bibfnamefont {Q.}~\bibnamefont {Zhang}}, \bibinfo
  {author} {\bibfnamefont {P.}~\bibnamefont {Wang}}, \bibinfo {author}
  {\bibfnamefont {H.}~\bibnamefont {Sun}}, \bibinfo {author} {\bibfnamefont
  {J.}~\bibnamefont {Wang}}, \bibinfo {author} {\bibfnamefont {X.}~\bibnamefont
  {Rong}}, \bibinfo {author} {\bibfnamefont {M.}~\bibnamefont {Chen}}, \bibinfo
  {author} {\bibfnamefont {C.}~\bibnamefont {Ju}}, \bibinfo {author}
  {\bibfnamefont {F.}~\bibnamefont {Reinhard}}, \bibinfo {author}
  {\bibfnamefont {H.}~\bibnamefont {Chen}}, \bibinfo {author} {\bibfnamefont
  {J.}~\bibnamefont {Wrachtrup}}, \bibinfo {author} {\bibfnamefont
  {J.}~\bibnamefont {Wang}}, \ and\ \bibinfo {author} {\bibfnamefont
  {J.}~\bibnamefont {Du}},\ }\href {\doibase 10.1126/science.aaa2253}
  {\bibfield  {journal} {\bibinfo  {journal} {Science}\ }\textbf {\bibinfo
  {volume} {347}},\ \bibinfo {pages} {1135} (\bibinfo {year}
  {2015})}\BibitemShut {NoStop}%
\bibitem [{\citenamefont {M{\"u}ller}\ \emph {et~al.}(2014)\citenamefont
  {M{\"u}ller}, \citenamefont {Kong}, \citenamefont {Cai}, \citenamefont
  {Melentijevi{\'{c}}}, \citenamefont {Stacey}, \citenamefont {Markham},
  \citenamefont {Twitchen}, \citenamefont {Isoya}, \citenamefont {Pezzagna},
  \citenamefont {Meijer}, \citenamefont {Du}, \citenamefont {Plenio},
  \citenamefont {Naydenov}, \citenamefont {McGuinness},\ and\ \citenamefont
  {Jelezko}}]{Muller14}%
  \BibitemOpen
  \bibfield  {author} {\bibinfo {author} {\bibfnamefont {C.}~\bibnamefont
  {M{\"u}ller}}, \bibinfo {author} {\bibfnamefont {X.}~\bibnamefont {Kong}},
  \bibinfo {author} {\bibfnamefont {J.~M.}\ \bibnamefont {Cai}}, \bibinfo
  {author} {\bibfnamefont {K.}~\bibnamefont {Melentijevi{\'{c}}}}, \bibinfo
  {author} {\bibfnamefont {A.}~\bibnamefont {Stacey}}, \bibinfo {author}
  {\bibfnamefont {M.}~\bibnamefont {Markham}}, \bibinfo {author} {\bibfnamefont
  {D.}~\bibnamefont {Twitchen}}, \bibinfo {author} {\bibfnamefont
  {J.}~\bibnamefont {Isoya}}, \bibinfo {author} {\bibfnamefont
  {S.}~\bibnamefont {Pezzagna}}, \bibinfo {author} {\bibfnamefont
  {J.}~\bibnamefont {Meijer}}, \bibinfo {author} {\bibfnamefont {J.~F.}\
  \bibnamefont {Du}}, \bibinfo {author} {\bibfnamefont {M.~B.}\ \bibnamefont
  {Plenio}}, \bibinfo {author} {\bibfnamefont {B.}~\bibnamefont {Naydenov}},
  \bibinfo {author} {\bibfnamefont {L.~P.}\ \bibnamefont {McGuinness}}, \ and\
  \bibinfo {author} {\bibfnamefont {F.}~\bibnamefont {Jelezko}},\ }\href
  {\doibase 10.1038/ncomms5703
  https://www.nature.com/articles/ncomms5703#supplementary-information}
  {\bibfield  {journal} {\bibinfo  {journal} {Nature Communications}\ }\textbf
  {\bibinfo {volume} {5}},\ \bibinfo {pages} {4703} (\bibinfo {year}
  {2014})}\BibitemShut {NoStop}%
\bibitem [{\citenamefont {Sushkov}\ \emph {et~al.}(2014)\citenamefont
  {Sushkov}, \citenamefont {Lovchinsky}, \citenamefont {Chisholm},
  \citenamefont {Walsworth}, \citenamefont {Park},\ and\ \citenamefont
  {Lukin}}]{Sus14}%
  \BibitemOpen
  \bibfield  {author} {\bibinfo {author} {\bibfnamefont {A.~O.}\ \bibnamefont
  {Sushkov}}, \bibinfo {author} {\bibfnamefont {I.}~\bibnamefont {Lovchinsky}},
  \bibinfo {author} {\bibfnamefont {N.}~\bibnamefont {Chisholm}}, \bibinfo
  {author} {\bibfnamefont {R.~L.}\ \bibnamefont {Walsworth}}, \bibinfo {author}
  {\bibfnamefont {H.}~\bibnamefont {Park}}, \ and\ \bibinfo {author}
  {\bibfnamefont {M.~D.}\ \bibnamefont {Lukin}},\ }\href {\doibase
  10.1103/PhysRevLett.113.197601} {\bibfield  {journal} {\bibinfo  {journal}
  {Phys. Rev. Lett.}\ }\textbf {\bibinfo {volume} {113}},\ \bibinfo {pages}
  {197601} (\bibinfo {year} {2014})}\BibitemShut {NoStop}%
\bibitem [{\citenamefont {Staudacher}\ \emph {et~al.}(2013)\citenamefont
  {Staudacher}, \citenamefont {Shi}, \citenamefont {Pezzagna}, \citenamefont
  {Meijer}, \citenamefont {Du}, \citenamefont {Meriles}, \citenamefont
  {Reinhard},\ and\ \citenamefont {Wrachtrup}}]{Stau13}%
  \BibitemOpen
  \bibfield  {author} {\bibinfo {author} {\bibfnamefont {T.}~\bibnamefont
  {Staudacher}}, \bibinfo {author} {\bibfnamefont {F.}~\bibnamefont {Shi}},
  \bibinfo {author} {\bibfnamefont {S.}~\bibnamefont {Pezzagna}}, \bibinfo
  {author} {\bibfnamefont {J.}~\bibnamefont {Meijer}}, \bibinfo {author}
  {\bibfnamefont {J.}~\bibnamefont {Du}}, \bibinfo {author} {\bibfnamefont
  {C.~A.}\ \bibnamefont {Meriles}}, \bibinfo {author} {\bibfnamefont
  {F.}~\bibnamefont {Reinhard}}, \ and\ \bibinfo {author} {\bibfnamefont
  {J.}~\bibnamefont {Wrachtrup}},\ }\href {\doibase 10.1126/science.1231675}
  {\bibfield  {journal} {\bibinfo  {journal} {Science}\ }\textbf {\bibinfo
  {volume} {339}},\ \bibinfo {pages} {561} (\bibinfo {year}
  {2013})}\BibitemShut {NoStop}%
\bibitem [{\citenamefont {Mamin}\ \emph {et~al.}(2013)\citenamefont {Mamin},
  \citenamefont {Kim}, \citenamefont {Sherwood}, \citenamefont {Rettner},
  \citenamefont {Ohno}, \citenamefont {Awschalom},\ and\ \citenamefont
  {Rugar}}]{Mamin13}%
  \BibitemOpen
  \bibfield  {author} {\bibinfo {author} {\bibfnamefont {H.~J.}\ \bibnamefont
  {Mamin}}, \bibinfo {author} {\bibfnamefont {M.}~\bibnamefont {Kim}}, \bibinfo
  {author} {\bibfnamefont {M.~H.}\ \bibnamefont {Sherwood}}, \bibinfo {author}
  {\bibfnamefont {C.~T.}\ \bibnamefont {Rettner}}, \bibinfo {author}
  {\bibfnamefont {K.}~\bibnamefont {Ohno}}, \bibinfo {author} {\bibfnamefont
  {D.~D.}\ \bibnamefont {Awschalom}}, \ and\ \bibinfo {author} {\bibfnamefont
  {D.}~\bibnamefont {Rugar}},\ }\href {\doibase 10.1126/science.1231540}
  {\bibfield  {journal} {\bibinfo  {journal} {Science}\ }\textbf {\bibinfo
  {volume} {339}},\ \bibinfo {pages} {557} (\bibinfo {year}
  {2013})}\BibitemShut {NoStop}%
\bibitem [{\citenamefont {Cai}\ \emph {et~al.}(2014)\citenamefont {Cai},
  \citenamefont {Jelezko},\ and\ \citenamefont {Plenio}}]{Cai2014}%
  \BibitemOpen
  \bibfield  {author} {\bibinfo {author} {\bibfnamefont {J.}~\bibnamefont
  {Cai}}, \bibinfo {author} {\bibfnamefont {F.}~\bibnamefont {Jelezko}}, \ and\
  \bibinfo {author} {\bibfnamefont {M.~B.}\ \bibnamefont {Plenio}},\ }\href
  {https://doi.org/10.1038/ncomms5065} {\bibfield  {journal} {\bibinfo
  {journal} {Nature Communications}\ }\textbf {\bibinfo {volume} {5}},\
  \bibinfo {pages} {4065} (\bibinfo {year} {2014})}\BibitemShut {NoStop}%
\bibitem [{\citenamefont {Zhang}\ \emph {et~al.}(2018)\citenamefont {Zhang},
  \citenamefont {Liu}, \citenamefont {Leong}, \citenamefont {Liu},
  \citenamefont {Kwok}, \citenamefont {Ngai}, \citenamefont {Liu},\ and\
  \citenamefont {Li}}]{Zhang2018_NC}%
  \BibitemOpen
  \bibfield  {author} {\bibinfo {author} {\bibfnamefont {T.}~\bibnamefont
  {Zhang}}, \bibinfo {author} {\bibfnamefont {G.-Q.}\ \bibnamefont {Liu}},
  \bibinfo {author} {\bibfnamefont {W.-H.}\ \bibnamefont {Leong}}, \bibinfo
  {author} {\bibfnamefont {C.-F.}\ \bibnamefont {Liu}}, \bibinfo {author}
  {\bibfnamefont {M.-H.}\ \bibnamefont {Kwok}}, \bibinfo {author}
  {\bibfnamefont {T.}~\bibnamefont {Ngai}}, \bibinfo {author} {\bibfnamefont
  {R.-B.}\ \bibnamefont {Liu}}, \ and\ \bibinfo {author} {\bibfnamefont
  {Q.}~\bibnamefont {Li}},\ }\href
  {https://www.nature.com/articles/s41467-018-05673-9/} {\bibfield  {journal}
  {\bibinfo  {journal} {Nature Communications}\ }\textbf {\bibinfo {volume}
  {9}},\ \bibinfo {pages} {3188} (\bibinfo {year} {2018})}\BibitemShut
  {NoStop}%
\bibitem [{\citenamefont {Doherty}\ \emph {et~al.}(2013)\citenamefont
  {Doherty}, \citenamefont {Manson}, \citenamefont {Delaney}, \citenamefont
  {Jelezko}, \citenamefont {Wrachtrup},\ and\ \citenamefont
  {Hollenberg}}]{Doherty13}%
  \BibitemOpen
  \bibfield  {author} {\bibinfo {author} {\bibfnamefont {M.~W.}\ \bibnamefont
  {Doherty}}, \bibinfo {author} {\bibfnamefont {N.~B.}\ \bibnamefont {Manson}},
  \bibinfo {author} {\bibfnamefont {P.}~\bibnamefont {Delaney}}, \bibinfo
  {author} {\bibfnamefont {F.}~\bibnamefont {Jelezko}}, \bibinfo {author}
  {\bibfnamefont {J.}~\bibnamefont {Wrachtrup}}, \ and\ \bibinfo {author}
  {\bibfnamefont {L.~C.}\ \bibnamefont {Hollenberg}},\ }\href {\doibase
  https://doi.org/10.1016/j.physrep.2013.02.001} {\bibfield  {journal}
  {\bibinfo  {journal} {Physics Reports}\ }\textbf {\bibinfo {volume} {528}},\
  \bibinfo {pages} {1 } (\bibinfo {year} {2013})}\BibitemShut {NoStop}%
\bibitem [{\citenamefont {Schirhagl}\ \emph {et~al.}(2014)\citenamefont
  {Schirhagl}, \citenamefont {Chang}, \citenamefont {Loretz},\ and\
  \citenamefont {Degen}}]{Schir14}%
  \BibitemOpen
  \bibfield  {author} {\bibinfo {author} {\bibfnamefont {R.}~\bibnamefont
  {Schirhagl}}, \bibinfo {author} {\bibfnamefont {K.}~\bibnamefont {Chang}},
  \bibinfo {author} {\bibfnamefont {M.}~\bibnamefont {Loretz}}, \ and\ \bibinfo
  {author} {\bibfnamefont {C.~L.}\ \bibnamefont {Degen}},\ }\href {\doibase
  10.1146/annurev-physchem-040513-103659} {\bibfield  {journal} {\bibinfo
  {journal} {Annual Review of Physical Chemistry}\ }\textbf {\bibinfo {volume}
  {65}},\ \bibinfo {pages} {83} (\bibinfo {year} {2014})}\BibitemShut {NoStop}%
\bibitem [{\citenamefont {Wu}\ \emph {et~al.}(2016)\citenamefont {Wu},
  \citenamefont {Jelezko}, \citenamefont {Plenio},\ and\ \citenamefont
  {Weil}}]{Wu2016}%
  \BibitemOpen
  \bibfield  {author} {\bibinfo {author} {\bibfnamefont {Y.}~\bibnamefont
  {Wu}}, \bibinfo {author} {\bibfnamefont {F.}~\bibnamefont {Jelezko}},
  \bibinfo {author} {\bibfnamefont {M.~B.}\ \bibnamefont {Plenio}}, \ and\
  \bibinfo {author} {\bibfnamefont {T.}~\bibnamefont {Weil}},\ }\href {\doibase
  10.1002/anie.201506556} {\bibfield  {journal} {\bibinfo  {journal}
  {Angewandte Chemie International Edition}\ }\textbf {\bibinfo {volume}
  {55}},\ \bibinfo {pages} {6586} (\bibinfo {year} {2016})}\BibitemShut
  {NoStop}%
\bibitem [{\citenamefont {Tisler}\ \emph {et~al.}(2009)\citenamefont {Tisler},
  \citenamefont {Balasubramanian}, \citenamefont {Naydenov}, \citenamefont
  {Kolesov}, \citenamefont {Grotz}, \citenamefont {Reuter}, \citenamefont
  {Boudou}, \citenamefont {Curmi}, \citenamefont {Sennour}, \citenamefont
  {Thorel}, \citenamefont {B{\"o}rsch}, \citenamefont {Aulenbacher},
  \citenamefont {Erdmann}, \citenamefont {Hemmer}, \citenamefont {Jelezko},\
  and\ \citenamefont {Wrachtrup}}]{Tisler09}%
  \BibitemOpen
  \bibfield  {author} {\bibinfo {author} {\bibfnamefont {J.}~\bibnamefont
  {Tisler}}, \bibinfo {author} {\bibfnamefont {G.}~\bibnamefont
  {Balasubramanian}}, \bibinfo {author} {\bibfnamefont {B.}~\bibnamefont
  {Naydenov}}, \bibinfo {author} {\bibfnamefont {R.}~\bibnamefont {Kolesov}},
  \bibinfo {author} {\bibfnamefont {B.}~\bibnamefont {Grotz}}, \bibinfo
  {author} {\bibfnamefont {R.}~\bibnamefont {Reuter}}, \bibinfo {author}
  {\bibfnamefont {J.-P.}\ \bibnamefont {Boudou}}, \bibinfo {author}
  {\bibfnamefont {P.~A.}\ \bibnamefont {Curmi}}, \bibinfo {author}
  {\bibfnamefont {M.}~\bibnamefont {Sennour}}, \bibinfo {author} {\bibfnamefont
  {A.}~\bibnamefont {Thorel}}, \bibinfo {author} {\bibfnamefont
  {M.}~\bibnamefont {B{\"o}rsch}}, \bibinfo {author} {\bibfnamefont
  {K.}~\bibnamefont {Aulenbacher}}, \bibinfo {author} {\bibfnamefont
  {R.}~\bibnamefont {Erdmann}}, \bibinfo {author} {\bibfnamefont {P.~R.}\
  \bibnamefont {Hemmer}}, \bibinfo {author} {\bibfnamefont {F.}~\bibnamefont
  {Jelezko}}, \ and\ \bibinfo {author} {\bibfnamefont {J.}~\bibnamefont
  {Wrachtrup}},\ }\href {\doibase 10.1021/nn9003617} {\bibfield  {journal}
  {\bibinfo  {journal} {ACS Nano}\ }\textbf {\bibinfo {volume} {3}},\ \bibinfo
  {pages} {1959} (\bibinfo {year} {2009})}\BibitemShut {NoStop}%
\bibitem [{\citenamefont {Rosskopf}\ \emph {et~al.}(2014)\citenamefont
  {Rosskopf}, \citenamefont {Dussaux}, \citenamefont {Ohashi}, \citenamefont
  {Loretz}, \citenamefont {Schirhagl}, \citenamefont {Watanabe}, \citenamefont
  {Shikata}, \citenamefont {Itoh},\ and\ \citenamefont {Degen}}]{Rosskopf2014}%
  \BibitemOpen
  \bibfield  {author} {\bibinfo {author} {\bibfnamefont {T.}~\bibnamefont
  {Rosskopf}}, \bibinfo {author} {\bibfnamefont {A.}~\bibnamefont {Dussaux}},
  \bibinfo {author} {\bibfnamefont {K.}~\bibnamefont {Ohashi}}, \bibinfo
  {author} {\bibfnamefont {M.}~\bibnamefont {Loretz}}, \bibinfo {author}
  {\bibfnamefont {R.}~\bibnamefont {Schirhagl}}, \bibinfo {author}
  {\bibfnamefont {H.}~\bibnamefont {Watanabe}}, \bibinfo {author}
  {\bibfnamefont {S.}~\bibnamefont {Shikata}}, \bibinfo {author} {\bibfnamefont
  {K.~M.}\ \bibnamefont {Itoh}}, \ and\ \bibinfo {author} {\bibfnamefont
  {C.~L.}\ \bibnamefont {Degen}},\ }\href {\doibase
  10.1103/PhysRevLett.112.147602} {\bibfield  {journal} {\bibinfo  {journal}
  {Phys. Rev. Lett.}\ }\textbf {\bibinfo {volume} {112}},\ \bibinfo {pages}
  {147602} (\bibinfo {year} {2014})}\BibitemShut {NoStop}%
\bibitem [{\citenamefont {Myers}\ \emph {et~al.}(2014)\citenamefont {Myers},
  \citenamefont {Das}, \citenamefont {Dartiailh}, \citenamefont {Ohno},
  \citenamefont {Awschalom},\ and\ \citenamefont
  {Bleszynski~Jayich}}]{Myers2014}%
  \BibitemOpen
  \bibfield  {author} {\bibinfo {author} {\bibfnamefont {B.~A.}\ \bibnamefont
  {Myers}}, \bibinfo {author} {\bibfnamefont {A.}~\bibnamefont {Das}}, \bibinfo
  {author} {\bibfnamefont {M.~C.}\ \bibnamefont {Dartiailh}}, \bibinfo {author}
  {\bibfnamefont {K.}~\bibnamefont {Ohno}}, \bibinfo {author} {\bibfnamefont
  {D.~D.}\ \bibnamefont {Awschalom}}, \ and\ \bibinfo {author} {\bibfnamefont
  {A.~C.}\ \bibnamefont {Bleszynski~Jayich}},\ }\href {\doibase
  10.1103/PhysRevLett.113.027602} {\bibfield  {journal} {\bibinfo  {journal}
  {Phys. Rev. Lett.}\ }\textbf {\bibinfo {volume} {113}},\ \bibinfo {pages}
  {027602} (\bibinfo {year} {2014})}\BibitemShut {NoStop}%
\bibitem [{\citenamefont {Kim}\ \emph {et~al.}(2015)\citenamefont {Kim},
  \citenamefont {Mamin}, \citenamefont {Sherwood}, \citenamefont {Ohno},
  \citenamefont {Awschalom},\ and\ \citenamefont {Rugar}}]{Kim2015}%
  \BibitemOpen
  \bibfield  {author} {\bibinfo {author} {\bibfnamefont {M.}~\bibnamefont
  {Kim}}, \bibinfo {author} {\bibfnamefont {H.~J.}\ \bibnamefont {Mamin}},
  \bibinfo {author} {\bibfnamefont {M.~H.}\ \bibnamefont {Sherwood}}, \bibinfo
  {author} {\bibfnamefont {K.}~\bibnamefont {Ohno}}, \bibinfo {author}
  {\bibfnamefont {D.~D.}\ \bibnamefont {Awschalom}}, \ and\ \bibinfo {author}
  {\bibfnamefont {D.}~\bibnamefont {Rugar}},\ }\href {\doibase
  10.1103/PhysRevLett.115.087602} {\bibfield  {journal} {\bibinfo  {journal}
  {Phys. Rev. Lett.}\ }\textbf {\bibinfo {volume} {115}},\ \bibinfo {pages}
  {087602} (\bibinfo {year} {2015})}\BibitemShut {NoStop}%
\bibitem [{\citenamefont {Laird}\ \emph {et~al.}(2015)\citenamefont {Laird},
  \citenamefont {Kuemmeth}, \citenamefont {Steele}, \citenamefont
  {Grove-Rasmussen}, \citenamefont {Nyg\aa{}rd}, \citenamefont {Flensberg},\
  and\ \citenamefont {Kouwenhoven}}]{Laird2015}%
  \BibitemOpen
  \bibfield  {author} {\bibinfo {author} {\bibfnamefont {E.~A.}\ \bibnamefont
  {Laird}}, \bibinfo {author} {\bibfnamefont {F.}~\bibnamefont {Kuemmeth}},
  \bibinfo {author} {\bibfnamefont {G.~A.}\ \bibnamefont {Steele}}, \bibinfo
  {author} {\bibfnamefont {K.}~\bibnamefont {Grove-Rasmussen}}, \bibinfo
  {author} {\bibfnamefont {J.}~\bibnamefont {Nyg\aa{}rd}}, \bibinfo {author}
  {\bibfnamefont {K.}~\bibnamefont {Flensberg}}, \ and\ \bibinfo {author}
  {\bibfnamefont {L.~P.}\ \bibnamefont {Kouwenhoven}},\ }\href {\doibase
  10.1103/RevModPhys.87.703} {\bibfield  {journal} {\bibinfo  {journal} {Rev.
  Mod. Phys.}\ }\textbf {\bibinfo {volume} {87}},\ \bibinfo {pages} {703}
  (\bibinfo {year} {2015})}\BibitemShut {NoStop}%
\bibitem [{\citenamefont {Rohling}\ and\ \citenamefont
  {Burkard}(2012)}]{Burkard2012}%
  \BibitemOpen
  \bibfield  {author} {\bibinfo {author} {\bibfnamefont {N.}~\bibnamefont
  {Rohling}}\ and\ \bibinfo {author} {\bibfnamefont {G.}~\bibnamefont
  {Burkard}},\ }\href {http://stacks.iop.org/1367-2630/14/i=8/a=083008}
  {\bibfield  {journal} {\bibinfo  {journal} {New Journal of Physics}\ }\textbf
  {\bibinfo {volume} {14}},\ \bibinfo {pages} {083008} (\bibinfo {year}
  {2012})}\BibitemShut {NoStop}%
\bibitem [{\citenamefont {P\'alyi}\ and\ \citenamefont
  {Burkard}(2009)}]{Burkard2009}%
  \BibitemOpen
  \bibfield  {author} {\bibinfo {author} {\bibfnamefont {A.}~\bibnamefont
  {P\'alyi}}\ and\ \bibinfo {author} {\bibfnamefont {G.}~\bibnamefont
  {Burkard}},\ }\href {\doibase 10.1103/PhysRevB.80.201404} {\bibfield
  {journal} {\bibinfo  {journal} {Phys. Rev. B}\ }\textbf {\bibinfo {volume}
  {80}},\ \bibinfo {pages} {201404} (\bibinfo {year} {2009})}\BibitemShut
  {NoStop}%
\bibitem [{\citenamefont {Chorley}\ \emph {et~al.}(2011)\citenamefont
  {Chorley}, \citenamefont {Giavaras}, \citenamefont {Wabnig}, \citenamefont
  {Jones}, \citenamefont {Smith}, \citenamefont {Briggs},\ and\ \citenamefont
  {Buitelaar}}]{Briggs2011}%
  \BibitemOpen
  \bibfield  {author} {\bibinfo {author} {\bibfnamefont {S.~J.}\ \bibnamefont
  {Chorley}}, \bibinfo {author} {\bibfnamefont {G.}~\bibnamefont {Giavaras}},
  \bibinfo {author} {\bibfnamefont {J.}~\bibnamefont {Wabnig}}, \bibinfo
  {author} {\bibfnamefont {G.~A.~C.}\ \bibnamefont {Jones}}, \bibinfo {author}
  {\bibfnamefont {C.~G.}\ \bibnamefont {Smith}}, \bibinfo {author}
  {\bibfnamefont {G.~A.~D.}\ \bibnamefont {Briggs}}, \ and\ \bibinfo {author}
  {\bibfnamefont {M.~R.}\ \bibnamefont {Buitelaar}},\ }\href {\doibase
  10.1103/PhysRevLett.106.206801} {\bibfield  {journal} {\bibinfo  {journal}
  {Phys. Rev. Lett.}\ }\textbf {\bibinfo {volume} {106}},\ \bibinfo {pages}
  {206801} (\bibinfo {year} {2011})}\BibitemShut {NoStop}%
\bibitem [{\citenamefont {Churchill}\ \emph
  {et~al.}(2009{\natexlab{a}})\citenamefont {Churchill}, \citenamefont
  {Kuemmeth}, \citenamefont {Harlow}, \citenamefont {Bestwick}, \citenamefont
  {Rashba}, \citenamefont {Flensberg}, \citenamefont {Stwertka}, \citenamefont
  {Taychatanapat}, \citenamefont {Watson},\ and\ \citenamefont
  {Marcus}}]{Marcus2009}%
  \BibitemOpen
  \bibfield  {author} {\bibinfo {author} {\bibfnamefont {H.~O.~H.}\
  \bibnamefont {Churchill}}, \bibinfo {author} {\bibfnamefont {F.}~\bibnamefont
  {Kuemmeth}}, \bibinfo {author} {\bibfnamefont {J.~W.}\ \bibnamefont
  {Harlow}}, \bibinfo {author} {\bibfnamefont {A.~J.}\ \bibnamefont
  {Bestwick}}, \bibinfo {author} {\bibfnamefont {E.~I.}\ \bibnamefont
  {Rashba}}, \bibinfo {author} {\bibfnamefont {K.}~\bibnamefont {Flensberg}},
  \bibinfo {author} {\bibfnamefont {C.~H.}\ \bibnamefont {Stwertka}}, \bibinfo
  {author} {\bibfnamefont {T.}~\bibnamefont {Taychatanapat}}, \bibinfo {author}
  {\bibfnamefont {S.~K.}\ \bibnamefont {Watson}}, \ and\ \bibinfo {author}
  {\bibfnamefont {C.~M.}\ \bibnamefont {Marcus}},\ }\href {\doibase
  10.1103/PhysRevLett.102.166802} {\bibfield  {journal} {\bibinfo  {journal}
  {Phys. Rev. Lett.}\ }\textbf {\bibinfo {volume} {102}},\ \bibinfo {pages}
  {166802} (\bibinfo {year} {2009}{\natexlab{a}})}\BibitemShut {NoStop}%
\bibitem [{\citenamefont {Pei}\ \emph {et~al.}(2012)\citenamefont {Pei},
  \citenamefont {Laird}, \citenamefont {Steele},\ and\ \citenamefont
  {Kouwenhoven}}]{Laird2012}%
  \BibitemOpen
  \bibfield  {author} {\bibinfo {author} {\bibfnamefont {F.}~\bibnamefont
  {Pei}}, \bibinfo {author} {\bibfnamefont {E.~A.}\ \bibnamefont {Laird}},
  \bibinfo {author} {\bibfnamefont {G.~A.}\ \bibnamefont {Steele}}, \ and\
  \bibinfo {author} {\bibfnamefont {L.~P.}\ \bibnamefont {Kouwenhoven}},\
  }\href {\doibase 10.1038/nnano.2012.160
  https://www.nature.com/articles/nnano.2012.160#supplementary-information}
  {\bibfield  {journal} {\bibinfo  {journal} {Nature Nanotechnology}\ }\textbf
  {\bibinfo {volume} {7}},\ \bibinfo {pages} {630} (\bibinfo {year}
  {2012})}\BibitemShut {NoStop}%
\bibitem [{\citenamefont {Moriyama}\ \emph {et~al.}(2005)\citenamefont
  {Moriyama}, \citenamefont {Fuse}, \citenamefont {Suzuki}, \citenamefont
  {Aoyagi},\ and\ \citenamefont {Ishibashi}}]{Ishibashi2005}%
  \BibitemOpen
  \bibfield  {author} {\bibinfo {author} {\bibfnamefont {S.}~\bibnamefont
  {Moriyama}}, \bibinfo {author} {\bibfnamefont {T.}~\bibnamefont {Fuse}},
  \bibinfo {author} {\bibfnamefont {M.}~\bibnamefont {Suzuki}}, \bibinfo
  {author} {\bibfnamefont {Y.}~\bibnamefont {Aoyagi}}, \ and\ \bibinfo {author}
  {\bibfnamefont {K.}~\bibnamefont {Ishibashi}},\ }\href {\doibase
  10.1103/PhysRevLett.94.186806} {\bibfield  {journal} {\bibinfo  {journal}
  {Phys. Rev. Lett.}\ }\textbf {\bibinfo {volume} {94}},\ \bibinfo {pages}
  {186806} (\bibinfo {year} {2005})}\BibitemShut {NoStop}%
\bibitem [{\citenamefont {Grove-Rasmussen}\ \emph {et~al.}(2012)\citenamefont
  {Grove-Rasmussen}, \citenamefont {Grap}, \citenamefont {Paaske},
  \citenamefont {Flensberg}, \citenamefont {Andergassen}, \citenamefont
  {Meden}, \citenamefont {J\o{}rgensen}, \citenamefont {Muraki},\ and\
  \citenamefont {Fujisawa}}]{Fujisawa2012}%
  \BibitemOpen
  \bibfield  {author} {\bibinfo {author} {\bibfnamefont {K.}~\bibnamefont
  {Grove-Rasmussen}}, \bibinfo {author} {\bibfnamefont {S.}~\bibnamefont
  {Grap}}, \bibinfo {author} {\bibfnamefont {J.}~\bibnamefont {Paaske}},
  \bibinfo {author} {\bibfnamefont {K.}~\bibnamefont {Flensberg}}, \bibinfo
  {author} {\bibfnamefont {S.}~\bibnamefont {Andergassen}}, \bibinfo {author}
  {\bibfnamefont {V.}~\bibnamefont {Meden}}, \bibinfo {author} {\bibfnamefont
  {H.~I.}\ \bibnamefont {J\o{}rgensen}}, \bibinfo {author} {\bibfnamefont
  {K.}~\bibnamefont {Muraki}}, \ and\ \bibinfo {author} {\bibfnamefont
  {T.}~\bibnamefont {Fujisawa}},\ }\href {\doibase
  10.1103/PhysRevLett.108.176802} {\bibfield  {journal} {\bibinfo  {journal}
  {Phys. Rev. Lett.}\ }\textbf {\bibinfo {volume} {108}},\ \bibinfo {pages}
  {176802} (\bibinfo {year} {2012})}\BibitemShut {NoStop}%
\bibitem [{\citenamefont {Koppens}\ \emph {et~al.}(2006)\citenamefont
  {Koppens}, \citenamefont {Buizert}, \citenamefont {Tielrooij}, \citenamefont
  {Vink}, \citenamefont {Nowack}, \citenamefont {Meunier}, \citenamefont
  {Kouwenhoven},\ and\ \citenamefont {Vandersypen}}]{Vander2006}%
  \BibitemOpen
  \bibfield  {author} {\bibinfo {author} {\bibfnamefont {F.~H.~L.}\
  \bibnamefont {Koppens}}, \bibinfo {author} {\bibfnamefont {C.}~\bibnamefont
  {Buizert}}, \bibinfo {author} {\bibfnamefont {K.~J.}\ \bibnamefont
  {Tielrooij}}, \bibinfo {author} {\bibfnamefont {I.~T.}\ \bibnamefont {Vink}},
  \bibinfo {author} {\bibfnamefont {K.~C.}\ \bibnamefont {Nowack}}, \bibinfo
  {author} {\bibfnamefont {T.}~\bibnamefont {Meunier}}, \bibinfo {author}
  {\bibfnamefont {L.~P.}\ \bibnamefont {Kouwenhoven}}, \ and\ \bibinfo {author}
  {\bibfnamefont {L.~M.~K.}\ \bibnamefont {Vandersypen}},\ }\href {\doibase
  10.1038/nature05065
  https://www.nature.com/articles/nature05065#supplementary-information}
  {\bibfield  {journal} {\bibinfo  {journal} {Nature}\ }\textbf {\bibinfo
  {volume} {442}},\ \bibinfo {pages} {766} (\bibinfo {year}
  {2006})}\BibitemShut {NoStop}%
\bibitem [{\citenamefont {Thiele}\ \emph {et~al.}(2014)\citenamefont {Thiele},
  \citenamefont {Balestro}, \citenamefont {Ballou}, \citenamefont {Klyatskaya},
  \citenamefont {Ruben},\ and\ \citenamefont {Wernsdorfer}}]{Wern2014}%
  \BibitemOpen
  \bibfield  {author} {\bibinfo {author} {\bibfnamefont {S.}~\bibnamefont
  {Thiele}}, \bibinfo {author} {\bibfnamefont {F.}~\bibnamefont {Balestro}},
  \bibinfo {author} {\bibfnamefont {R.}~\bibnamefont {Ballou}}, \bibinfo
  {author} {\bibfnamefont {S.}~\bibnamefont {Klyatskaya}}, \bibinfo {author}
  {\bibfnamefont {M.}~\bibnamefont {Ruben}}, \ and\ \bibinfo {author}
  {\bibfnamefont {W.}~\bibnamefont {Wernsdorfer}},\ }\href {\doibase
  10.1126/science.1249802} {\bibfield  {journal} {\bibinfo  {journal}
  {Science}\ }\textbf {\bibinfo {volume} {344}},\ \bibinfo {pages} {1135}
  (\bibinfo {year} {2014})}\BibitemShut {NoStop}%
\bibitem [{\citenamefont {Flensberg}\ and\ \citenamefont
  {Marcus}(2010)}]{Marcus2010}%
  \BibitemOpen
  \bibfield  {author} {\bibinfo {author} {\bibfnamefont {K.}~\bibnamefont
  {Flensberg}}\ and\ \bibinfo {author} {\bibfnamefont {C.~M.}\ \bibnamefont
  {Marcus}},\ }\href {\doibase 10.1103/PhysRevB.81.195418} {\bibfield
  {journal} {\bibinfo  {journal} {Phys. Rev. B}\ }\textbf {\bibinfo {volume}
  {81}},\ \bibinfo {pages} {195418} (\bibinfo {year} {2010})}\BibitemShut
  {NoStop}%
\bibitem [{\citenamefont {Laird}\ \emph {et~al.}(2013)\citenamefont {Laird},
  \citenamefont {Pei},\ and\ \citenamefont {Kouwenhoven}}]{Laird2013}%
  \BibitemOpen
  \bibfield  {author} {\bibinfo {author} {\bibfnamefont {E.~A.}\ \bibnamefont
  {Laird}}, \bibinfo {author} {\bibfnamefont {F.}~\bibnamefont {Pei}}, \ and\
  \bibinfo {author} {\bibfnamefont {L.~P.}\ \bibnamefont {Kouwenhoven}},\
  }\href {\doibase 10.1038/nnano.2013.140
  https://www.nature.com/articles/nnano.2013.140#supplementary-information}
  {\bibfield  {journal} {\bibinfo  {journal} {Nature Nanotechnology}\ }\textbf
  {\bibinfo {volume} {8}},\ \bibinfo {pages} {565} (\bibinfo {year}
  {2013})}\BibitemShut {NoStop}%
\bibitem [{\citenamefont {Koppens}\ \emph {et~al.}(2005)\citenamefont
  {Koppens}, \citenamefont {Folk}, \citenamefont {Elzerman}, \citenamefont
  {Hanson}, \citenamefont {van Beveren}, \citenamefont {Vink}, \citenamefont
  {Tranitz}, \citenamefont {Wegscheider}, \citenamefont {Kouwenhoven},\ and\
  \citenamefont {Vandersypen}}]{Koppens2005}%
  \BibitemOpen
  \bibfield  {author} {\bibinfo {author} {\bibfnamefont {F.~H.~L.}\
  \bibnamefont {Koppens}}, \bibinfo {author} {\bibfnamefont {J.~A.}\
  \bibnamefont {Folk}}, \bibinfo {author} {\bibfnamefont {J.~M.}\ \bibnamefont
  {Elzerman}}, \bibinfo {author} {\bibfnamefont {R.}~\bibnamefont {Hanson}},
  \bibinfo {author} {\bibfnamefont {L.~H.~W.}\ \bibnamefont {van Beveren}},
  \bibinfo {author} {\bibfnamefont {I.~T.}\ \bibnamefont {Vink}}, \bibinfo
  {author} {\bibfnamefont {H.~P.}\ \bibnamefont {Tranitz}}, \bibinfo {author}
  {\bibfnamefont {W.}~\bibnamefont {Wegscheider}}, \bibinfo {author}
  {\bibfnamefont {L.~P.}\ \bibnamefont {Kouwenhoven}}, \ and\ \bibinfo {author}
  {\bibfnamefont {L.~M.~K.}\ \bibnamefont {Vandersypen}},\ }\href {\doibase
  10.1126/science.1113719} {\bibfield  {journal} {\bibinfo  {journal}
  {Science}\ }\textbf {\bibinfo {volume} {309}},\ \bibinfo {pages} {1346}
  (\bibinfo {year} {2005})}\BibitemShut {NoStop}%
\bibitem [{\citenamefont {Jouravlev}\ and\ \citenamefont
  {Nazarov}(2006)}]{Nazarov2006}%
  \BibitemOpen
  \bibfield  {author} {\bibinfo {author} {\bibfnamefont {O.~N.}\ \bibnamefont
  {Jouravlev}}\ and\ \bibinfo {author} {\bibfnamefont {Y.~V.}\ \bibnamefont
  {Nazarov}},\ }\href {\doibase 10.1103/PhysRevLett.96.176804} {\bibfield
  {journal} {\bibinfo  {journal} {Phys. Rev. Lett.}\ }\textbf {\bibinfo
  {volume} {96}},\ \bibinfo {pages} {176804} (\bibinfo {year}
  {2006})}\BibitemShut {NoStop}%
\bibitem [{\citenamefont {Grove-Rasmussen}\ \emph {et~al.}(2008)\citenamefont
  {Grove-Rasmussen}, \citenamefont {J{\o}rgensen}, \citenamefont {Hayashi},
  \citenamefont {Lindelof},\ and\ \citenamefont {Fujisawa}}]{Fujisawa2008}%
  \BibitemOpen
  \bibfield  {author} {\bibinfo {author} {\bibfnamefont {K.}~\bibnamefont
  {Grove-Rasmussen}}, \bibinfo {author} {\bibfnamefont {H.~I.}\ \bibnamefont
  {J{\o}rgensen}}, \bibinfo {author} {\bibfnamefont {T.}~\bibnamefont
  {Hayashi}}, \bibinfo {author} {\bibfnamefont {P.~E.}\ \bibnamefont
  {Lindelof}}, \ and\ \bibinfo {author} {\bibfnamefont {T.}~\bibnamefont
  {Fujisawa}},\ }\href {\doibase 10.1021/nl072948y} {\bibfield  {journal}
  {\bibinfo  {journal} {Nano Letters}\ }\textbf {\bibinfo {volume} {8}},\
  \bibinfo {pages} {1055} (\bibinfo {year} {2008})}\BibitemShut {NoStop}%
\bibitem [{\citenamefont {Lu}(1995)}]{Lu1995}%
  \BibitemOpen
  \bibfield  {author} {\bibinfo {author} {\bibfnamefont {J.~P.}\ \bibnamefont
  {Lu}},\ }\href {\doibase 10.1103/PhysRevLett.74.1123} {\bibfield  {journal}
  {\bibinfo  {journal} {Phys. Rev. Lett.}\ }\textbf {\bibinfo {volume} {74}},\
  \bibinfo {pages} {1123} (\bibinfo {year} {1995})}\BibitemShut {NoStop}%
\bibitem [{\citenamefont {Sz\'echenyi}\ and\ \citenamefont
  {P\'alyi}(2015)}]{Palyi2015}%
  \BibitemOpen
  \bibfield  {author} {\bibinfo {author} {\bibfnamefont {G.}~\bibnamefont
  {Sz\'echenyi}}\ and\ \bibinfo {author} {\bibfnamefont {A.}~\bibnamefont
  {P\'alyi}},\ }\href {\doibase 10.1103/PhysRevB.91.045431} {\bibfield
  {journal} {\bibinfo  {journal} {Phys. Rev. B}\ }\textbf {\bibinfo {volume}
  {91}},\ \bibinfo {pages} {045431} (\bibinfo {year} {2015})}\BibitemShut
  {NoStop}%
\bibitem [{\citenamefont {Sz\'echenyi}\ and\ \citenamefont
  {P\'alyi}(2017)}]{Palyi2017}%
  \BibitemOpen
  \bibfield  {author} {\bibinfo {author} {\bibfnamefont {G.}~\bibnamefont
  {Sz\'echenyi}}\ and\ \bibinfo {author} {\bibfnamefont {A.}~\bibnamefont
  {P\'alyi}},\ }\href {\doibase 10.1103/PhysRevB.95.035431} {\bibfield
  {journal} {\bibinfo  {journal} {Phys. Rev. B}\ }\textbf {\bibinfo {volume}
  {95}},\ \bibinfo {pages} {035431} (\bibinfo {year} {2017})}\BibitemShut
  {NoStop}%
\bibitem [{\citenamefont {Li}\ \emph {et~al.}(2014)\citenamefont {Li},
  \citenamefont {Benjamin}, \citenamefont {Briggs},\ and\ \citenamefont
  {Laird}}]{Laird2014}%
  \BibitemOpen
  \bibfield  {author} {\bibinfo {author} {\bibfnamefont {Y.}~\bibnamefont
  {Li}}, \bibinfo {author} {\bibfnamefont {S.~C.}\ \bibnamefont {Benjamin}},
  \bibinfo {author} {\bibfnamefont {G.~A.~D.}\ \bibnamefont {Briggs}}, \ and\
  \bibinfo {author} {\bibfnamefont {E.~A.}\ \bibnamefont {Laird}},\ }\href
  {\doibase 10.1103/PhysRevB.90.195440} {\bibfield  {journal} {\bibinfo
  {journal} {Phys. Rev. B}\ }\textbf {\bibinfo {volume} {90}},\ \bibinfo
  {pages} {195440} (\bibinfo {year} {2014})}\BibitemShut {NoStop}%
\bibitem [{\citenamefont {Kuemmeth}\ \emph {et~al.}(2008)\citenamefont
  {Kuemmeth}, \citenamefont {Ilani}, \citenamefont {Ralph},\ and\ \citenamefont
  {McEuen}}]{Kuemmeth2008}%
  \BibitemOpen
  \bibfield  {author} {\bibinfo {author} {\bibfnamefont {F.}~\bibnamefont
  {Kuemmeth}}, \bibinfo {author} {\bibfnamefont {S.}~\bibnamefont {Ilani}},
  \bibinfo {author} {\bibfnamefont {D.~C.}\ \bibnamefont {Ralph}}, \ and\
  \bibinfo {author} {\bibfnamefont {P.~L.}\ \bibnamefont {McEuen}},\ }\href
  {\doibase 10.1038/nature06822
  https://www.nature.com/articles/nature06822#supplementary-information}
  {\bibfield  {journal} {\bibinfo  {journal} {Nature}\ }\textbf {\bibinfo
  {volume} {452}},\ \bibinfo {pages} {448} (\bibinfo {year}
  {2008})}\BibitemShut {NoStop}%
\bibitem [{\citenamefont {P\'alyi}\ and\ \citenamefont
  {Burkard}(2010)}]{Burkard2010}%
  \BibitemOpen
  \bibfield  {author} {\bibinfo {author} {\bibfnamefont {A.}~\bibnamefont
  {P\'alyi}}\ and\ \bibinfo {author} {\bibfnamefont {G.}~\bibnamefont
  {Burkard}},\ }\href {\doibase 10.1103/PhysRevB.82.155424} {\bibfield
  {journal} {\bibinfo  {journal} {Phys. Rev. B}\ }\textbf {\bibinfo {volume}
  {82}},\ \bibinfo {pages} {155424} (\bibinfo {year} {2010})}\BibitemShut
  {NoStop}%
\bibitem [{\citenamefont {Hels}\ \emph {et~al.}(2016)\citenamefont {Hels},
  \citenamefont {Braunecker}, \citenamefont {Grove-Rasmussen},\ and\
  \citenamefont {Nyg\aa{}rd}}]{Hels2016}%
  \BibitemOpen
  \bibfield  {author} {\bibinfo {author} {\bibfnamefont {M.~C.}\ \bibnamefont
  {Hels}}, \bibinfo {author} {\bibfnamefont {B.}~\bibnamefont {Braunecker}},
  \bibinfo {author} {\bibfnamefont {K.}~\bibnamefont {Grove-Rasmussen}}, \ and\
  \bibinfo {author} {\bibfnamefont {J.}~\bibnamefont {Nyg\aa{}rd}},\ }\href
  {\doibase 10.1103/PhysRevLett.117.276802} {\bibfield  {journal} {\bibinfo
  {journal} {Phys. Rev. Lett.}\ }\textbf {\bibinfo {volume} {117}},\ \bibinfo
  {pages} {276802} (\bibinfo {year} {2016})}\BibitemShut {NoStop}%
\bibitem [{\citenamefont {Cai}\ \emph {et~al.}(2012)\citenamefont {Cai},
  \citenamefont {Naydenov}, \citenamefont {Pfeiffer}, \citenamefont
  {McGuinness}, \citenamefont {Jahnke}, \citenamefont {Jelezko}, \citenamefont
  {Plenio},\ and\ \citenamefont {Retzker}}]{Cai_NJP_2012}%
  \BibitemOpen
  \bibfield  {author} {\bibinfo {author} {\bibfnamefont {J.-M.}\ \bibnamefont
  {Cai}}, \bibinfo {author} {\bibfnamefont {B.}~\bibnamefont {Naydenov}},
  \bibinfo {author} {\bibfnamefont {R.}~\bibnamefont {Pfeiffer}}, \bibinfo
  {author} {\bibfnamefont {L.~P.}\ \bibnamefont {McGuinness}}, \bibinfo
  {author} {\bibfnamefont {K.~D.}\ \bibnamefont {Jahnke}}, \bibinfo {author}
  {\bibfnamefont {F.}~\bibnamefont {Jelezko}}, \bibinfo {author} {\bibfnamefont
  {M.~B.}\ \bibnamefont {Plenio}}, \ and\ \bibinfo {author} {\bibfnamefont
  {A.}~\bibnamefont {Retzker}},\ }\href
  {http://stacks.iop.org/1367-2630/14/i=11/a=113023} {\bibfield  {journal}
  {\bibinfo  {journal} {New Journal of Physics}\ }\textbf {\bibinfo {volume}
  {14}},\ \bibinfo {pages} {113023} (\bibinfo {year} {2012})}\BibitemShut
  {NoStop}%
\bibitem [{\citenamefont {Hanson}\ \emph {et~al.}(2007)\citenamefont {Hanson},
  \citenamefont {Kouwenhoven}, \citenamefont {Petta}, \citenamefont {Tarucha},\
  and\ \citenamefont {Vandersypen}}]{Hanson2007}%
  \BibitemOpen
  \bibfield  {author} {\bibinfo {author} {\bibfnamefont {R.}~\bibnamefont
  {Hanson}}, \bibinfo {author} {\bibfnamefont {L.~P.}\ \bibnamefont
  {Kouwenhoven}}, \bibinfo {author} {\bibfnamefont {J.~R.}\ \bibnamefont
  {Petta}}, \bibinfo {author} {\bibfnamefont {S.}~\bibnamefont {Tarucha}}, \
  and\ \bibinfo {author} {\bibfnamefont {L.~M.~K.}\ \bibnamefont
  {Vandersypen}},\ }\href {\doibase 10.1103/RevModPhys.79.1217} {\bibfield
  {journal} {\bibinfo  {journal} {Rev. Mod. Phys.}\ }\textbf {\bibinfo {volume}
  {79}},\ \bibinfo {pages} {1217} (\bibinfo {year} {2007})}\BibitemShut
  {NoStop}%
\bibitem [{\citenamefont {Churchill}\ \emph
  {et~al.}(2009{\natexlab{b}})\citenamefont {Churchill}, \citenamefont
  {Kuemmeth}, \citenamefont {Harlow}, \citenamefont {Bestwick}, \citenamefont
  {Rashba}, \citenamefont {Flensberg}, \citenamefont {Stwertka}, \citenamefont
  {Taychatanapat}, \citenamefont {Watson},\ and\ \citenamefont
  {Marcus}}]{Churchill2009}%
  \BibitemOpen
  \bibfield  {author} {\bibinfo {author} {\bibfnamefont {H.~O.~H.}\
  \bibnamefont {Churchill}}, \bibinfo {author} {\bibfnamefont {F.}~\bibnamefont
  {Kuemmeth}}, \bibinfo {author} {\bibfnamefont {J.~W.}\ \bibnamefont
  {Harlow}}, \bibinfo {author} {\bibfnamefont {A.~J.}\ \bibnamefont
  {Bestwick}}, \bibinfo {author} {\bibfnamefont {E.~I.}\ \bibnamefont
  {Rashba}}, \bibinfo {author} {\bibfnamefont {K.}~\bibnamefont {Flensberg}},
  \bibinfo {author} {\bibfnamefont {C.~H.}\ \bibnamefont {Stwertka}}, \bibinfo
  {author} {\bibfnamefont {T.}~\bibnamefont {Taychatanapat}}, \bibinfo {author}
  {\bibfnamefont {S.~K.}\ \bibnamefont {Watson}}, \ and\ \bibinfo {author}
  {\bibfnamefont {C.~M.}\ \bibnamefont {Marcus}},\ }\href {\doibase
  10.1103/PhysRevLett.102.166802} {\bibfield  {journal} {\bibinfo  {journal}
  {Phys. Rev. Lett.}\ }\textbf {\bibinfo {volume} {102}},\ \bibinfo {pages}
  {166802} (\bibinfo {year} {2009}{\natexlab{b}})}\BibitemShut {NoStop}%
\bibitem [{\citenamefont {Bulaev}\ \emph {et~al.}(2008)\citenamefont {Bulaev},
  \citenamefont {Trauzettel},\ and\ \citenamefont {Loss}}]{Bulaev2008}%
  \BibitemOpen
  \bibfield  {author} {\bibinfo {author} {\bibfnamefont {D.~V.}\ \bibnamefont
  {Bulaev}}, \bibinfo {author} {\bibfnamefont {B.}~\bibnamefont {Trauzettel}},
  \ and\ \bibinfo {author} {\bibfnamefont {D.}~\bibnamefont {Loss}},\ }\href
  {\doibase 10.1103/PhysRevB.77.235301} {\bibfield  {journal} {\bibinfo
  {journal} {Phys. Rev. B}\ }\textbf {\bibinfo {volume} {77}},\ \bibinfo
  {pages} {235301} (\bibinfo {year} {2008})}\BibitemShut {NoStop}%
\bibitem [{\citenamefont {Rudner}\ and\ \citenamefont
  {Rashba}(2010)}]{Rudner2010}%
  \BibitemOpen
  \bibfield  {author} {\bibinfo {author} {\bibfnamefont {M.~S.}\ \bibnamefont
  {Rudner}}\ and\ \bibinfo {author} {\bibfnamefont {E.~I.}\ \bibnamefont
  {Rashba}},\ }\href {\doibase 10.1103/PhysRevB.81.125426} {\bibfield
  {journal} {\bibinfo  {journal} {Phys. Rev. B}\ }\textbf {\bibinfo {volume}
  {81}},\ \bibinfo {pages} {125426} (\bibinfo {year} {2010})}\BibitemShut
  {NoStop}%
\bibitem [{\citenamefont {Freeman}\ \emph {et~al.}(2016)\citenamefont
  {Freeman}, \citenamefont {Schoenfield},\ and\ \citenamefont
  {Jiang}}]{Jiang2016}%
  \BibitemOpen
  \bibfield  {author} {\bibinfo {author} {\bibfnamefont {B.~M.}\ \bibnamefont
  {Freeman}}, \bibinfo {author} {\bibfnamefont {J.~S.}\ \bibnamefont
  {Schoenfield}}, \ and\ \bibinfo {author} {\bibfnamefont {H.}~\bibnamefont
  {Jiang}},\ }\href {\doibase 10.1063/1.4954700} {\bibfield  {journal}
  {\bibinfo  {journal} {Applied Physics Letters}\ }\textbf {\bibinfo {volume}
  {108}},\ \bibinfo {pages} {253108} (\bibinfo {year} {2016})}\BibitemShut
  {NoStop}%
\bibitem [{\citenamefont {Coish}\ and\ \citenamefont
  {Qassemi}(2011)}]{Coish_2011_PRB}%
  \BibitemOpen
  \bibfield  {author} {\bibinfo {author} {\bibfnamefont {W.~A.}\ \bibnamefont
  {Coish}}\ and\ \bibinfo {author} {\bibfnamefont {F.}~\bibnamefont
  {Qassemi}},\ }\href {\doibase 10.1103/PhysRevB.84.245407} {\bibfield
  {journal} {\bibinfo  {journal} {Phys. Rev. B}\ }\textbf {\bibinfo {volume}
  {84}},\ \bibinfo {pages} {245407} (\bibinfo {year} {2011})}\BibitemShut
  {NoStop}%
\bibitem [{\citenamefont {Lai}\ \emph {et~al.}(2011)\citenamefont {Lai},
  \citenamefont {Lim}, \citenamefont {Yang}, \citenamefont {Zwanenburg},
  \citenamefont {Coish}, \citenamefont {Qassemi}, \citenamefont {Morello},\
  and\ \citenamefont {Dzurak}}]{Lai_2011_SR1}%
  \BibitemOpen
  \bibfield  {author} {\bibinfo {author} {\bibfnamefont {N.~S.}\ \bibnamefont
  {Lai}}, \bibinfo {author} {\bibfnamefont {W.~H.}\ \bibnamefont {Lim}},
  \bibinfo {author} {\bibfnamefont {C.~H.}\ \bibnamefont {Yang}}, \bibinfo
  {author} {\bibfnamefont {F.~A.}\ \bibnamefont {Zwanenburg}}, \bibinfo
  {author} {\bibfnamefont {W.~A.}\ \bibnamefont {Coish}}, \bibinfo {author}
  {\bibfnamefont {F.}~\bibnamefont {Qassemi}}, \bibinfo {author} {\bibfnamefont
  {A.}~\bibnamefont {Morello}}, \ and\ \bibinfo {author} {\bibfnamefont
  {A.~S.}\ \bibnamefont {Dzurak}},\ }\href {\doibase 10.1038/srep00110}
  {\bibfield  {journal} {\bibinfo  {journal} {Scientific Reports}\ }\textbf
  {\bibinfo {volume} {1}},\ \bibinfo {pages} {110} (\bibinfo {year}
  {2011})}\BibitemShut {NoStop}%
\bibitem [{\citenamefont {Hanson}\ \emph {et~al.}(2004)\citenamefont {Hanson},
  \citenamefont {Vandersypen}, \citenamefont {van Beveren}, \citenamefont
  {Elzerman}, \citenamefont {Vink},\ and\ \citenamefont
  {Kouwenhoven}}]{Hanson2004}%
  \BibitemOpen
  \bibfield  {author} {\bibinfo {author} {\bibfnamefont {R.}~\bibnamefont
  {Hanson}}, \bibinfo {author} {\bibfnamefont {L.~M.~K.}\ \bibnamefont
  {Vandersypen}}, \bibinfo {author} {\bibfnamefont {L.~H.~W.}\ \bibnamefont
  {van Beveren}}, \bibinfo {author} {\bibfnamefont {J.~M.}\ \bibnamefont
  {Elzerman}}, \bibinfo {author} {\bibfnamefont {I.~T.}\ \bibnamefont {Vink}},
  \ and\ \bibinfo {author} {\bibfnamefont {L.~P.}\ \bibnamefont
  {Kouwenhoven}},\ }\href {\doibase 10.1103/PhysRevB.70.241304} {\bibfield
  {journal} {\bibinfo  {journal} {Phys. Rev. B}\ }\textbf {\bibinfo {volume}
  {70}},\ \bibinfo {pages} {241304} (\bibinfo {year} {2004})}\BibitemShut
  {NoStop}%
\bibitem [{\citenamefont {Tu}\ and\ \citenamefont {Zheng}(2008)}]{FCN1}%
  \BibitemOpen
  \bibfield  {author} {\bibinfo {author} {\bibfnamefont {X.}~\bibnamefont
  {Tu}}\ and\ \bibinfo {author} {\bibfnamefont {M.}~\bibnamefont {Zheng}},\
  }\href {\doibase 10.1007/s12274-008-8022-7} {\bibfield  {journal} {\bibinfo
  {journal} {Nano Research}\ }\textbf {\bibinfo {volume} {1}},\ \bibinfo
  {pages} {185} (\bibinfo {year} {2008})}\BibitemShut {NoStop}%
\bibitem [{\citenamefont {Moon}\ \emph {et~al.}(2008)\citenamefont {Moon},
  \citenamefont {Chang}, \citenamefont {Lee},\ and\ \citenamefont
  {Choi}}]{FCN2}%
  \BibitemOpen
  \bibfield  {author} {\bibinfo {author} {\bibfnamefont {H.~K.}\ \bibnamefont
  {Moon}}, \bibinfo {author} {\bibfnamefont {C.~I.}\ \bibnamefont {Chang}},
  \bibinfo {author} {\bibfnamefont {D.-K.}\ \bibnamefont {Lee}}, \ and\
  \bibinfo {author} {\bibfnamefont {H.~C.}\ \bibnamefont {Choi}},\ }\href
  {\doibase 10.1007/s12274-008-8038-z} {\bibfield  {journal} {\bibinfo
  {journal} {Nano Research}\ }\textbf {\bibinfo {volume} {1}},\ \bibinfo
  {pages} {351} (\bibinfo {year} {2008})}\BibitemShut {NoStop}%
\bibitem [{\citenamefont {Gurvitz}\ and\ \citenamefont
  {Prager}(1996)}]{Gurvitz1996}%
  \BibitemOpen
  \bibfield  {author} {\bibinfo {author} {\bibfnamefont {S.~A.}\ \bibnamefont
  {Gurvitz}}\ and\ \bibinfo {author} {\bibfnamefont {Y.~S.}\ \bibnamefont
  {Prager}},\ }\href {\doibase 10.1103/PhysRevB.53.15932} {\bibfield  {journal}
  {\bibinfo  {journal} {Phys. Rev. B}\ }\textbf {\bibinfo {volume} {53}},\
  \bibinfo {pages} {15932} (\bibinfo {year} {1996})}\BibitemShut {NoStop}%
\bibitem [{\citenamefont {Li}\ \emph {et~al.}(2005)\citenamefont {Li},
  \citenamefont {Luo}, \citenamefont {Yang}, \citenamefont {Cui},\ and\
  \citenamefont {Yan}}]{Li2005}%
  \BibitemOpen
  \bibfield  {author} {\bibinfo {author} {\bibfnamefont {X.-Q.}\ \bibnamefont
  {Li}}, \bibinfo {author} {\bibfnamefont {J.}~\bibnamefont {Luo}}, \bibinfo
  {author} {\bibfnamefont {Y.-G.}\ \bibnamefont {Yang}}, \bibinfo {author}
  {\bibfnamefont {P.}~\bibnamefont {Cui}}, \ and\ \bibinfo {author}
  {\bibfnamefont {Y.}~\bibnamefont {Yan}},\ }\href {\doibase
  10.1103/PhysRevB.71.205304} {\bibfield  {journal} {\bibinfo  {journal} {Phys.
  Rev. B}\ }\textbf {\bibinfo {volume} {71}},\ \bibinfo {pages} {205304}
  (\bibinfo {year} {2005})}\BibitemShut {NoStop}%
\end{thebibliography}

%

\end{document}